\documentclass[11pt,a4paper]{article}
\usepackage{longtable}
\usepackage{amsmath}
\usepackage{amssymb}
\usepackage{graphicx}
\usepackage{bm}% bold math
\usepackage{footmisc}
\usepackage{mathtools}
\usepackage[section]{placeins}

\newcommand*{\bea}{\begin{eqnarray}}
\newcommand*{\eea}{\end{eqnarray}}
\newcommand*{\be}{\begin{equation}}
\newcommand*{\ee}{\end{equation}}

\newcommand{\bma}{\begin{pmatrix}}
\newcommand{\ema}{\end{pmatrix}}

\usepackage[square,comma,sort&compress,numbers]{natbib}

\title{Exploring Higgs boson Yukawa interactions with a scalar singlet using Dyson-Schwinger equations}
\author{Tajdar Mufti \footnote{tajdar.mufti@gmail.com, tajdar.mufti@lums.edu.pk} \\ Lahore University of Management Sciences\\ Sector U, D.H.A, Lahore Cantt, 54792, Pakistan}
\date{}
\begin{document}
\title{Exploring Higgs boson Yukawa interactions with a scalar singlet using Dyson Schwinger equations}
% \title{Dyson Schwinger equations for Higgs doublet scalar singlet Yukawa interactions}
\author{Tajdar Mufti \footnote{tajdar.mufti@gmail.com, tajdar.mufti@lums.edu.pk} \\ Lahore University of Management Sciences\\ Sector U, D.H.A, Lahore Cantt, 54792, Pakistan}
\maketitle
% **********************************************************
% **********************************************************
% **********************************************************
% ************************ Abstract ************************
% **********************************************************
% **********************************************************
% **********************************************************
\begin{abstract}
Higgs, being the first discovery of a fundamental scalar field in the standard model (SM), opens the possibility of existence of other scalar or pseudo scalar particles in nature. Though it does not conclusively fulfill the role of inflaton, it does provide motivation for experimental searches involving interactions of scalar (or pseudo scalar) particles as well as implications of such particles' existence from the perspective of theoretical understanding. Yukawa interactions are among the possible interactions. The considered model addresses Yukawa interaction among the Higgs (and Higgs bar) \footnote{Higgs bar is referred to $h^{\dagger}$, with h being the Higgs field, throughout this paper.} and a real singlet scalar field using Dyson Schwinger equations (DSEs) over a range of bare coupling values relevant to inflation-related physics for several scalar bare masses and cutoff values. The Higgs mass is kept fixed at two values for two different scenarios. It is found that Higgs propagators are tree level dominated while scalar propagators, despite their dependence on parameter region, do not degenerate over the whole parameter space. The interaction vertices show significant deviations from tree level expression for a number of parameters and exhibit two distinct behaviors. Furthermore, the theory does not show any conclusive sign of triviality over the explored parameter space, despite suppressed vertices for some parameters, particularly for small coupling values in a certain region of parameter space.
\end{abstract}
% **************************************************************
% **************************************************************
% **************************************************************
% ************************ Introduction ************************
% **************************************************************
% **************************************************************
% **************************************************************
\section{Introduction} \label{sec:intr}
Higgs \cite{pdg,Maas:2017wzi,Carena:2002es} sector has taken the status of a cornerstone in the standard model \cite{Schwartz:2013pla,Kaku:1993ym,Barnett:1996yz} after Higgs' experimental finding \cite{Aad:2012tfa,Chatrchyan:2012xdj}. Its very existence was indeed expected since it renders several particles in the SM, electroweak interaction bosons in particular, massive in a renormalizable manner \cite{Schwartz:2013pla,Kaku:1993ym}. However, implications of its existence may very well reach far beyond low energy phenomenology of the standard model \cite{Kaku:1993ym}. Supersymmetry \cite{Martin:1997ns} and cosmology \cite{Mukhanov:2005sc,Guth:1980zm} are two of these examples. It also opens a possibility of existence of a whole scalar sector to be experimentally discovered and studied from the perspectives of both theory and phenomenology.
\par
In supersymmetry \cite{Martin:1997ns}, implications of scalar fields as fundamental degrees of freedom are beyond a mass generating mechanism which was indeed one of the reasons for frantic SM Higgs-related searches over past few decades. A plethora of reports, such as \cite{Ellis:2000ig}, suggest a number of scalars providing various extensions to the already known SM \cite{Ellis:2007wa}, and even to cosmological scenarios \cite{Jungman:1995df}. In particular, in minimal supersymmetric models Higgs boson is expected to be smaller than 135 GeV \cite{Haber:1990aw,Ellis:1990nz,Okada:1990vk,Carena:2002es,Ellis:1992cp} which turns out to be the correct prediction \cite{Aad:2012tfa,Chatrchyan:2012xdj}. However, as experimental searches \cite{Aaboud:2018doq,Aaboud:2017leg} are yet to confirm their existence, the status of supersymmetry yet remains inconclusive.
\par
In cosmology, Higgs was suspected to cause inflation \cite{Starobinsky:1980te,Sato:1980yn,Guth:1980zm}, which revived interest in Higgs related cosmology \cite{Enckell:2018kkc,Bezrukov:2013fka} in recent years. However, the quartic self interaction coupling of Higgs is found \cite{Ferreira:2017ynu} significantly different than what was expected to produce slow rolling during cosmic inflation \cite{Hakim:1984oya,Mukhanov:2005sc,Ferreira:2017ynu}. Hence, it is reasonable to expect at least one more scalar field required by current understanding of inflation \cite{Linde:2005ht}. As the experimental discovery \cite{Aad:2012tfa,Chatrchyan:2012xdj} places Higgs as the only fundamental scalar field in the standard model, it presents an opportunity to explore interactions among scalar fields using experimentally known results. There have also been several studies of Higgs interactions from other perspectives, see for example \cite{Lahanas:1998wf,Zarikas:1995qb,Aliferis:2014ofa}.
\par
Similarly, interactions of a scalar singlet field have also been studied at length from various perspectives \cite{Sauli:2002qa}, particularly dark matter physics \cite{Hasenfratz:1988kr,Gliozzi:1997ve,Bertolami:2016ywc,Bento:2000ah,Burgess:2000yq,Bento:2001yk}. Though, the $\phi^{4}$ theory is found to be a trivial theory \cite{Jora:2015yga,Beg:1984yh,Aizenman:1981zz,Dashen:1983ts,Wolff:2009ke,Weisz:2010xx,Siefert:2014ela,Hogervorst:2011zw}, it has not been yet established if its interactions with other fields also render the theories trivial. For the case of Higgs, which is also a (complex doublet) scalar field, its interactions with gauge fields \cite{Maas:2013aia} have not shown any conclusive sign of triviality. It supports assuming scalar interactions involving Higgs non-trivial unless proven otherwise for a model.
\par
In this paper Yukawa interaction between Higgs and a real scalar singlet is studied using Dyson Schwinger equations \footnote{Throughout this paper, the Higgs is referred to the doublet complex scalar field, and scalar field is reserved for scalar singlet field.} \cite{Dyson:1949bp,Schwinger:1951ex,Schwinger:1951hq,Swanson:2010pw,Rivers:1987hi}. The study is conducted under the paradigm of quantum field theory with flat background and coupling values of the order relevant to physics related to inflation \cite{Ferreira:2017ynu}. The theory is explored in terms of propagators, vertices, and cutoff effects. Higgs (renormalized) masses are set at $125.09$ GeV \cite{Aaboud:2018xdt,Aad:2012tfa,Chatrchyan:2012xdj,MalbertionbehalfoftheCMSCollaboration:2018eqs}, and 160 GeV \cite{Gies:2017zwf}. The theory is studied with different scalar (bare) masses from electroweak to TeV regime \footnote{Throughout the paper, electroweak regime is taken as $m_{s} \leq $ 600 GeV, while TeV regime refers to $m_{s} > $ 600 GeV.}. There is a companion paper which takes into account other renormalizable interaction vertices \cite{tajdar:2018lat1} covering a larger parameter space \footnote{This study is being conducted in the presence of four point interactions, hence containing the current parameter space as a sub space.}, though using a different numerical approach \cite{Ruthe:2008rut}, and a paper which addresses phenomenology and further generalized results of the same theory \cite{tajdar:2018yukl2d}.
\par
At this point, the theory does not contain any four point self interactions in the Lagrangian. As the Yukawa interaction term in the theory can also produce 4 point self interactions for both Higgs and scalar fields, these self interactions are not included in the Lagrangian in favor of three point Yukawa interaction. Inclusion of these four point self interactions are considered somewhere else \cite{tajdar:2018lat1}.
\par
The model considered here is a variant of Wick-Cutkosky model \cite{Darewych:1998mb}, investigated using the approach of DSEs \cite{Swanson:2010pw,Roberts:1994dr}, which has been studied in different contexts \cite{Nugaev:2016uqd,Darewych:2009wk,Weber:2000dp}. A significant part of such studies takes into account real, massive as well as massless, singlet scalar self interacting fields via a three point interaction vertex \cite{Sauli:2002qa}. Two scalar fields under the same model have also been studied under the same model \cite{Efimov:2003hs}. However, most of these studies have various kinds of assumptions used to either study different aspects or applications of the model or solve the theory exactly. This paper presents results from the study which involve no additional assumptions in numerically extracting the correlation functions which are to be used to calculate further quantities related to Higgs phenomenology \cite{tajdar:2018yukl2d}. The only prominent constraints are renormalization conditions on the field propagators, which are among the typical features of, particularly phenomenology related, quantum field theories, and a locally implemented condition to suppress local numerical fluctuations and extracting stable correlation functions, as further mentioned below.
\section{Technical Details} \label{sec:tech}
The theory is explored in Euclidean space. The (bare form of) Lagrangian is given by
\begin{equation} \label{Lagrangian}
 L= \delta^{\mu \nu} \partial_{\mu} h^{\dagger} \partial_{\nu} h + m_{h}^{2} h^{\dagger} h + \frac{1}{2} \delta^{\mu \nu} \partial_{\mu} \phi \partial_{\nu} \phi + \frac{1}{2} m_{s}^{2} \phi^{2}+ \lambda  \Pi \phi h^{\dagger} h
\end{equation}
with Higgs fields (h) with SU(2) symmetry and $\phi$ a real singlet scalar field. $\Pi$ is a dimensionful parameter used to render $\lambda$ dimensionless. For the current investigation $\Pi$ is set to 1 GeV. Since $\Pi$ here is a non-dynamic parameter, $\lambda$ is used throughout this paper instead of the coupling $\lambda_{e}$ $(=\lambda \Pi)$ \footnote{From this point, the word \textit{coupling} is reserved for $\lambda$ throughout the current report.}.
\par
Dyson Schwinger equations (in unrenormalized form) for propagators, $H^{ij}(p)$ for Higgs and $S(p)$ for scalar singlet fields, respectively, in momentum space are given by
\begin{equation} \label{hpr1:dse}
 H^{ij}(p)^{-1} = \delta^{ij} (\ p^{2} + m_{h}^{2} )\  + 2 \lambda_{e} \int \frac{d^{4}q}{(2\pi)^{4}} S(q) \Gamma^{ik}(-p,p-q,q) H^{kj}(q-p)
\end{equation}
\begin{equation} \label{spr1:dse}
 S(p) ^{-1}  = p^{2} + m_{s}^{2} + \lambda_{e}  \int \frac{d^{4}q}{(2\pi)^{4}} H^{ik}(q) \Gamma^{kl}(q,p-q,-p) H^{li}(q-p)
\end{equation}
with $\Gamma^{kl}(u,v,w)$ being the three point Yukawa interaction vertex of Higgs, Higgs bar, and scalar fields with momentum u, v, and w, respectively. Higgs and Higgs bar fields have indices k and $\l$, respectively. The renormalization conditions for the propagators \cite{Roberts:1994dr} are
\begin{equation} \label{hpr:ren_condition}
H^{ij}(p)  |_{p^{2}=m_{h}^{2}}= \frac{\delta ^{ij}}{p^{2}+m_{h}^{2}} |_{p^{2}=m_{h}^{2}}
\end{equation}
\begin{equation} \label{spr:ren_condition}
S(p)  |_{p^{2}=m_{s}^{2}}= \frac{1}{p^{2}+m_{s}^{2}} |_{p^{2}=m_{s}^{2}}
\end{equation}
Details of renormalization procedure is given later in the subsection. The starting expressions of correlation functions for numerical computations are set to their tree level expressions. Newton Raphson's method is implemented during each update of either propagators or the vertex. Higgs propagators are also updated using the same method while scalar propagators are calculated directly from the respective DSE under the implemented renormalization procedure. Hence, under the boundary conditions the propagators are updated or calculated, and the vertex is updated to numerically evolve towards a solution such that both equations are satisfied within the preselected size of local error. Uniqueness of solutions is implicitly assumed.
\par
As there are three unknown correlation functions, a commonly used approach is to use a third DSE for the interaction vertex. However, it introduces (most likely) unknown, further higher correlation function(s) depending upon the theory. This never ending sequence is coped with truncations and assumptions \cite{Roberts:1994dr,Swanson:2010pw}. Several approaches, such as ladder approximation, are commonly used for studies involving DSEs. Some of these approaches may still be numerically suitable and economical, as the unknown higher correlation functions are replaced by suitable assumptions, such as modeling them in terms of other lower correlation functions. However, the resulting correlation functions may be effected by such truncations and modeling. On the contrary, the current investigation involves an approach to extract propagators and vertices by introducing renormalization conditions on propagators, symmetry properties, and certain measures to stabilize the vertex. Hence, from two non-linear coupled integral equations three unknown correlation functions are extracted without resorting to conventionally employed means stated above, and only the interplay of the quantities within the theory is made the most of.
\par
Stability of vertex is ensured by three additional measures. Firstly, it was required that local fluctuations never exceed by an order of magnitude in 4-momentum space of the corresponding (Euclidean) space-time, i.e. at every point in 4-momentum space the value of the vertex never exceeds by an order of magnitude relative to its immediate neighborhood \footnote{A detailed study of such constraint to study stability of vertex is to be reported elsewhere.}. The constraint is somewhat similar to Lipschitz condition, abundantly employed in literature for various studies \cite{Gupal1980,Gu2001,Aronsson1967,Zevin2011,Ezquerro2016}, applied locally to bound the local fluctuations of the vertex which is particularly needed in the absence of a DSE for vertex. As the tree level structure of vertex is kept as the starting point of each computation, during every iteration the vertex numerically departs from its tree level expression within the preset bound to avoid unnecessary fluctuations. In addition, updates of correlation functions is based on sum of the squared errors in the DSE of Higgs propagator, instead of local error \cite{Swanson:2010pw}. It slows the computation but contributes in extracting a numerically stable vertex. Furthermore, for the Higgs propagators the following polynomial representation is introduced.
\begin{equation} \label{hpr:coeff1}
 H^{ij}(p)= \delta^{ij} \frac{1}{c(p^{2}+m^{2}+ f(p)) }
\end{equation}
with 
\begin{equation} \label{hpr:func1}
f(p)=\frac{\displaystyle \sum_{i=0}^{N} a_{i} p^{2i}}{\displaystyle \sum_{l=0}^{N}b_{l}p^{2l}}
\end{equation}
The parameters $a_{i}$, $b_{l}$, and $c$ are updated numerically during computations. The advantage is that, while allowing freedom to Higgs propagators within the renormalization scheme and symmetries of the two DSEs, it further contributes in removing numerical fluctuations in the Higgs self energy terms and, hence, contributes in stabilizing the vertex. It is also time efficient in comparison to other numerical approaches, such as interpolation. However, no representation is used for either scalar propagator or the vertex as it was found to severely effect the numerical precision and time efficiency. Thus, starting with the tree level structures, above mentioned measures immensely facilitate during computation of the correlation functions in the theory.
\par
The parameters for computations have been chosen to either explore the dynamics in the theory in both perturbative and non-perturbative regimes in the context of studies of richer theories \cite{tajdar:2018lat1,tajdar:20182hdm1}, or their relevance to phenomenology. Higgs masses have been chosen with inspiration from the discovery of Higgs \cite{Aad:2012tfa,Chatrchyan:2012xdj} as well as studies of heavy Higgs scenarios \cite{Maas:2013aia,Maas:2014pba}. In total, 144 sets of parameters are studied with $10^{-3} \leq  \lambda \leq 2.0 $ and $1.0$ GeV $\leq m_{s} \leq 2.5$ TeV, for $m_{h}=125.09$ GeV and $m_{h}=160$ GeV.
\par
The order of coupling values have been chosen regarding the quartic coupling value for inflationary scenarios involving Higgs \cite{Ferreira:2017ynu}. As Feynman's box diagrams containing only three point Yukawa interaction can also represent four point self interactions for both Higgs and scalar fields, it is assumed that fourth power of Yukawa bare coupling can naively represent a bare 4 point self interaction coupling, and the above mentioned range of bare coupling values becomes natural choice for such explorations.
\par
The theory was investigated with 5 TeV, 40 TeV, 60 TeV, and 100 TeV as cutoff values in order to explore cutoff effects.
\par
The computations are performed to achieve correlation functions with less than $10^{-20}$ local uncertainties in results. All results are for the vertices formed by Higgs and scalar fields with momenta perpendicular to each other, while the other Higgs (bar) field carries the momentum obeying standard conservation principles. Gaussian quadrature algorithm is used for numerical integration.
\subsection{Renormalization Procedure}
The renormalization procedure implemented \cite{Das:2008zze} in the equation of scalar propagator is as follows: Upon adding the counter terms, the Lagrangian takes the following form.
\begin{equation} \label{renLagrangian}
\begin{split}
L & = (1+\alpha) \delta^{\mu \nu} \partial_{\mu} h^{\dagger} \partial_{\nu} h + (m_{h}^{2}+\beta) h^{\dagger} h + \frac{1}{2} (1+A) \delta^{\mu \nu} \partial_{\mu} \phi \partial_{\nu} \phi + \frac{1}{2} (1+B) m_{s}^{2} \phi^{2} \\ & + \lambda(1 + \frac{C}{\lambda})  \Pi \phi h^{\dagger} h
\end{split}
\end{equation}
Adding the counter terms in the Lagrangian in equation \ref{Lagrangian}, equation \ref{spr1:dse} becomes
\begin{equation} \label{spr2:dse}
 S_{r}(p) ^{-1}  = (1+A) p^{2} + m_{s}^{2} + B + \lambda  \int \frac{d^{4}q}{(2\pi)^{4}} H_{r}^{ik}(q) \Gamma_{r}^{kl}(q,p-q,-p) H_{r}^{li}(q-p) )\
\end{equation}
where A, and B are real constants, r is the subscript for renormalized terms, and
\begin{equation} \label{vtx:dressing1}
\Gamma_{r}^{kl}(q,p-q,-p)= (1+\frac{C}{\lambda}) \Gamma^{kl}(q,p-q,-p)
\end{equation}
with C also a real constant \footnote{There is no \textit{running} coupling in this case because the contributions in coupling beyond the bare value are absorbed in the definition of the vertex.}. Similarly, for the case of the Higgs propagator, equation \ref{hpr1:dse} takes the form
\begin{equation} \label{hpr2:dse}
 H^{ij}(p)^{-1} = \delta^{ij}( (1+\alpha) \ p^{2} + m_{h}^{2} + \beta )  + 2 \lambda \int \frac{d^{4}q}{(2\pi)^{4}} S(q) \Gamma_{r}^{ik}(-p,p-q,q) H^{kj}(q-p)
\end{equation}
with $\alpha$ and $\beta$ as the constants due to the counter terms. Without imposing any strong constraints, parameters B and $\beta$ \footnote{The definition of $\beta$ is included for the sake of completion. For the renormalized Higgs mass fixed at its experimentally known value, which is the case here, the definition of $\beta$ do not need to be implemented in the algorithms.} are defined as,
\begin{equation} \label{Bpara:dse}
\begin{split}
 B=A m_{s}^{2} + 2 \lambda (1+A) (1+\alpha) \sigma_{s} \\
 \beta=\alpha m_{h}^{2} + 2 \lambda (1+A) (1+\alpha) \sigma_{h}
\end{split}
\end{equation}
with $\sigma_{s}$ and $\sigma_{h}$ as the parameters to be determined during computations. $\Gamma_{r}^{ab} (p,-p-q,q)$ are also defined as
\begin{equation} \label{vtx:def}
 \Gamma_{r}^{ab} (p,-p-q,q) = (1+A)(1+\alpha)\tilde{\Gamma}_{r}^{ab} (p,-p-q,q)
\end{equation}
In addition, as mass of the Higgs boson is already known, the renormalized mass of the Higgs boson ($m^{2}_{h,r}$) is fixed at $125.09$ GeV, and $160.0$ GeV as a test case of heavy Higgs \footnote{Since Higgs mass is kept fixed at its renormalized value, notations $m_{h}$ and $m_{h,r}$ are used interchangeably as they do not cause any confusion.}. Beside studying the model from the perspective of phenomenology, an advantage is reduction in the unknown terms emerging during renormalization procedure \cite{Das:2008zze}. The renormalized (squared) Higgs mass is related to its bare (squared) mass by the equation given below.
\begin{equation} \label{hms2:def1}
 m_{h,r}^{2}= m_{h}^{2} + \beta = m_{h}^{2} + \alpha m_{h}^{2} + 2 \lambda (1+A) (1+\alpha) \sigma_{h}
\end{equation}
As the Higgs renormalized mass is fixed, $\sigma_{h}$ does not need to be explicitly calculated. \footnote{If required, it could be directly calculated using \ref{hms2:def1} for each value of the parameters $\alpha$ and $A$.}.
\par
Hence, the DSEs to be used for computations are
\begin{equation} \label{spr3:dse}
 S_{r}(p) ^{-1}  = (1+A) (\ p^{2} + m_{s}^{2} + 2 \lambda (1+\alpha) \sigma_{s} + (1+\alpha) I_{s}(p))\
\end{equation}
and
\begin{equation} \label{hpr3:dse}
 H_{r}^{ij}(p)^{-1} = (1+\alpha) ( \delta^{ij}( \ p^{2} + \frac{m_{h,r}^{2}}{1+\alpha} )  + (1+A) I^{ij}_{h}(p) )
\end{equation}
where
\begin{equation} \label{ssen:dse}
I_{s}(p)=\lambda \int \frac{d^{4}q}{(2\pi)^{4}} H_{r}^{ik}(q) \tilde{\Gamma}_{r}^{kl}(q,p-q,-p) H_{r}^{li}(q-p)
\end{equation}
\begin{equation} \label{hsen:dse}
I^{ij}_{h}(p)=2 \lambda \int \frac{d^{4}q}{(2\pi)^{4}} S_{r}(q) \tilde{\Gamma}_{r}^{ik}(-p,p-q,q) H_{r}^{kj}(q-p)
\end{equation}
Only the flavor diagonal Higgs propagators are assumed to be non-zero, as is the case for the tree level Higgs propagators.
\par
During each iteration, the sequence of updates is as follows: First, the parameter $\sigma_{s}$ is updated using Newton Raphson method, as mentioned before. It is followed by that of  parameters in equation \ref{hpr:coeff1} which also provides the parameter $\alpha$. It is followed by updating of the vertex at every momentum point. During each update of Higgs propagator and vertex, scalar propagator and its term $A$ is determined before calculating the sum of squared errors.
\par
The correlation functions are found to be stable under change of ordering among updates of Higgs propagator and the vertex up to the set precision. It was also found that slight deviations from the tree level structures, which were used as the starting forms of correlation functions, do not effect the results, favoring the robustness of the approach. As the correlation functions do exist at their tree level, it is assumed throughout the study that correlation functions exist in the explored parameter space of the theory.
% ******************************************************************************
% ******************************************************************************
% ******************************************************************************
% ************************ Higgs and scalar Propagators ************************
% ******************************************************************************
% ******************************************************************************
% ******************************************************************************
\section{Propagators} \label{sec:prop}
\subsection{Scalar Propagators} \label{sec:sprs}
\begin{figure}
\centering
\parbox{1.0\linewidth}
{
\includegraphics[width=\linewidth]{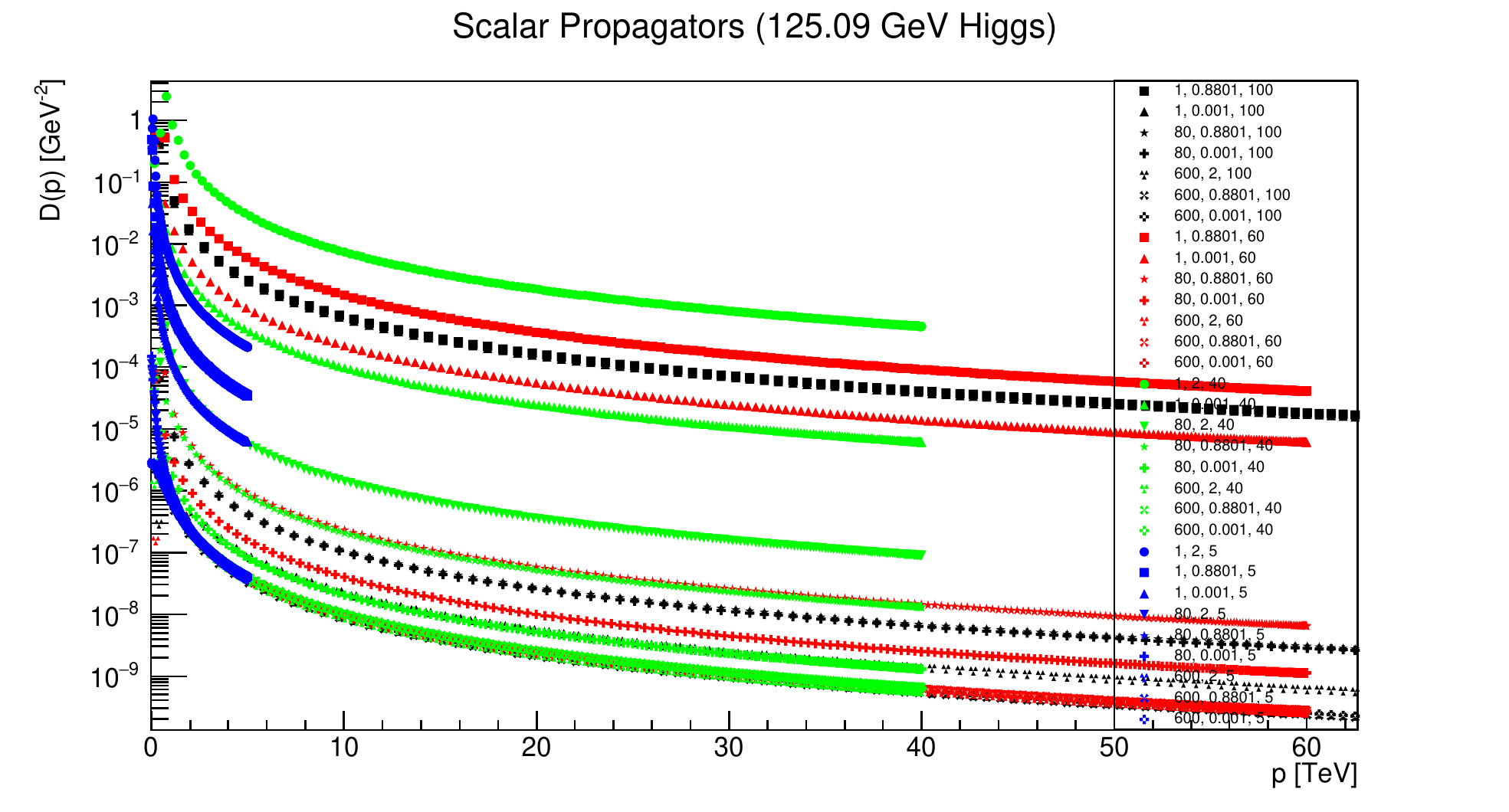}
\caption{\label{fig:sprs125ewlog} Scalar propagators (on logarithmic scale) for $m_{h}=125.09$ GeV are plotted (in electroweak regime) with 1 GeV $ \leq m_{s} \leq $ 600 GeV, 0.001 $\leq \lambda \leq$ 2.0, and 5 TeV $ \leq \Lambda \leq $ 60 TeV, shown as $(m_{s},\lambda,\Lambda)$ on the figure.}
}
\centering
\parbox{1.0\linewidth}
{
\includegraphics[width=\linewidth]{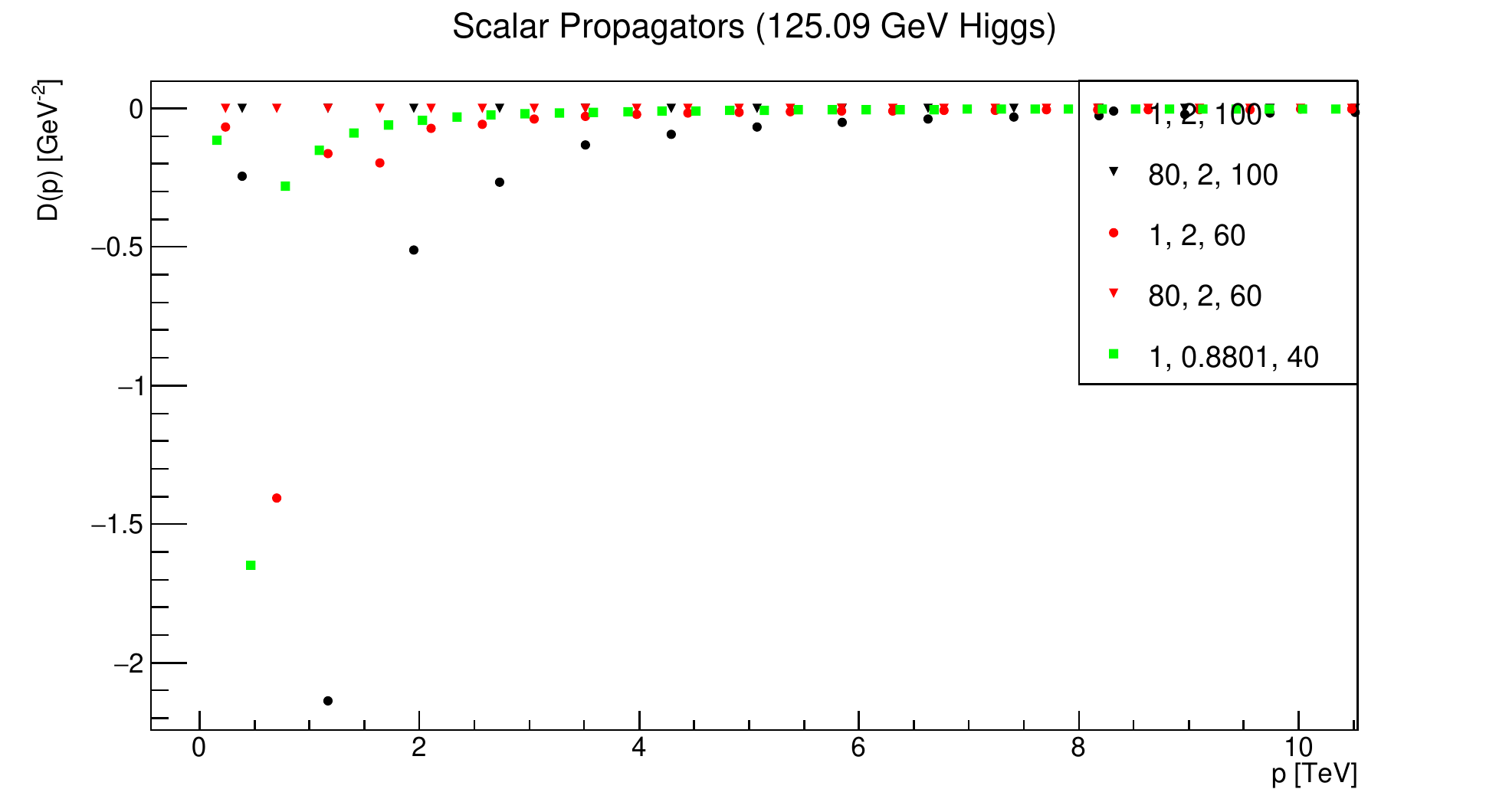}
\caption{\label{fig:sprs125ew1} Scalar propagators for $m_{h}=125.09$ GeV are plotted (in electroweak regime) with 1 GeV $ \leq m_{s} \leq $ 600 GeV, 0.001 $\leq \lambda \leq$ 2.0, and 5 TeV $ \leq \Lambda \leq $ 60 TeV, shown as $(m_{s},\lambda,\Lambda)$ on the figure.}
}
\end{figure}
\begin{figure}
\centering
\parbox{1.0\linewidth}
{
\includegraphics[width=\linewidth]{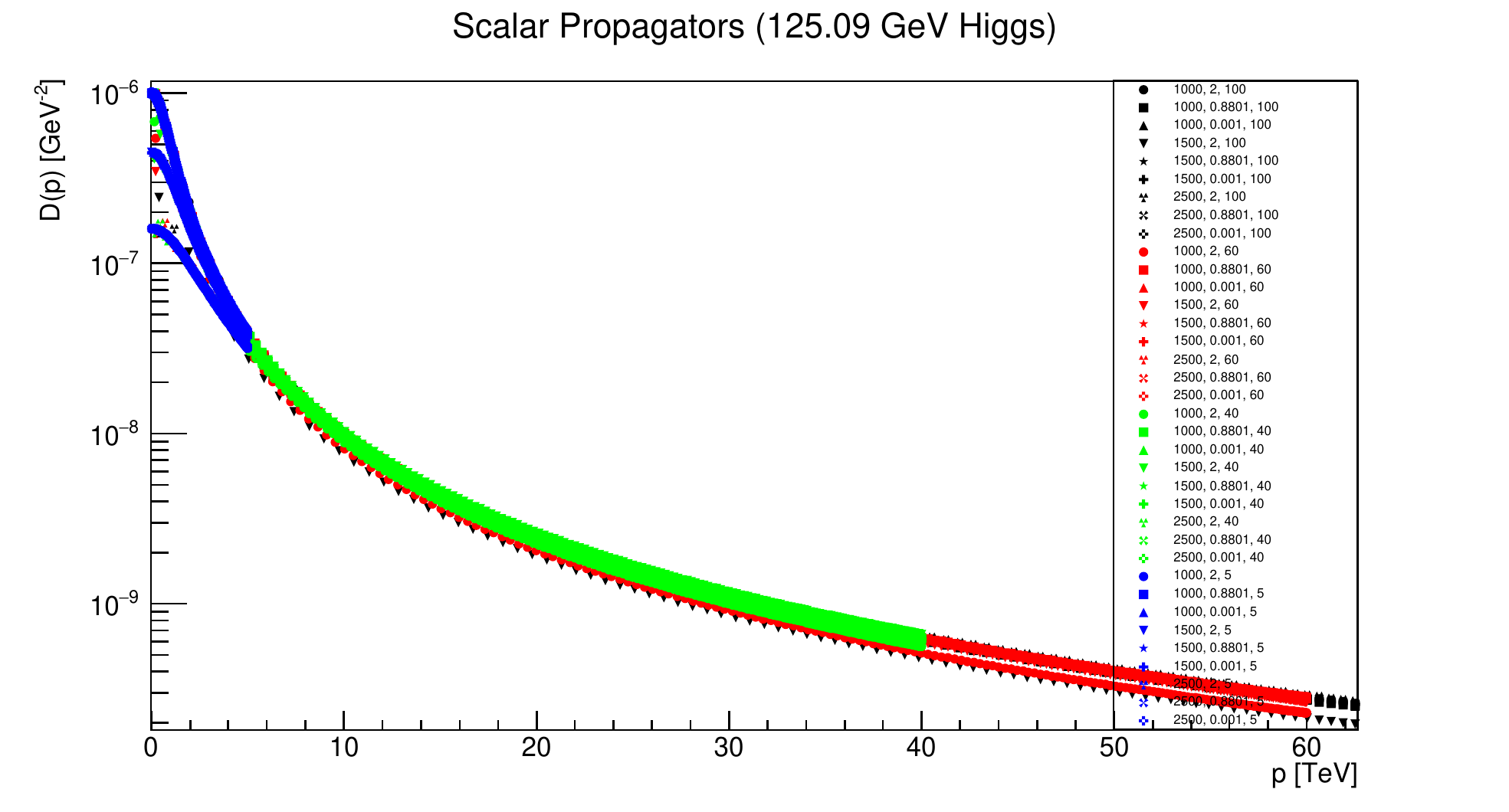}
\caption{\label{fig:sprs125tevlog} Scalar propagators (on logarithmic scale)  for $m_{h}=125.09$ GeV are plotted (in TeV regime) with 1000 GeV $ \leq m_{s} \leq $ 2500 GeV, 0.001 $\leq \lambda \leq$ 2.0, and 5 TeV $ \leq \Lambda \leq $ 60 TeV, shown as $(m_{s},\lambda,\Lambda)$ on the figure.}
}
\end{figure}
\begin{figure}
\centering
\parbox{1.0\linewidth}
{
\includegraphics[width=\linewidth]{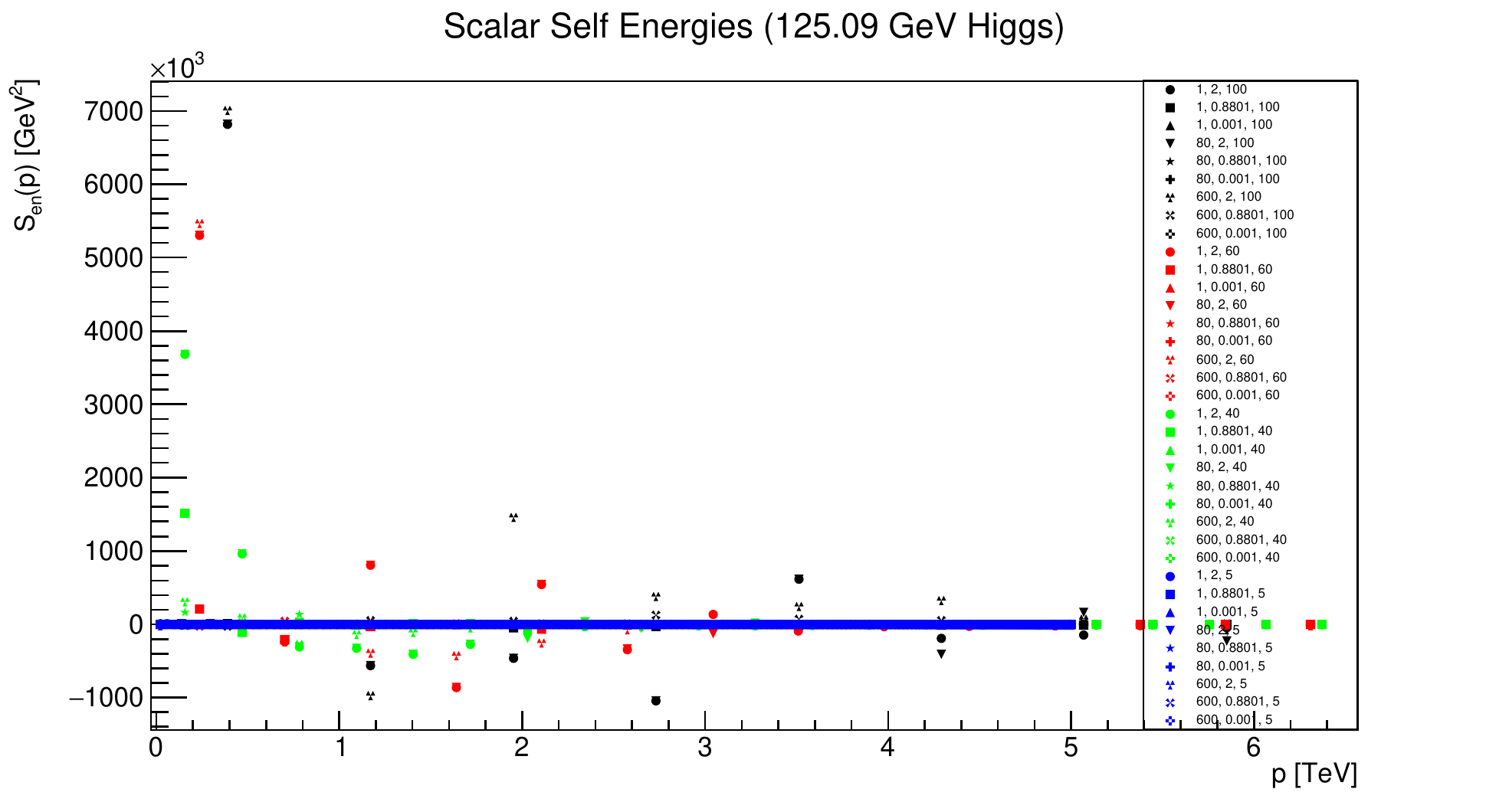}
\caption{\label{fig:ssen125ew} Scalar self energies for $m_{h}=125.09$ GeV are plotted (in electroweak regime) with 1 GeV $ \leq m_{s} \leq $ 600 GeV, 0.001 $\leq \lambda \leq$ 2.0, and 5 TeV $ \leq \Lambda \leq $ 60 TeV, shown as $(m_{s},\lambda,\Lambda)$ on the figure.}
}
\centering
\parbox{1.0\linewidth}
{
\includegraphics[width=\linewidth]{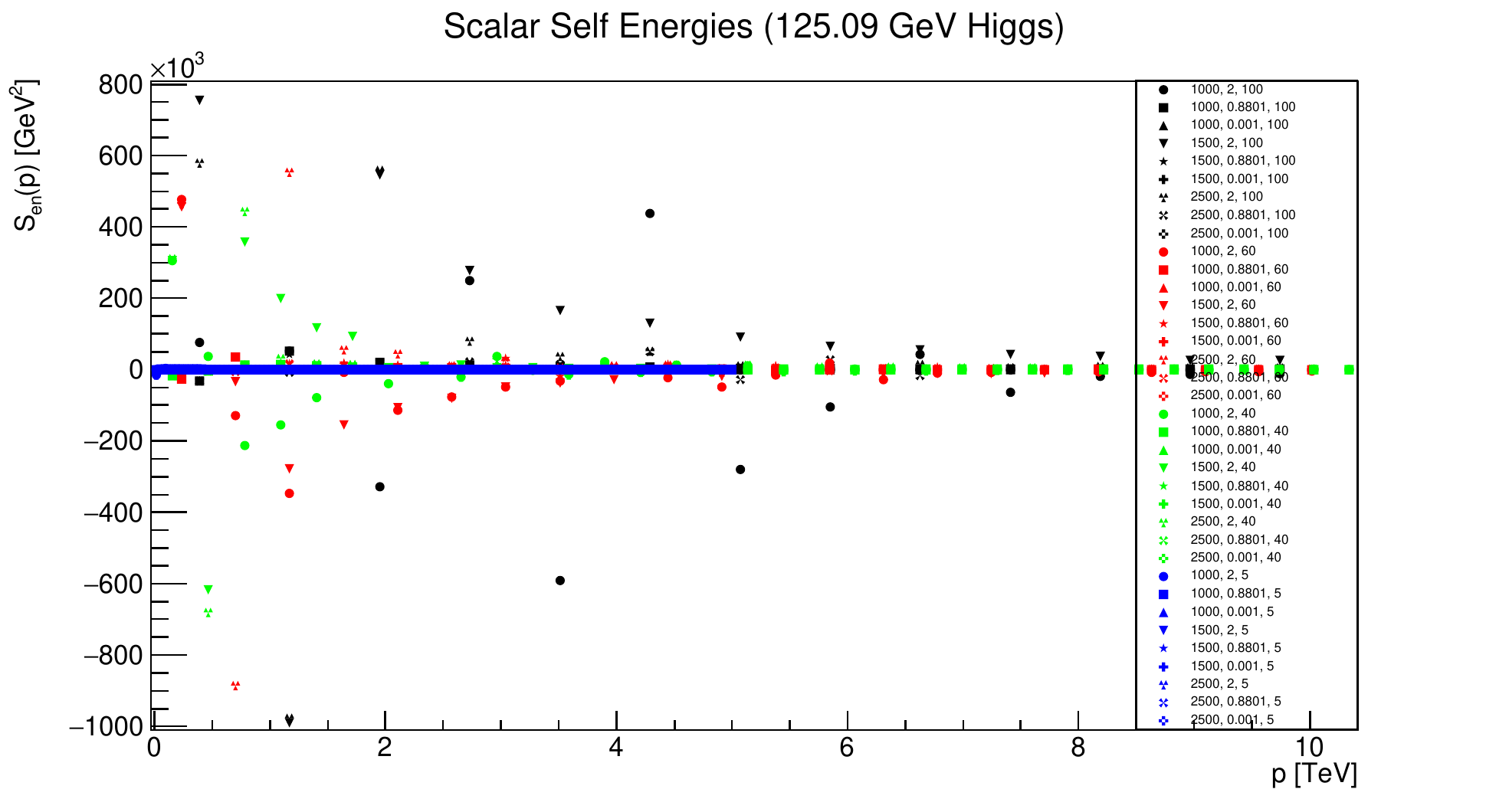}
\caption{\label{fig:ssen125tev} Scalar self energies for $m_{h}=125.09$ GeV are plotted (in TeV regime) with 1000 GeV $ \leq m_{s} \leq $ 2500 GeV, 0.001 $\leq \lambda \leq$ 2.0, and 5 TeV $ \leq \Lambda \leq $ 60 TeV, shown as $(m_{s},\lambda,\Lambda)$ on the figure.}
}
\end{figure}
\begin{figure}
\centering
\parbox{1.0\linewidth}
{
\includegraphics[width=\linewidth]{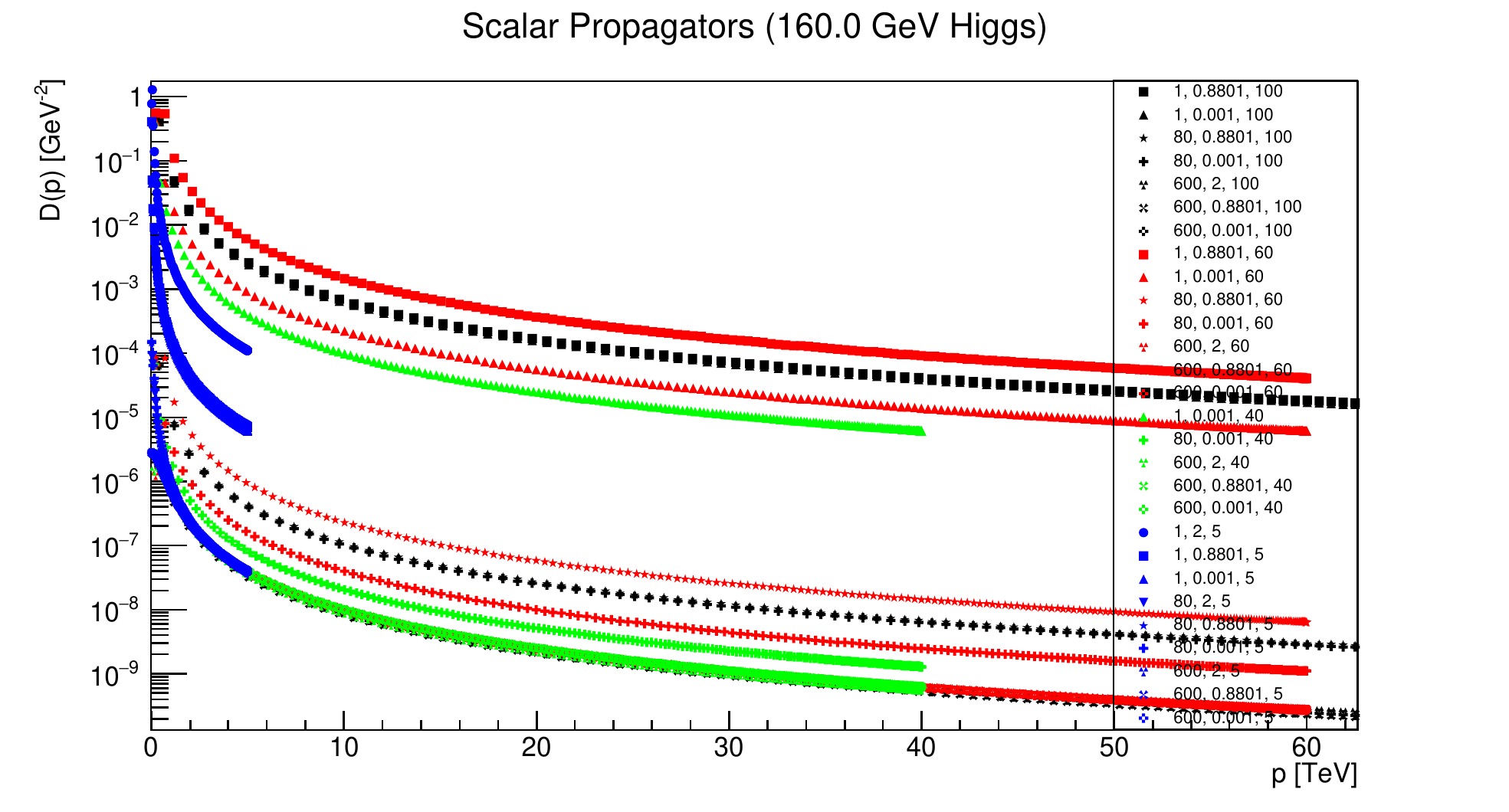}
\caption{\label{fig:sprs160ewlog} Scalar propagators (on logarithmic scale)  for $m_{h}=160.0$ GeV are plotted (in electroweak regime) with 1 GeV $ \leq m_{s} \leq $ 600 GeV, 0.001 $\leq \lambda \leq$ 2.0, and 5 TeV $ \leq \Lambda \leq $ 60 TeV, shown as $(m_{s},\lambda,\Lambda)$ on the figure.}
}
\centering
\parbox{1.0\linewidth}
{
\includegraphics[width=\linewidth]{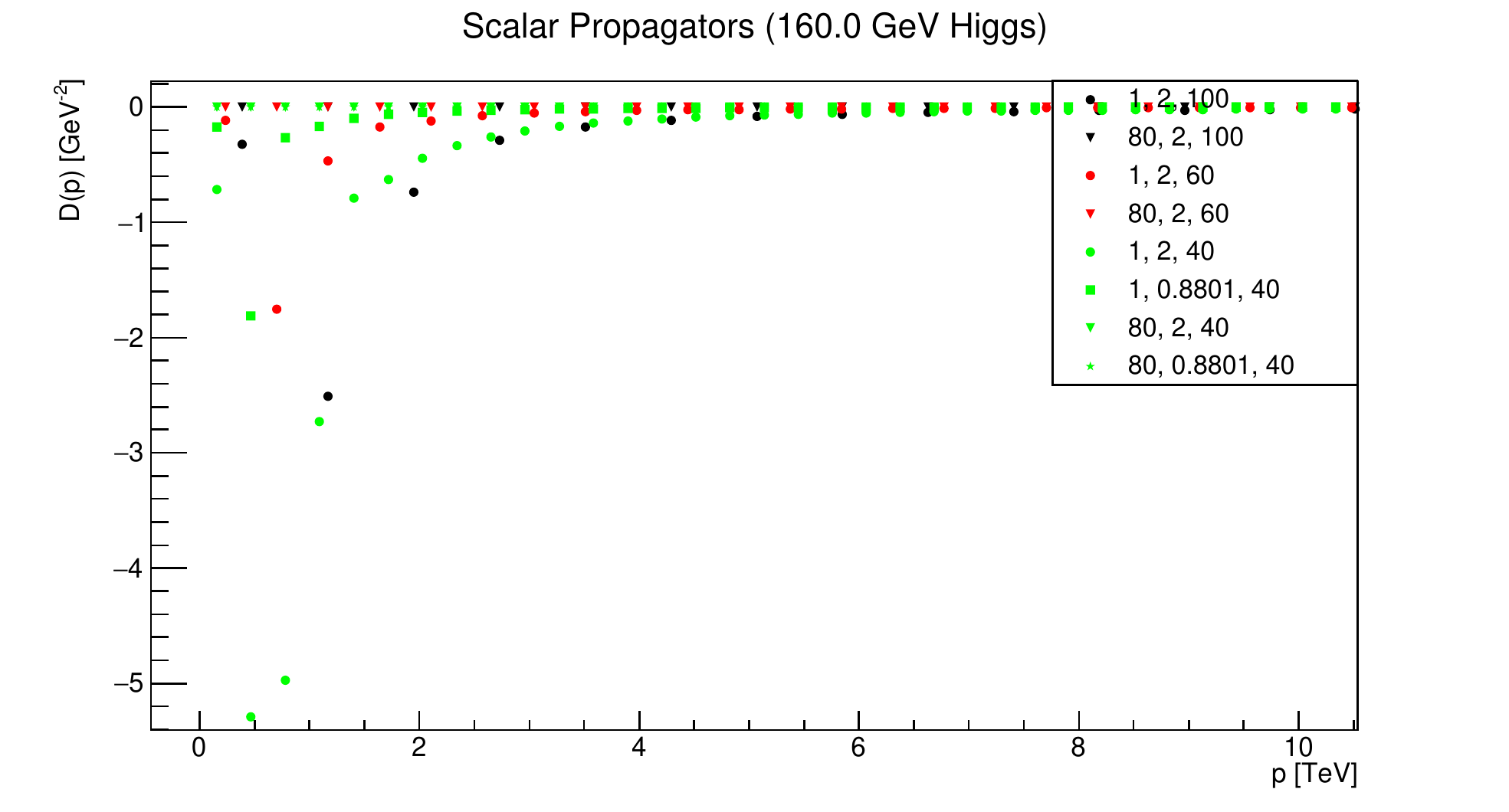}
\caption{\label{fig:sprs160ew1} Scalar propagators for $m_{h}=160.0$ GeV are plotted (in electroweak regime) with 1 GeV $ \leq m_{s} \leq $ 600 GeV, 0.001 $\leq \lambda \leq$ 2.0, and 5 TeV $ \leq \Lambda \leq $ 60 TeV, shown as $(m_{s},\lambda,\Lambda)$ on the figure.}
}
\end{figure}
\begin{figure}
\centering
\parbox{1.0\linewidth}
{
\includegraphics[width=\linewidth]{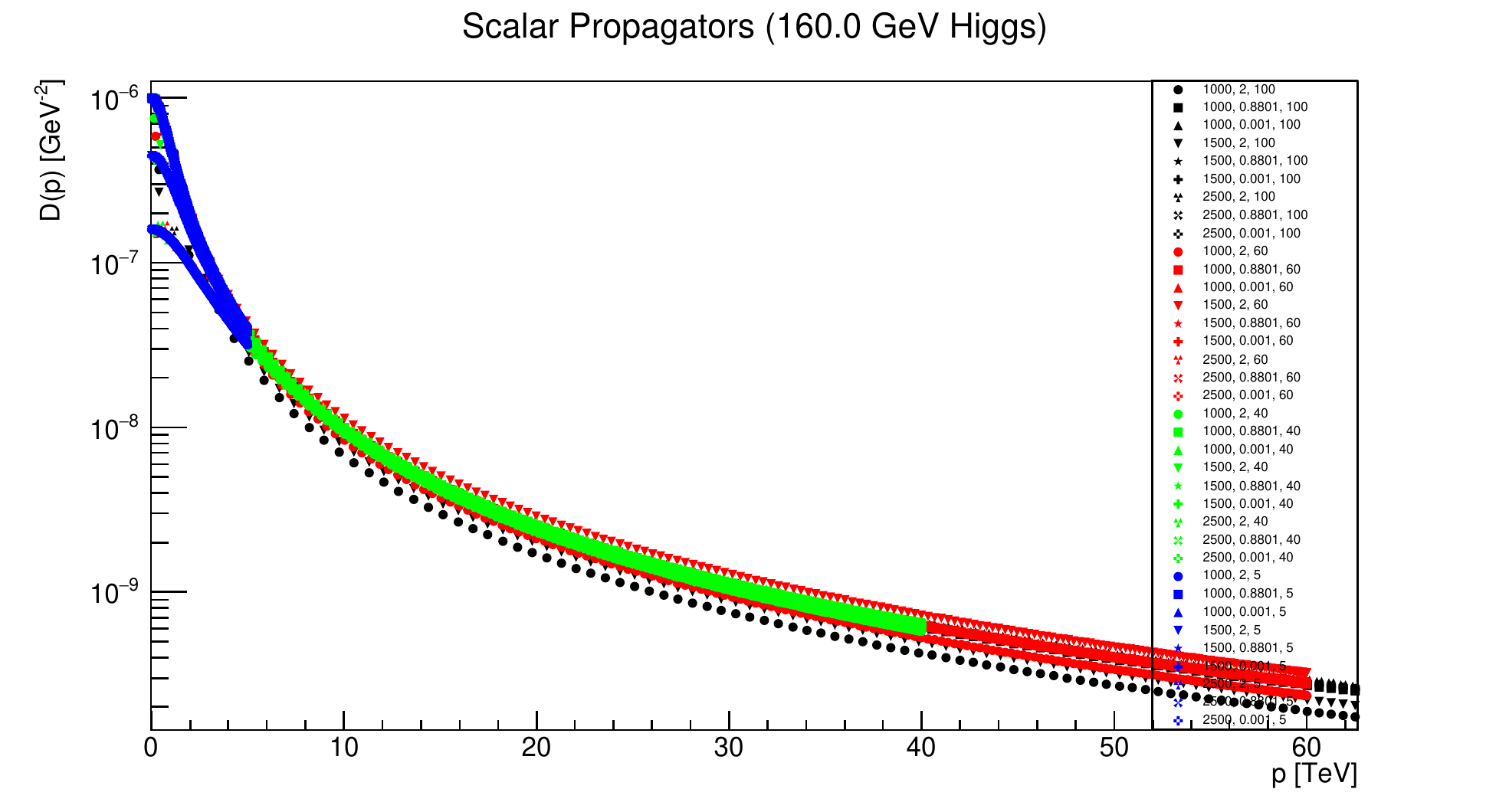}
\caption{\label{fig:sprs160tevlog} Scalar propagators (on logarithmic scale)  for $m_{h}=160.0$ GeV are plotted (in TeV regime) with 1000 GeV $ \leq m_{s} \leq $ 2500 GeV, 0.001 $\leq \lambda \leq$ 2.0, and 5 TeV $ \leq \Lambda \leq $ 60 TeV, shown as $(m_{s},\lambda,\Lambda)$ on the figure.}
}
\end{figure}
\begin{figure}
\centering
\parbox{1.0\linewidth}
{
\includegraphics[width=\linewidth]{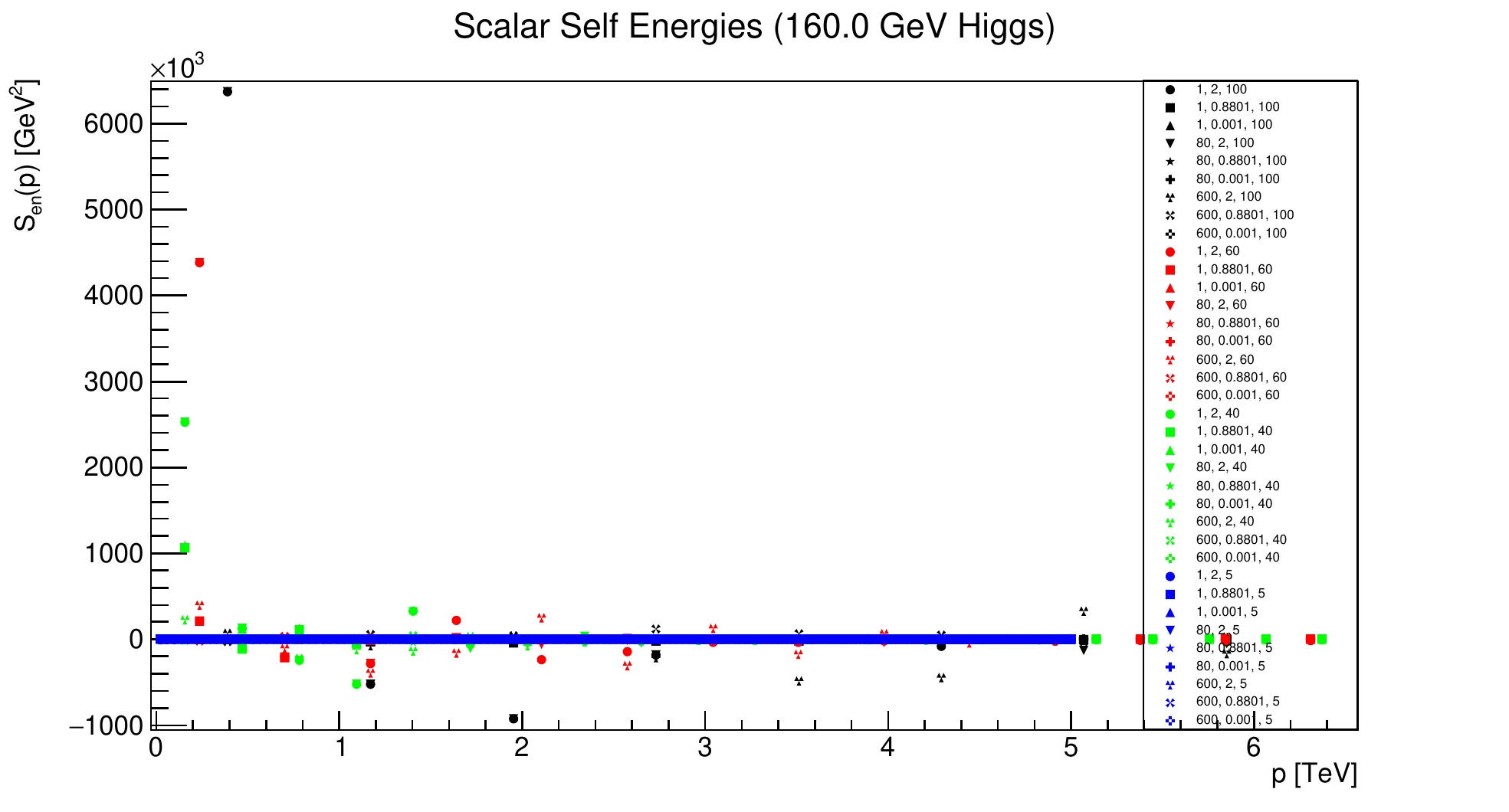}
\caption{\label{fig:ssen160ew} Scalar self energies for $m_{h}=160.0$ GeV are plotted (in electroweak regime) with 1 GeV $ \leq m_{s} \leq $ 600 GeV, 0.001 $\leq \lambda \leq$ 2.0, and 5 TeV $ \leq \Lambda \leq $ 60 TeV, shown as $(m_{s},\lambda,\Lambda)$ on the figure.}
}
\centering
\parbox{1.0\linewidth}
{
\includegraphics[width=\linewidth]{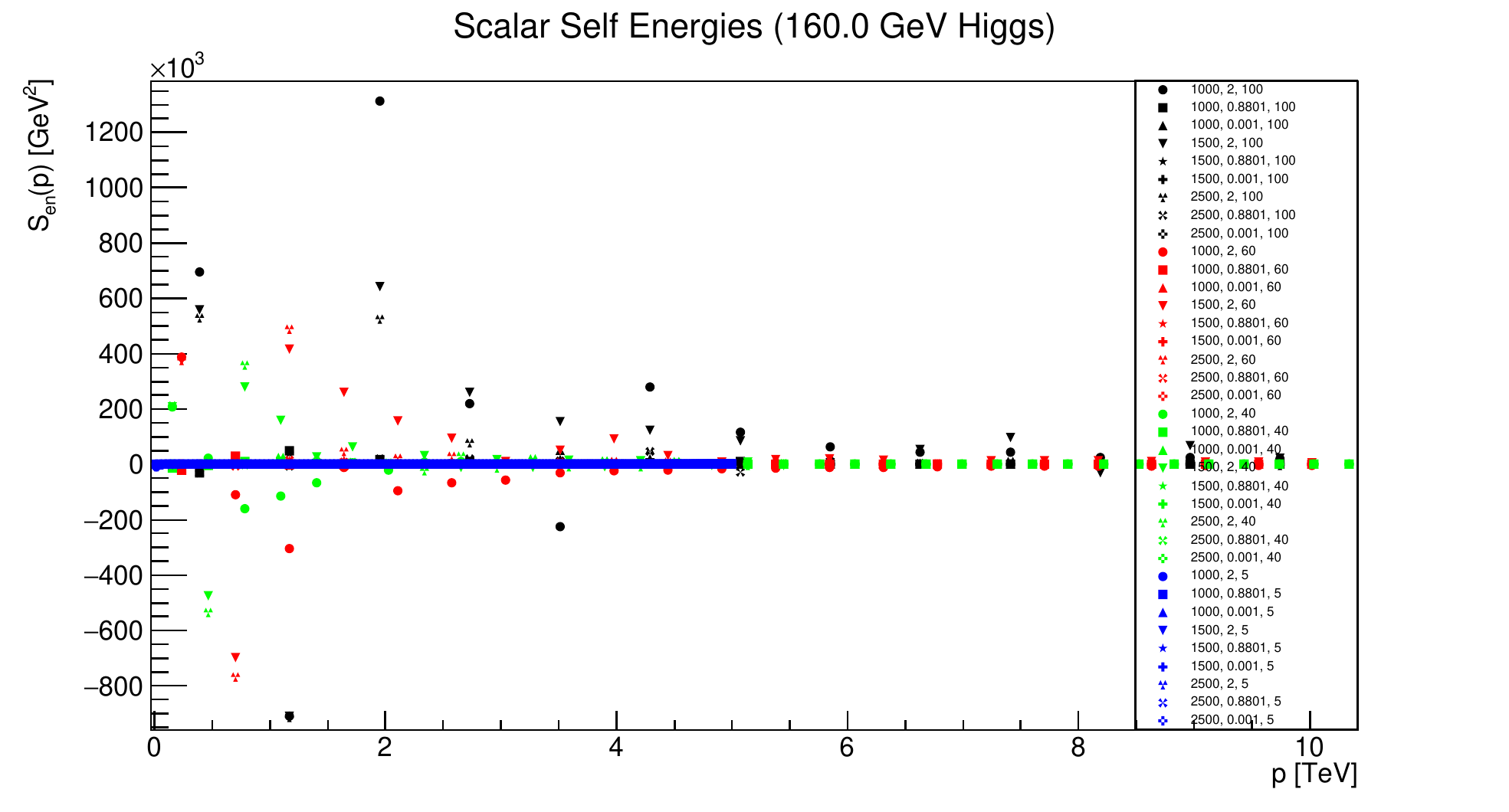}
\caption{\label{fig:ssen160tev} Scalar self energies for $m_{h}=160.0$ GeV are plotted (in TeV regime) with 1000 GeV $ \leq m_{s} \leq $ 2500 GeV, 0.001 $\leq \lambda \leq$ 2.0, and 5 TeV $ \leq \Lambda \leq $ 60 TeV, shown as $(m_{s},\lambda,\Lambda)$ on the figure.}
}
\end{figure}
\begin{figure}
\includegraphics[width=\linewidth]{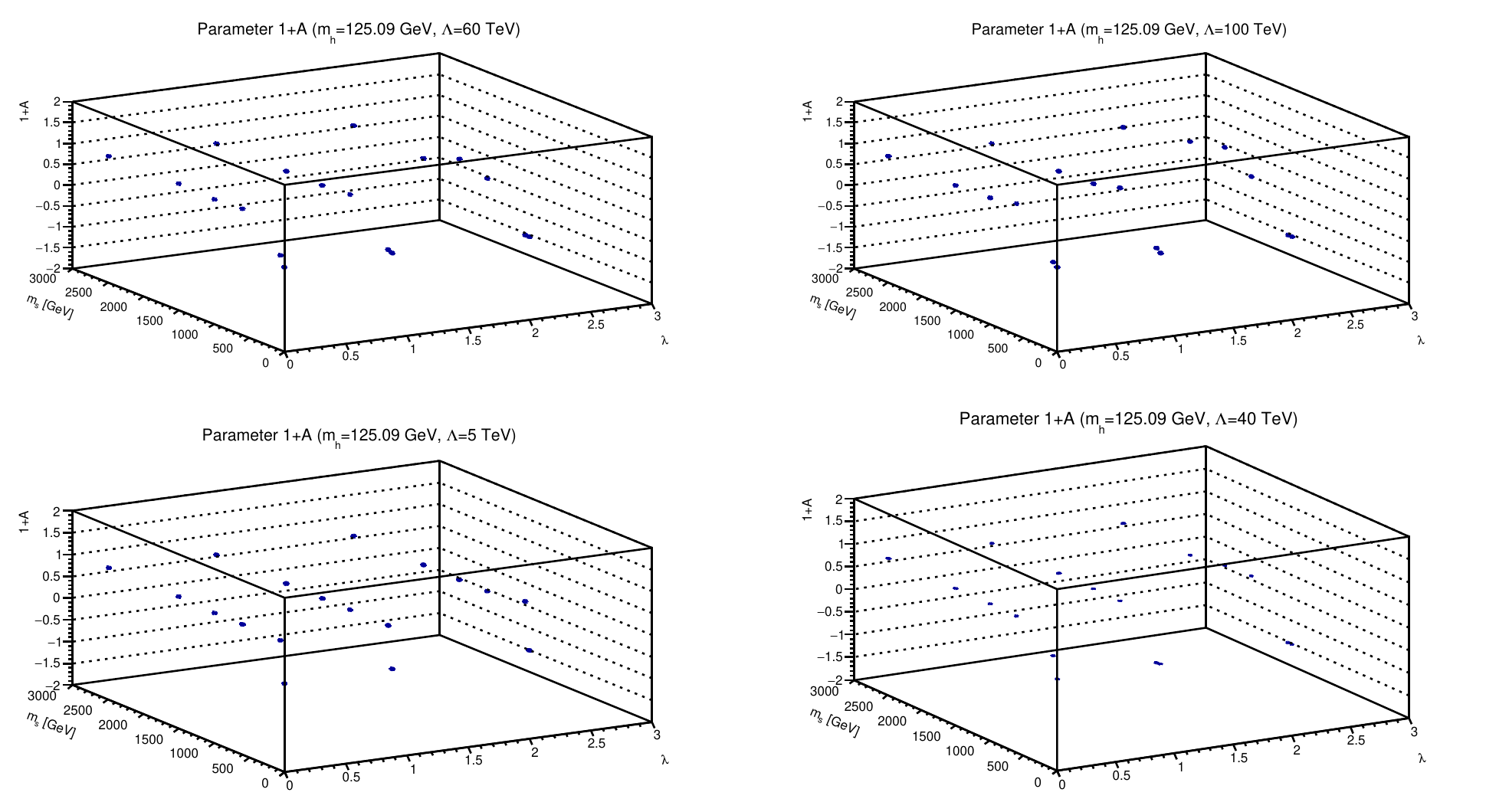}
\caption{\label{fig:stermlight} Renormalization quantity 1+A for scalar propagator with $m_{h}=125.09$ GeV are plotted for different scalar bare masses $m_{s}$ and bare couplings $\lambda$ at different cutoff values $\Lambda$.}
\end{figure}
\begin{figure}
\includegraphics[width=\linewidth]{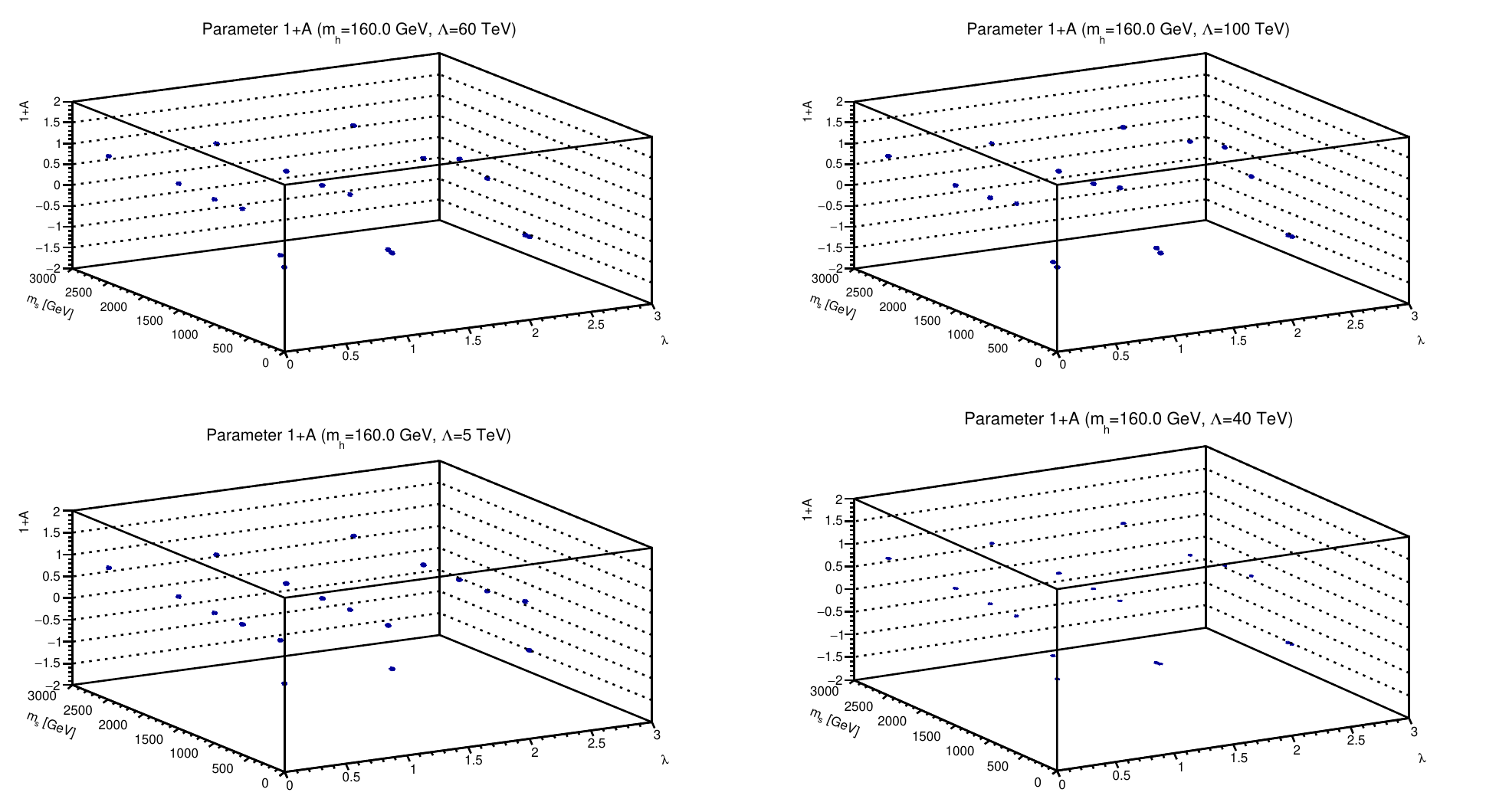}
\caption{\label{fig:stermheavy} Renormalization quantity 1+A for scalar propagator with $m_{h}=160.0$ GeV are plotted for different scalar bare masses $m_{s}$ and bare couplings $\lambda$ at different cutoff values $\Lambda$.}
\end{figure}
Scalar propagators, as they are calculated using Higgs propagators and Yukawa interaction vertex, are found to be highly informative in the model. Their DSE involves more constants to be known while implementing the renormalization procedure during numerical computations compared to Higgs propagators. The propagators are shown in figures \ref{fig:sprs125ewlog}, \ref{fig:sprs125ew1}, \ref{fig:sprs125tevlog}, \ref{fig:sprs160ewlog}, \ref{fig:sprs160ew1}, and \ref{fig:sprs160tevlog}.
\par
An immediate observation is that at the ultraviolet end scalar propagators have the same qualitative behavior for both of the physical masses of Higgs \footnote{Throughout the current paper, physical mass and renormalized mass are used interchangeably.}. It becomes clear that at this end scalar self energies, defined by equation \ref{ssen:dse}, should have lost any significant dependence for higher momentum while the quantitative differences appear due to the renormalization parameter $A$ in equation \ref{spr2:dse} which is corroborated by the the self energies plotted in figures \ref{fig:ssen125ew} and \ref{fig:ssen125tev} for $m_{h}=125.09$ GeV, and \ref{fig:ssen160ew} and \ref{fig:ssen160tev} for $m_{h}=160.0$ GeV, respectively.
\par
Furthermore, for higher coupling and the lowest scalar bare mass, $m_{s}=1$ GeV, the propagators exhibit a tendency of having a pole, see figures \ref{fig:sprs125ew1} and \ref{fig:sprs160ew1} in which the propagators are shown without a log scale, and scalar self energy terms in figures \ref{fig:ssen125ew} and \ref{fig:ssen160ew}. This peculiar behavior suggests that in the model it is possible for scalars to have a negative renormalized squared mass which, as in standard literature, is taken as a sign of symmetry breaking. For the corresponding parameters, scalar self energies also show relatively pronounced negative contributions. Several other parameters are also found to have this tendency, though it is mild and suppressed for lower cutoff values, which indicates that if such a phenomenon really exists in the model, it may take place for higher cutoff values \footnote{Further higher values of cutoff were not studied as it would compromise the resolution in momentum values.}.
\par
Another feature of $m_{s}$ in electroweak regime is the sensitivity on parameters and cutoff. Though, several of the scalar propagators are indeed found to have similar magnitude in the infrared region, the deviations are evident on the figures \ref{fig:sprs125ewlog}, \ref{fig:sprs125ew1}, \ref{fig:sprs160ewlog}, and \ref{fig:sprs160ew1}, which may arise due to different values of the $A$ parameters as mentioned above. It was also observed for a number of parameters in electroweak regime, particularly for 125.09 GeV Higgs and very low scalar bare mass, that lower the cutoff more suppressed the propagator may appear.
\par
For $m_{s}$ in TeV regime, the propagators manifest differently than in electroweak regime. First of all, instead of having a variety of propagators they are relatively closer to each other beyond (approximately) 10 TeV momentum. Hence, infrared region is more interesting than the ultraviolet region. In this region, instead of a wide variety as in case of electroweak regime, the propagators are found to accumulate in groups, see figure \ref{fig:sprs125tevlog}. For higher Higgs mass ($m_{h}=160$ GeV), scalar propagators are found to have slightly more enhanced dependence on parameter space, see figure \ref{fig:sprs160tevlog}, but qualitatively the propagators manifest in the similar way as for light Higgs case ($m_{h}=125.09$ GeV) due to the A parameter, see equation \ref{spr3:dse}. Their qualitatively similar behavior indicates less deviations among the magnitudes of physical scalar masses in the regime because the A parameter only rescales the propagators in compliance with the renormalization condition in equation \ref{spr:ren_condition}. A notable observation is suppression of propagators for higher masses and the highest cutoff value ($\Lambda = 60$ TeV). Qualitatively, scalar self energies for both of the selected Higgs masses in TeV regime share similar features, see figures \ref{fig:ssen125tev} and \ref{fig:ssen160tev}. Beyond around $10$ TeV, they are significantly suppressed which concurs with the quantitative behavior of scalar propagators for both of the Higgs masses.
\par
Scalar propagators for both Higgs masses are found to be qualitatively similar. In figures \ref{fig:stermlight} and \ref{fig:stermheavy}, renormalization parameter A for scalar propagator, in the form of the quantity $\tilde{A}=1+A$, is shown for various cutoff values and parameters of the theory. For higher values of $m_{s}$, irrespective of $m_{h,r}$, the parameter $\tilde{A}$ is close to 1.0 which indicates small contribution of A parameter during renormalization. It suggests that the deviations in scalar propagators in the infrared region is mostly due to the contributions from self energies. However, for the values of $m_{s}$ lower than $100$ GeV, deviations from unity is relatively more prominent. It suggests emergence of a different behavior peculiar for smaller scalar bare masses in contrast to TeV regime.
\par
Overall, beyond the infrared region where mutual (particularly qualitative) deviations between scalar propagators are less prominent, scalar propagators are generally found to have (decreasing) monotonic behavior which is an indicative of a physical particle in the theory.
\subsection{Higgs Propagators} \label{sec:hprs}
\begin{figure}
\centering
\parbox{1.0\linewidth}
{
\includegraphics[width=\linewidth]{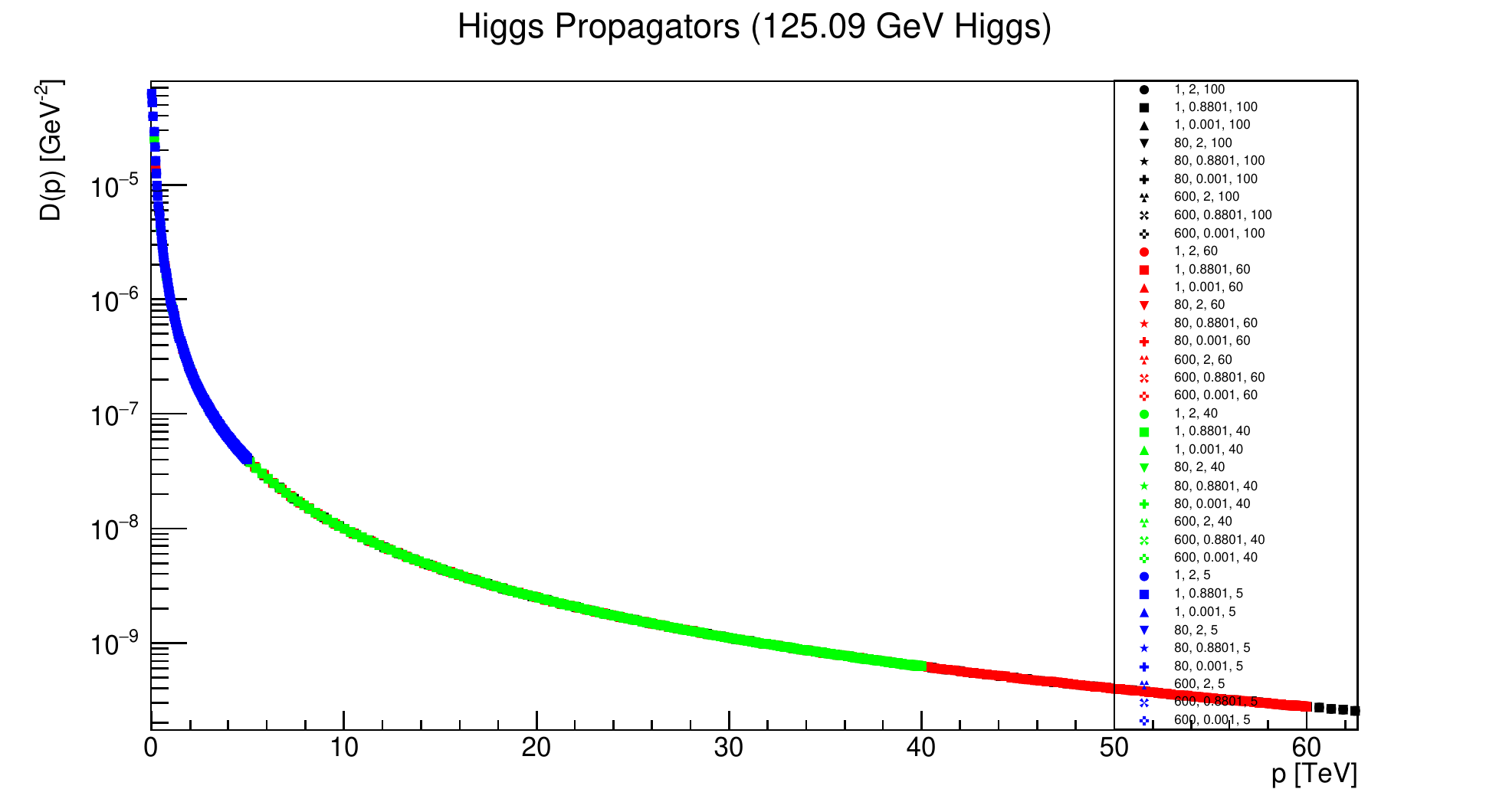}
\caption{\label{fig:hprs125ew} Higgs propagators (on logarithmic scale) for $m_{h}=125.09$ GeV are plotted (in electroweak regime) with 1 GeV $ \leq m_{s} \leq $ 600 GeV, 0.001 $\leq \lambda \leq$ 2.0, and 5 TeV $ \leq \Lambda \leq $ 60 TeV, shown as $(m_{s},\lambda,\Lambda)$ on the figure.}
}
\centering
\parbox{1.0\linewidth}
{
\includegraphics[width=\linewidth]{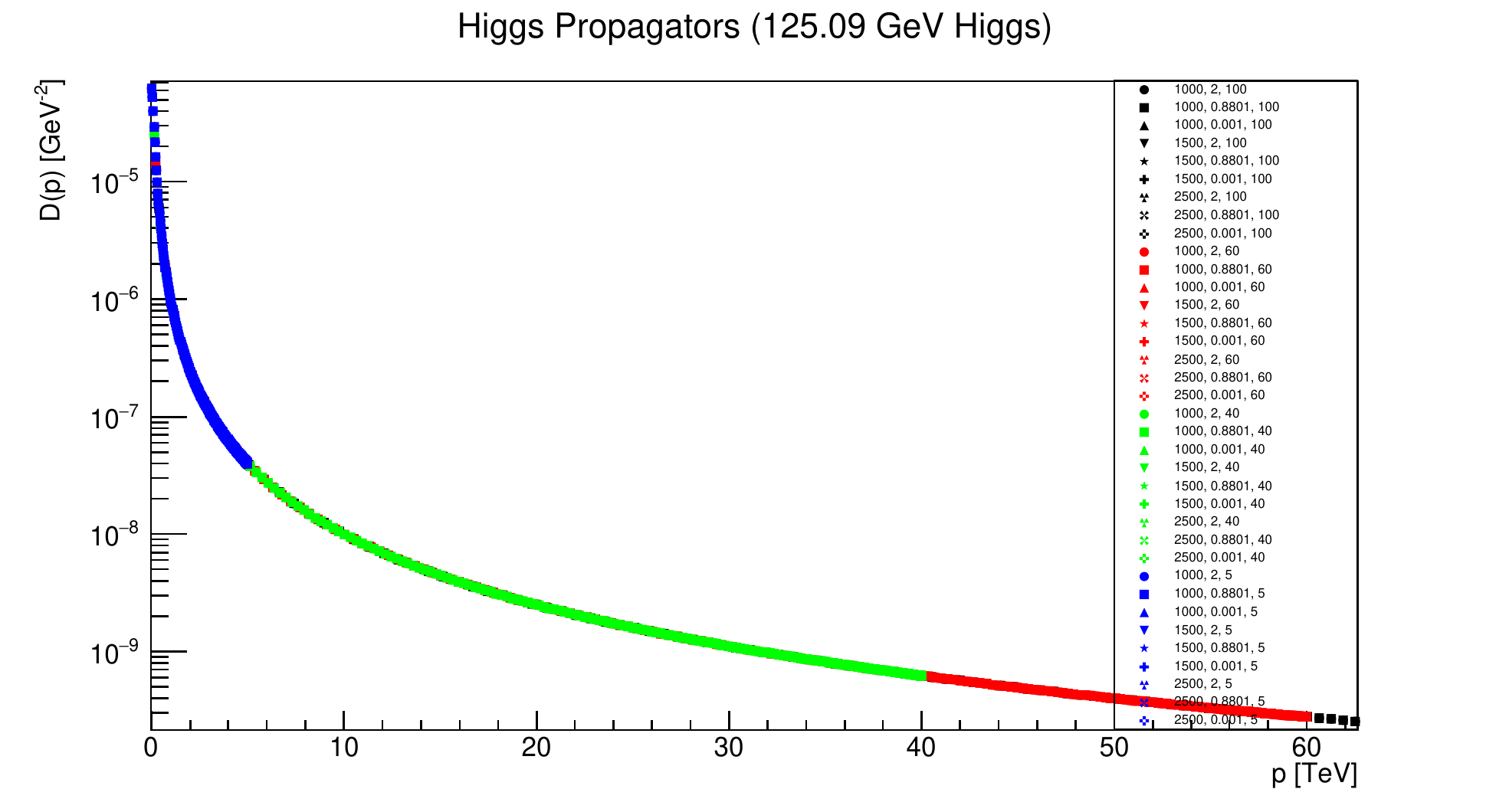}
\caption{\label{fig:hprs125tev} Higgs propagators (on logarithmic scale) for $m_{h}=125.09$ GeV are plotted (in TeV regime) with 1000 GeV $ \leq m_{s} \leq $ 2500 GeV, 0.001 $\leq \lambda \leq$ 2.0, and 5 TeV $ \leq \Lambda \leq $ 60 TeV, shown as $(m_{s},\lambda,\Lambda)$ on the figure.}
}
\end{figure}
\begin{figure}
\centering
\parbox{1.0\linewidth}
{
\includegraphics[width=\linewidth]{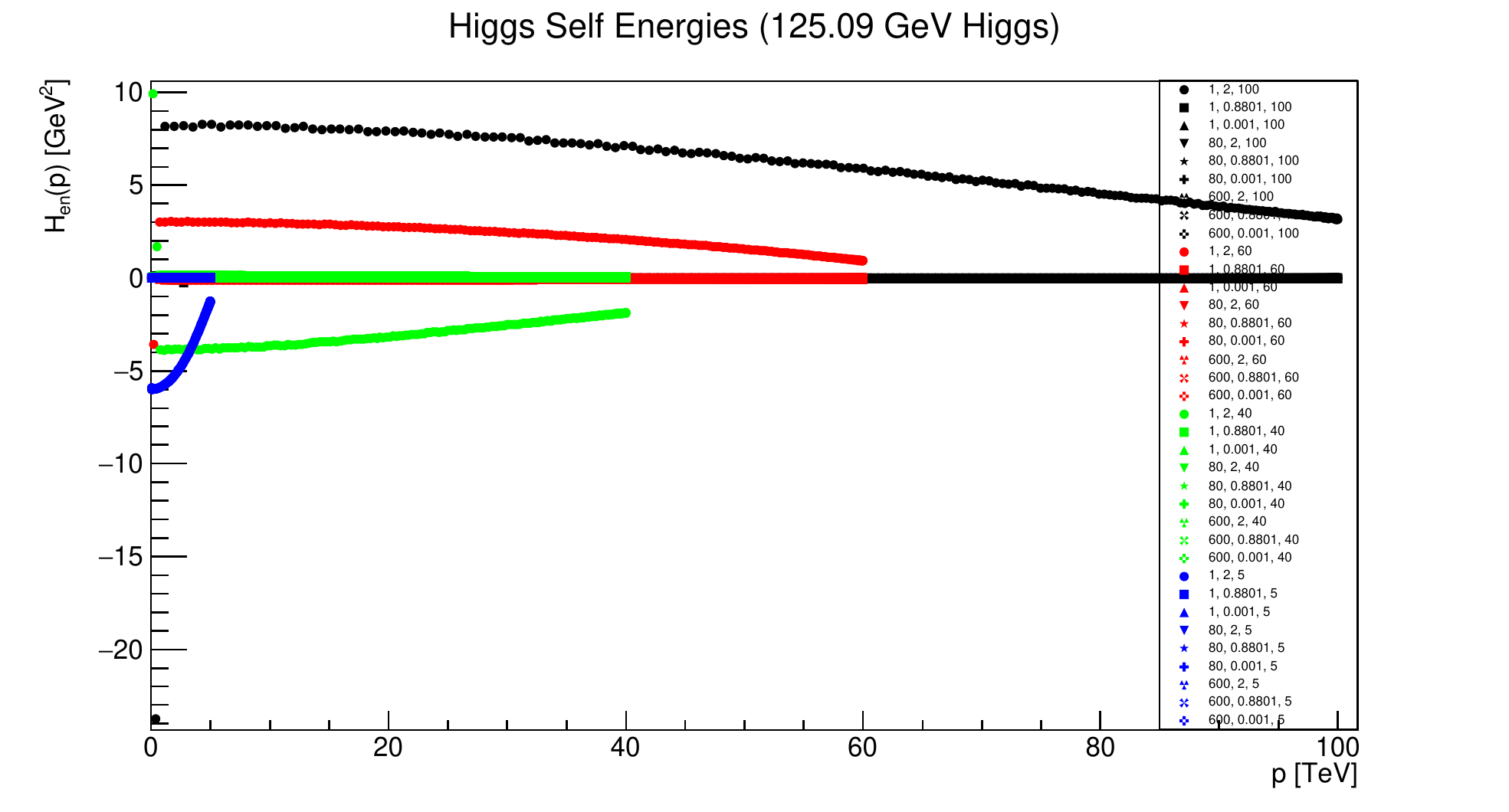}
\caption{\label{fig:hsen125ew} Higgs self energies for $m_{h}=125.09$ GeV are plotted (in electroweak regime) with 1 GeV $ \leq m_{s} \leq $ 600 GeV, 0.001 $\leq \lambda \leq$ 2.0, and 5 TeV $ \leq \Lambda \leq $ 60 TeV, shown as $(m_{s},\lambda,\Lambda)$ on the figure.}
}
\centering
\parbox{1.0\linewidth}
{
\includegraphics[width=\linewidth]{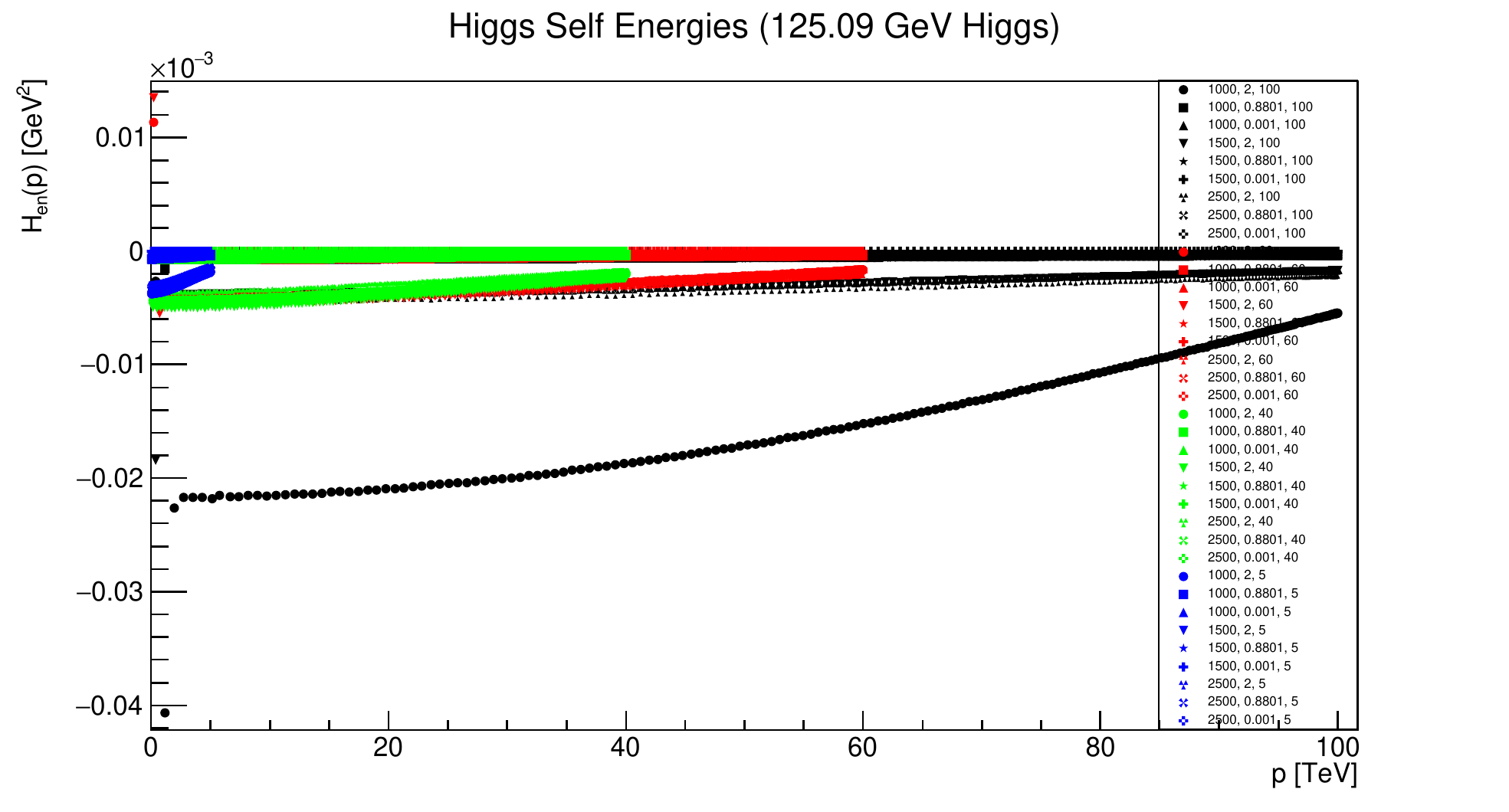}
\caption{\label{fig:hsen125tev} Higgs self energies for $m_{h}=125.09$ GeV are plotted (in TeV regime) with 1000 GeV $ \leq m_{s} \leq $ 2500 GeV, 0.001 $\leq \lambda \leq$ 2.0, and 5 TeV $ \leq \Lambda \leq $ 60 TeV, shown as $(m_{s},\lambda,\Lambda)$ on the figure.}
}
\end{figure}
\begin{figure}
\centering
\parbox{1.0\linewidth}
{
\includegraphics[width=\linewidth]{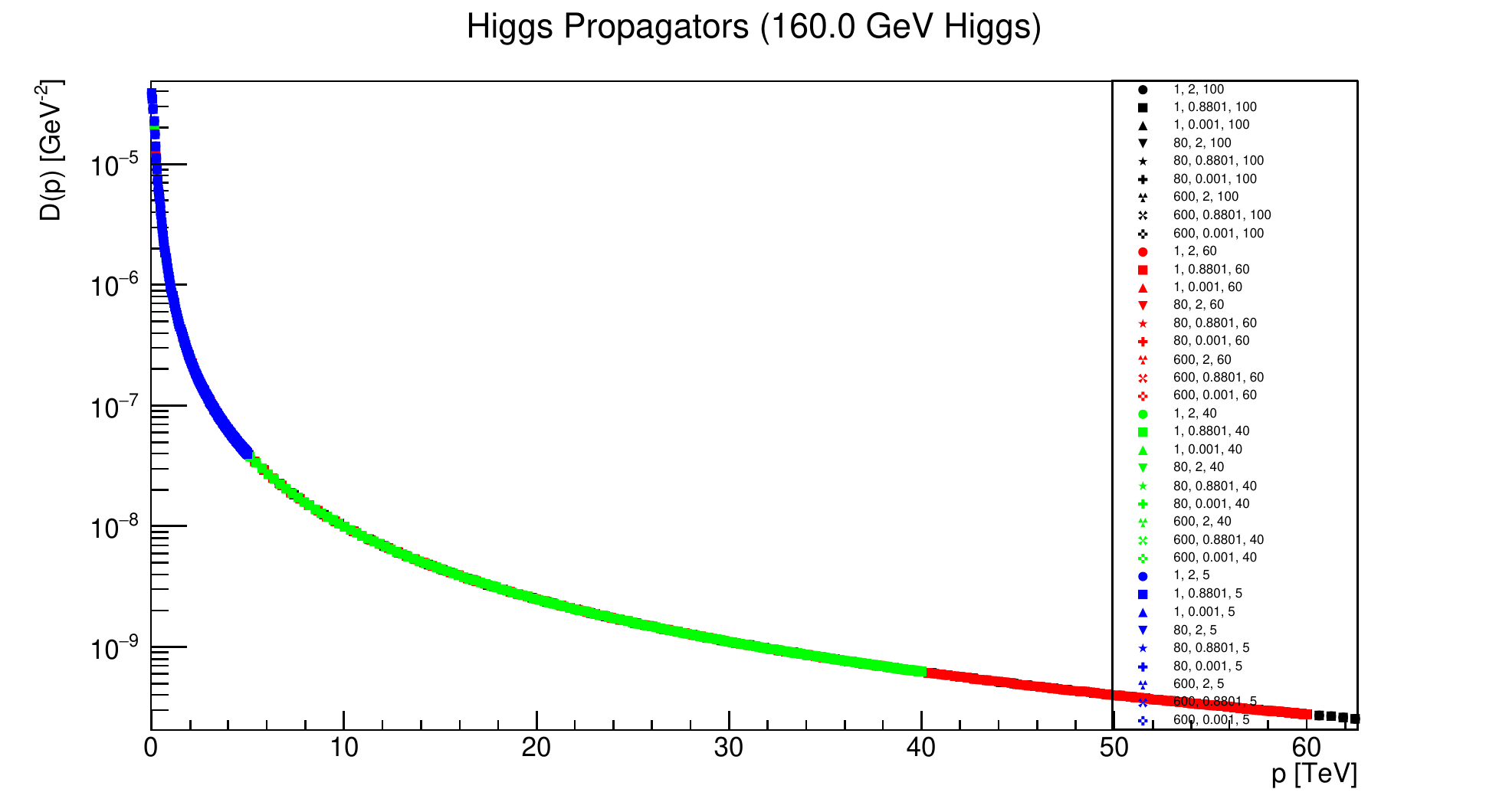}
\caption{\label{fig:hprs160ew} Higgs propagators (on logarithmic scale) for $m_{h}=160.0$ GeV are plotted (in electroweak regime) with 1 GeV $ \leq m_{s} \leq $ 600 GeV, 0.001 $\leq \lambda \leq$ 2.0, and 5 TeV $ \leq \Lambda \leq $ 60 TeV, shown as $(m_{s},\lambda,\Lambda)$ on the figure.}
}
\centering
\parbox{1.0\linewidth}
{
\includegraphics[width=\linewidth]{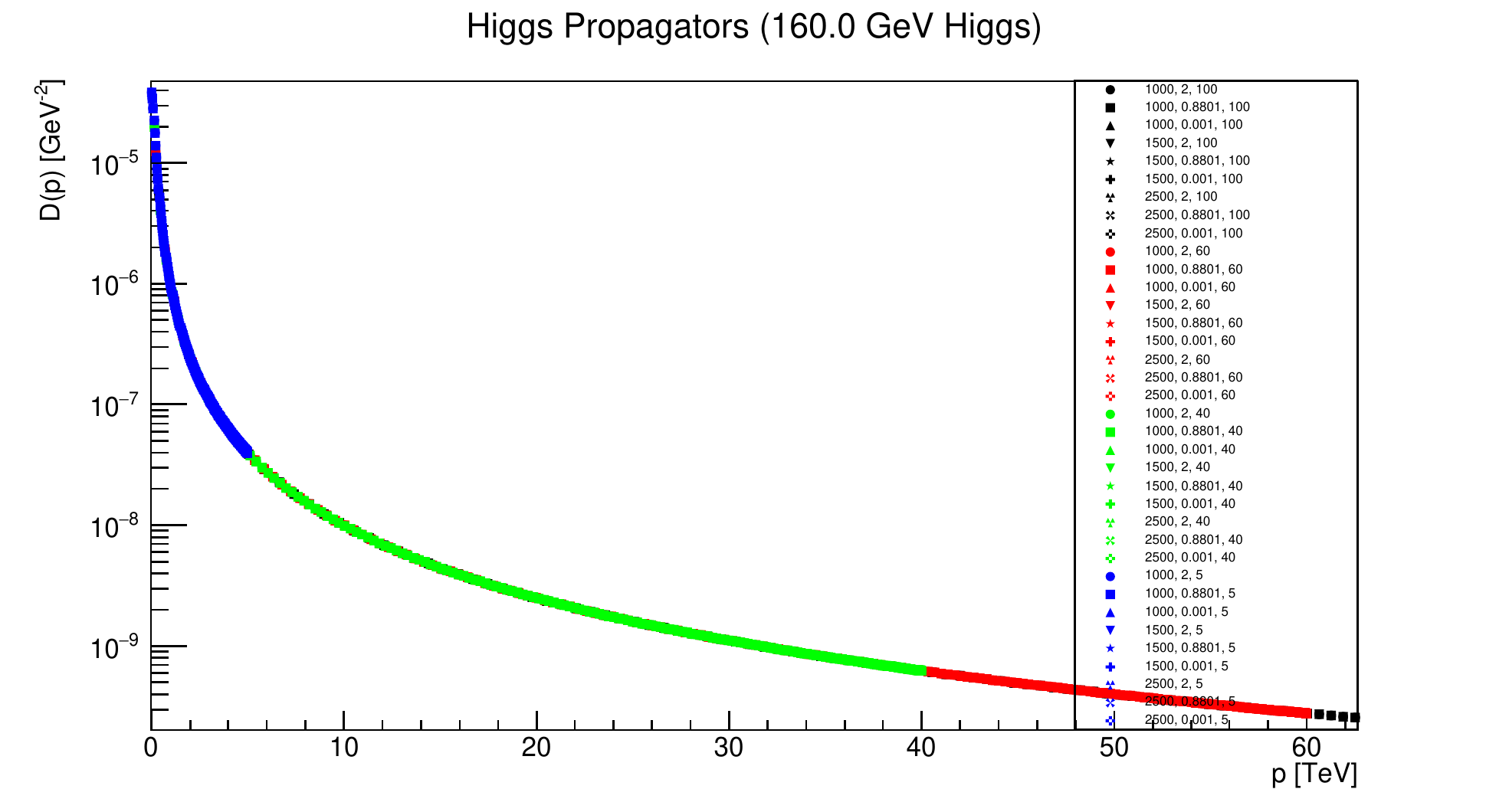}
\caption{\label{fig:hprs160tev} Higgs propagators (on logarithmic scale) for $m_{h}=160.0$ GeV are plotted (in TeV regime) with 1000 GeV $ \leq m_{s} \leq $ 2500 GeV, 0.001 $\leq \lambda \leq$ 2.0, and 5 TeV $ \leq \Lambda \leq $ 60 TeV, shown as $(m_{s},\lambda,\Lambda)$ on the figure.}
}
\end{figure}
\begin{figure}
\centering
\parbox{1.0\linewidth}
{
\includegraphics[width=\linewidth]{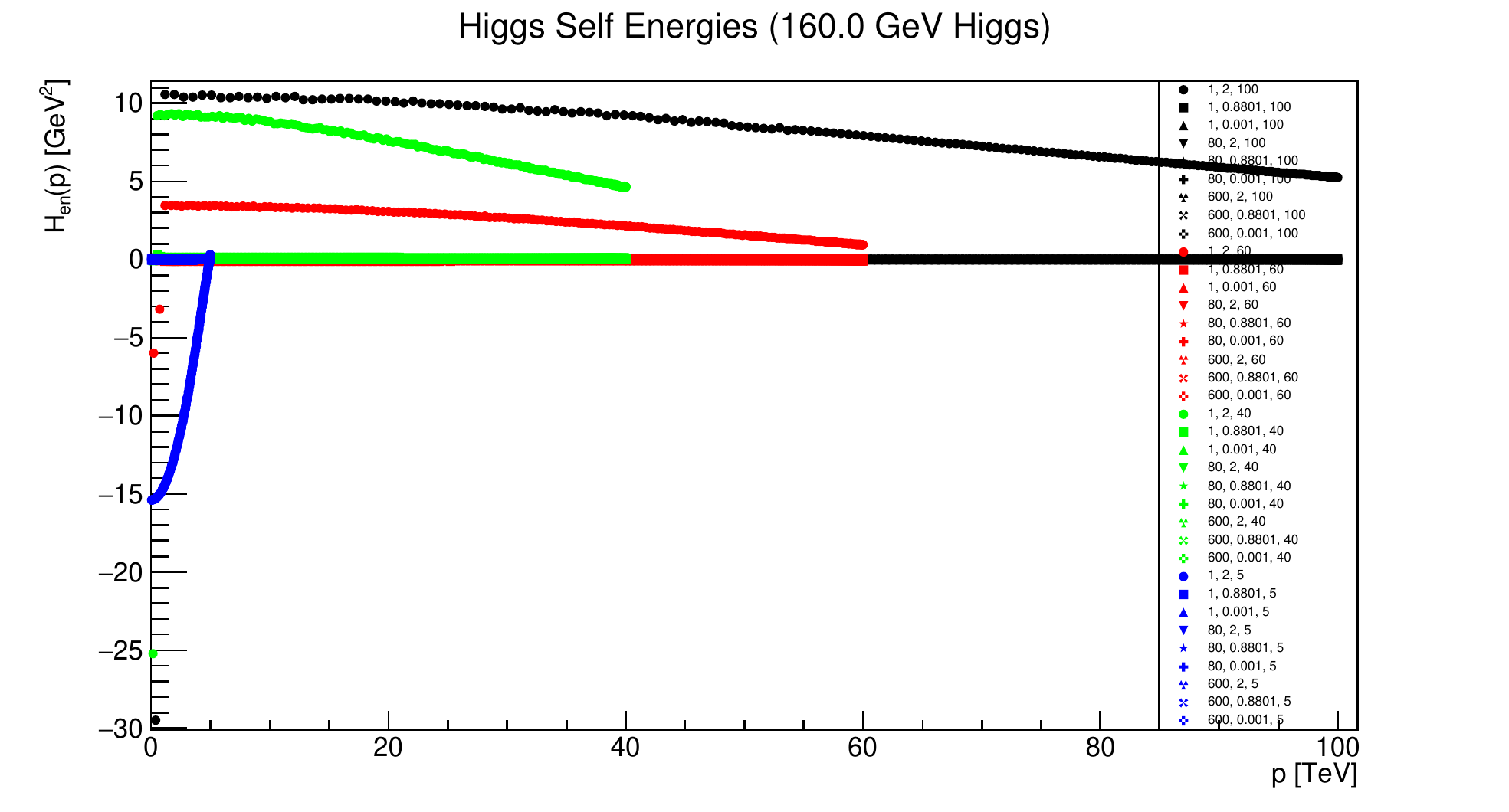}
\caption{\label{fig:hsen160ew} Higgs self energies for $m_{h}=160.0$ GeV are plotted (in electroweak regime) with 1 GeV $ \leq m_{s} \leq $ 600 GeV, 0.001 $\leq \lambda \leq$ 2.0, and 5 TeV $ \leq \Lambda \leq $ 60 TeV, shown as $(m_{s},\lambda,\Lambda)$ on the figure.}
}
\centering
\parbox{1.0\linewidth}
{
\includegraphics[width=\linewidth]{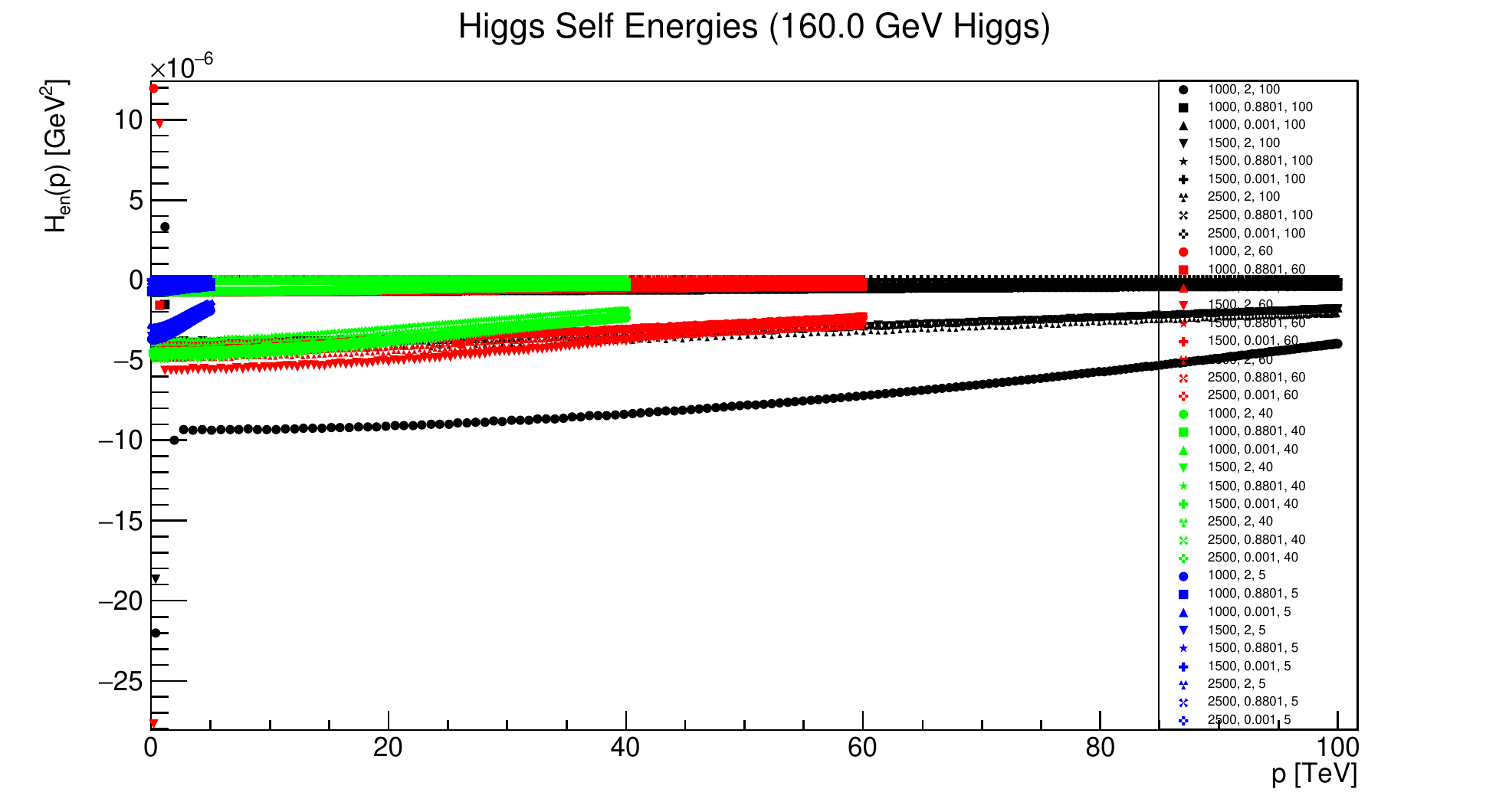}
\caption{\label{fig:hsen160tev} Higgs self energies for $m_{h}=160.0$ GeV are plotted (in TeV regime) with 1000 GeV $ \leq m_{s} \leq $ 2500 GeV, 0.001 $\leq \lambda \leq$ 2.0, and 5 TeV $ \leq \Lambda \leq $ 60 TeV, shown as $(m_{s},\lambda,\Lambda)$ on the figure.}
}
\end{figure}
\begin{figure}
\includegraphics[width=\linewidth]{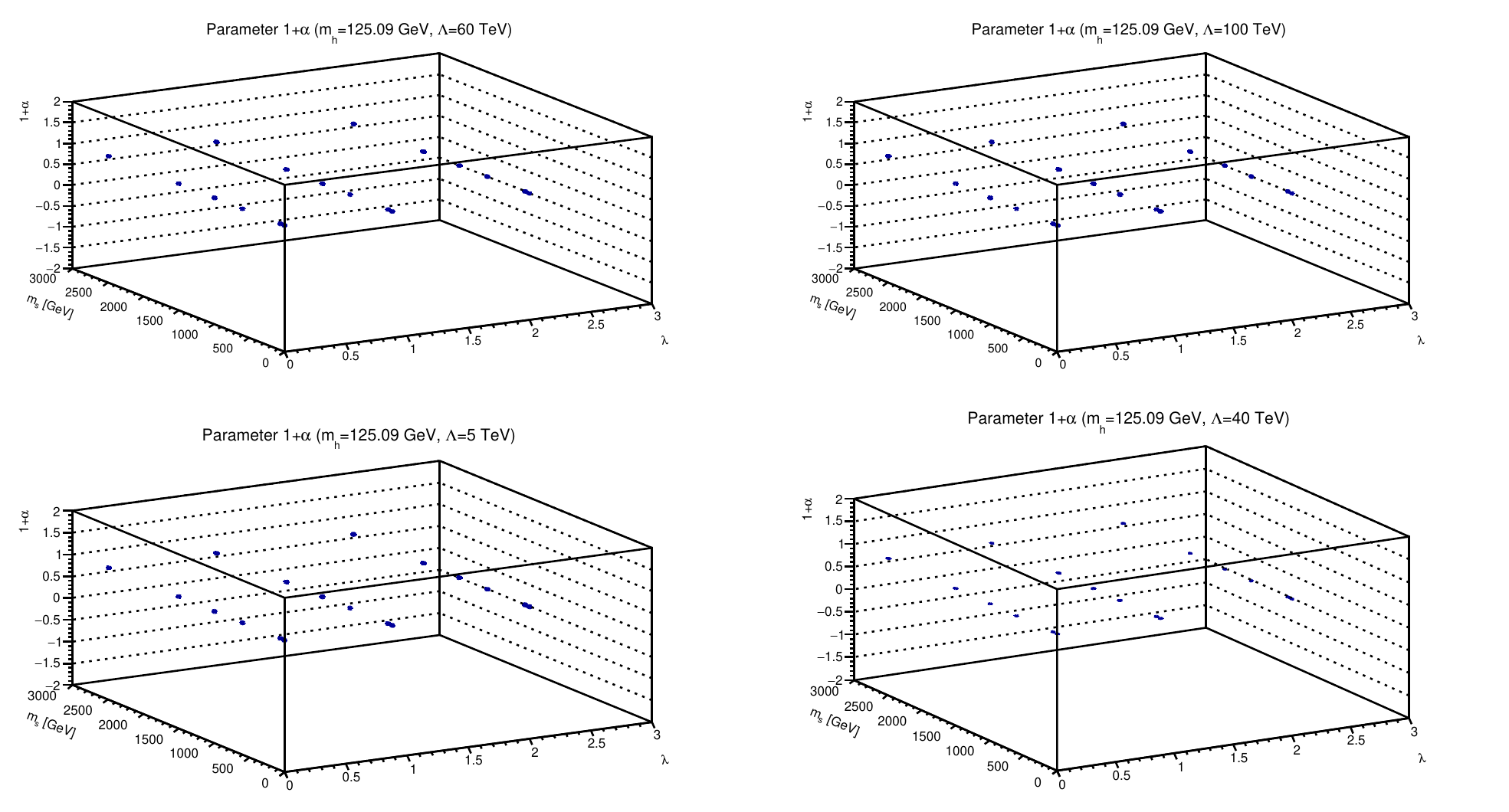}
\caption{\label{fig:htermlight} Renormalization quantity 1+$\alpha$ for scalar propagator with $m_{h}=125.09$ GeV are plotted for different scalar bare masses $m_{s}$ and bare couplings $\lambda$ at different cutoff values $\Lambda$.}
\end{figure}
\begin{figure}
\includegraphics[width=\linewidth]{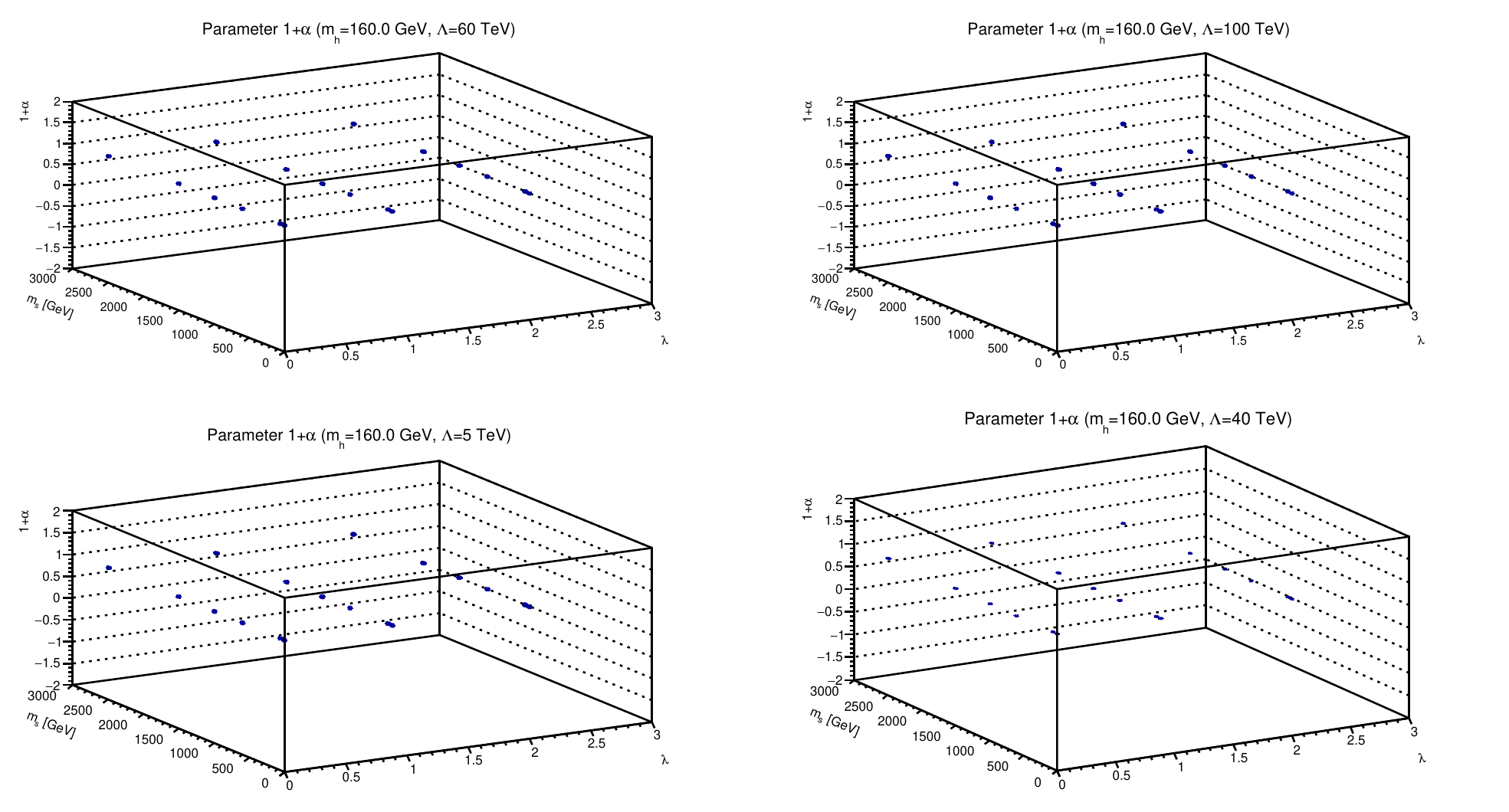}
\caption{\label{fig:htermheavy} Renormalization quantity 1+$\alpha$ for scalar propagator with $m_{h}=160.0$ GeV are plotted for different scalar bare masses $m_{s}$ and bare couplings $\lambda$ at different cutoff values $\Lambda$.}
\end{figure}
Unlike scalar propagators, Higgs propagators are independently updated with the renormalization condition in equation \ref{hpr:ren_condition} by updating the coefficients in the expansion of Higgs propagators given in equation \ref{hpr:coeff1}. Thus, with the Higgs renormalized mass fixed, as the constant $\alpha$ in equation \ref{hpr2:dse} is determined directly from equation \ref{hpr:coeff1}, the vertex remains the only unknown correlation function in equation \ref{hpr2:dse}, while the parameter A is calculated directly from the DSE of scalar propagator in equation \ref{spr3:dse}.
\par
Higgs propagators for $m_{h}=125.09$ GeV are shown in figures \ref{fig:hprs125ew} and \ref{fig:hprs125tev} for different parameters of the theory at 4 cutoff values. An immediate observation is strong insensitivity of Higgs propagators over bare scalar masses, couplings as well as cutoff values. As the propagators manifest very close to the tree level propagators, they are monotonically decreasing functions, which indicates a physical particle. Similar situation is observed for $m_{h}=160.0$ GeV as shown in figures \ref{fig:hprs160ew} and \ref{fig:hprs160tev}. This feature is also observed for the SM Higgs boson which has low contributions beyond tree level. However, what is peculiar in this model is significantly smaller contributions from self energies.
\par
Higgs self energy terms are plotted in figures \ref{fig:hsen125ew} and \ref{fig:hsen125tev} for $m_{h}=125.09$ GeV, and \ref{fig:hsen160ew} and \ref{fig:hsen160tev} for $m_{h}=160.0$ GeV. It is observed that, though there exists no competition between the contribution from self energies and the tree level structure ($p^{2}+m_{h}^{2}$), there is indeed a certain dependence which is relatively more distinct for higher coupling values and the scalar bare mass at 1.0 GeV, self energies for $\lambda=2.0$ are shown in the figures. Their quantitative behavior is relatively more prominent for higher cutoff values. The dependence is found to be stronger for momentum lower than $1$ TeV, where Higgs masses as well as the renormalization points lie. As for the case of scalar self energies, both positive and negative contributions from Higgs self energies are observed. The cutoff effects are particularly stronger below $\Lambda < 5$ TeV, which appear prominently on figures \ref{fig:hsen125ew}, \ref{fig:hsen125tev}, \ref{fig:hsen160ew}, and \ref{fig:hsen160tev}.
\par
The parameter $\alpha$, shown in figures \ref{fig:htermlight} and \ref{fig:htermheavy}, remains very close to 1 throughout the explored parameter space. It indicates that the deviations from the tree level expression may have features mostly due to a relatively stronger role of the self energy term, however small in magnitude, than the parameter $\alpha$.
\par
Given the observed less sensitivity of Higgs propagators in parameter space as well as cutoff values, and that this behavior has already been observed in other studies \cite{Gies:2017zwf}, it is speculated that even if Higgs mass was not fixed at its physical mass it may not have improved sensitivity of Higgs propagators in the considered region of the parameter space. Hence, the mass of Higgs may not have changed drastically from the bare mass value, particularly for a large bare mass value around electroweak scale, i.e. 100 GeV.
% *********************************************************************
% *********************************************************************
% *********************************************************************
% ************************ Higgs scalar vertex ************************
% *********************************************************************
% *********************************************************************
% *********************************************************************
\subsection{Higgs-scalar Vertex} \label{sec:vtx}
\begin{figure}
\centering
\parbox{1.0\linewidth}
{
\includegraphics[width=\linewidth]{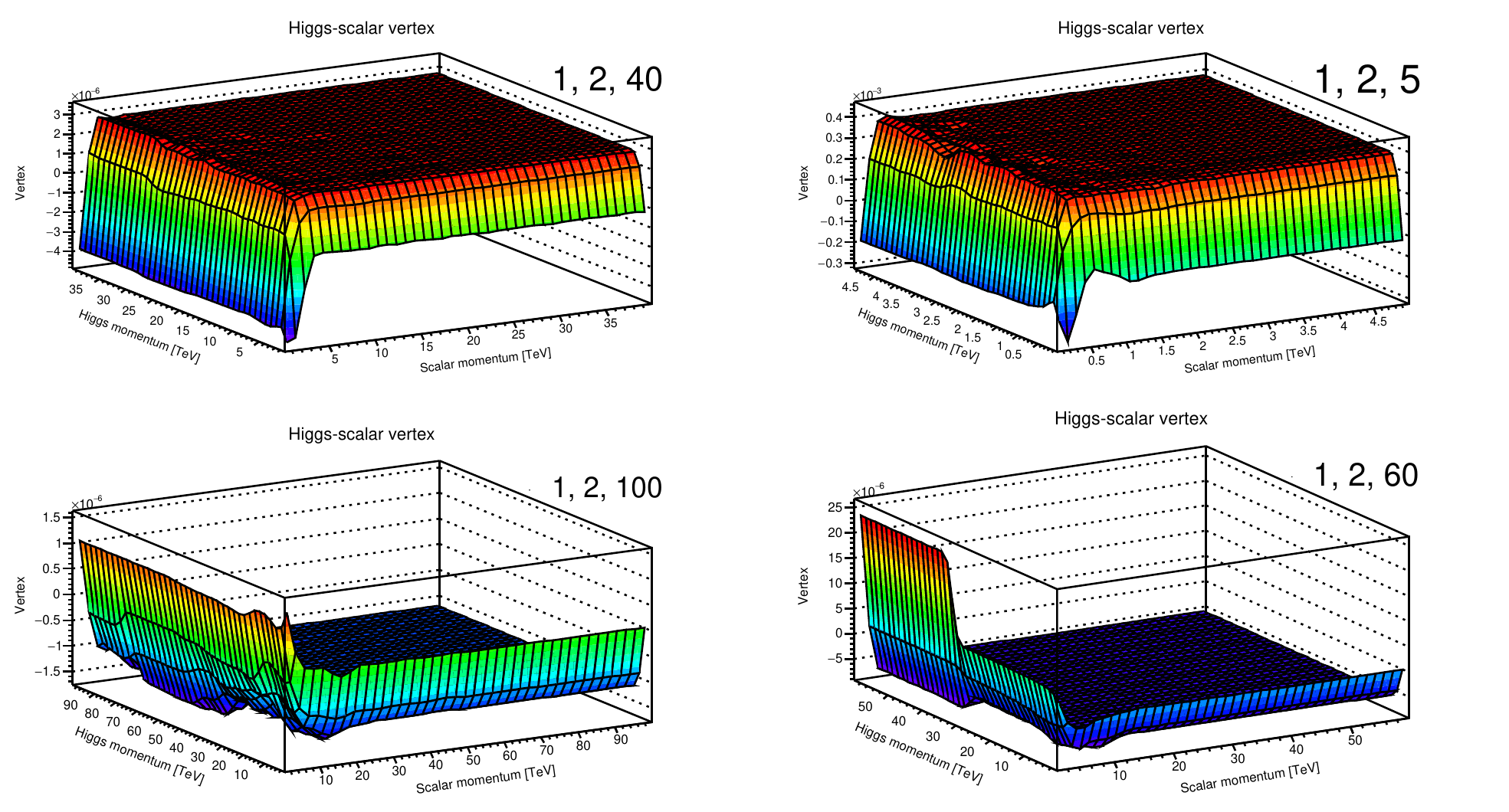}
\caption{\label{fig:vl12p0} Higgs-scalar vertex for $m_{h}=125.09$ GeV, $m_{s}=1$ GeV, coupling $\lambda=2.0$, for four cutoff values ($\Lambda$) given in TeV. The figures are labelled as $(m_{s},\lambda,\Lambda)$.}
}
\end{figure}
\begin{figure}
\centering
\parbox{1.0\linewidth}
{
\includegraphics[width=\linewidth]{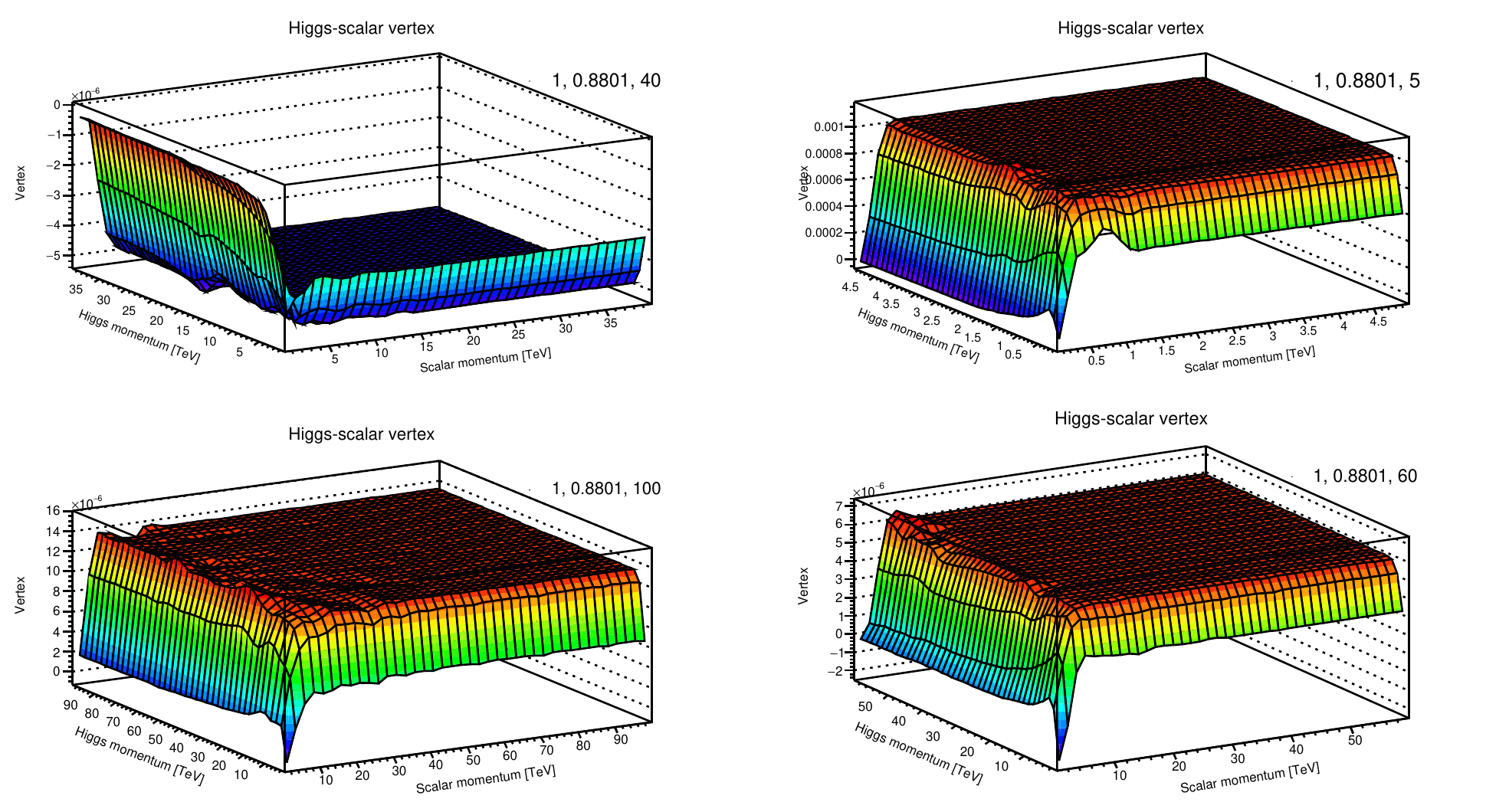}
\caption{\label{fig:vl10p8801} Higgs-scalar vertex for $m_{h}=125.09$ GeV, $m_{s}=1$ GeV, coupling $\lambda=0.8801$, for four cutoff values ($\Lambda$) given in TeV. The figures are labelled as $(m_{s},\lambda,\Lambda)$.}
}
\end{figure}
\begin{figure}
\centering
\parbox{1.0\linewidth}
{
\includegraphics[width=\linewidth]{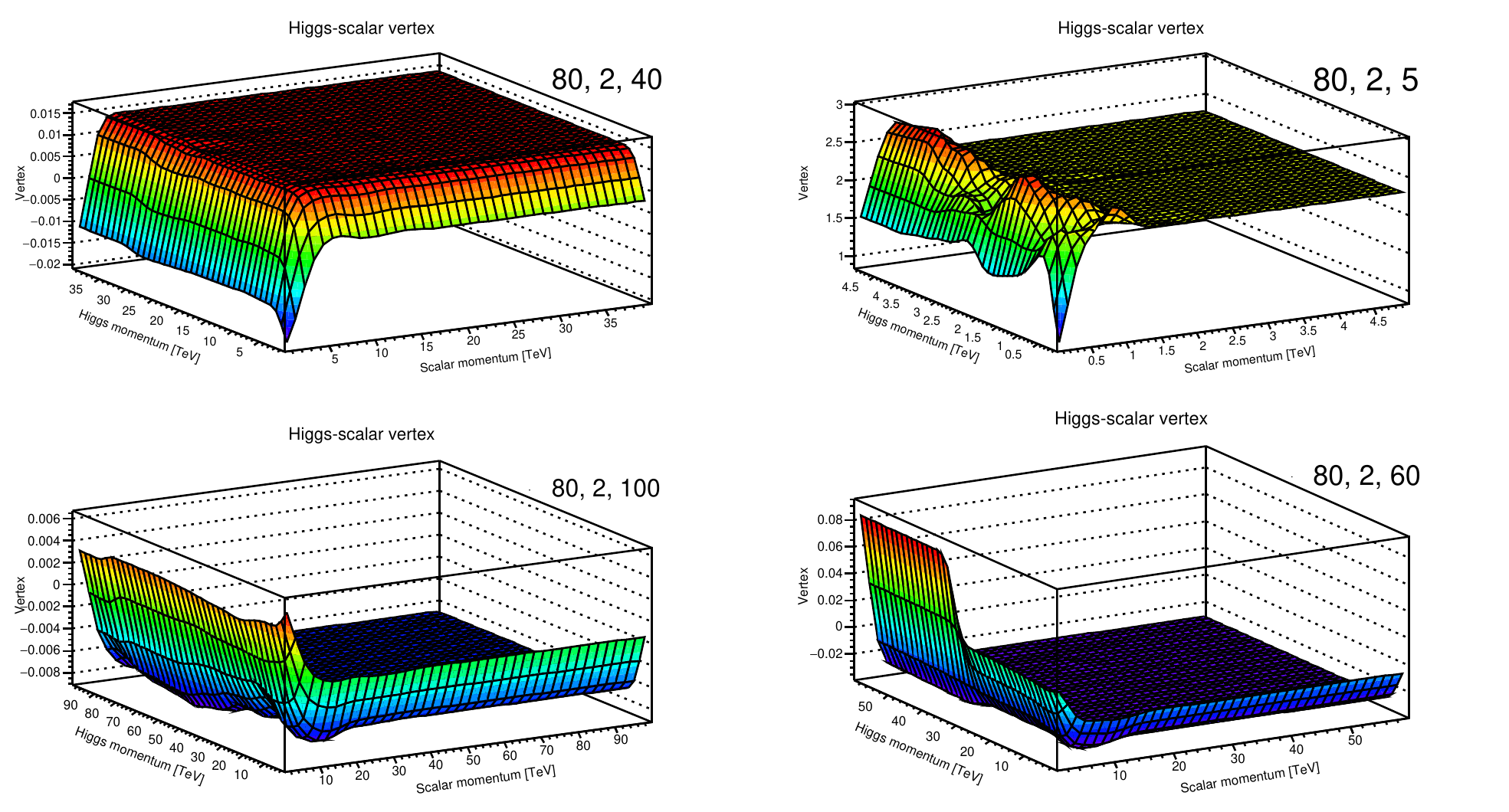}
\caption{\label{fig:vl802p0} Higgs-scalar vertex for $m_{h}=125.09$ GeV, $m_{s}=80$ GeV, coupling $\lambda=2.0$, for four cutoff values ($\Lambda$) given in TeV. The figures are labelled as $(m_{s},\lambda,\Lambda)$.}
}
\end{figure}
\begin{figure}
\centering
\parbox{1.0\linewidth}
{
\includegraphics[width=\linewidth]{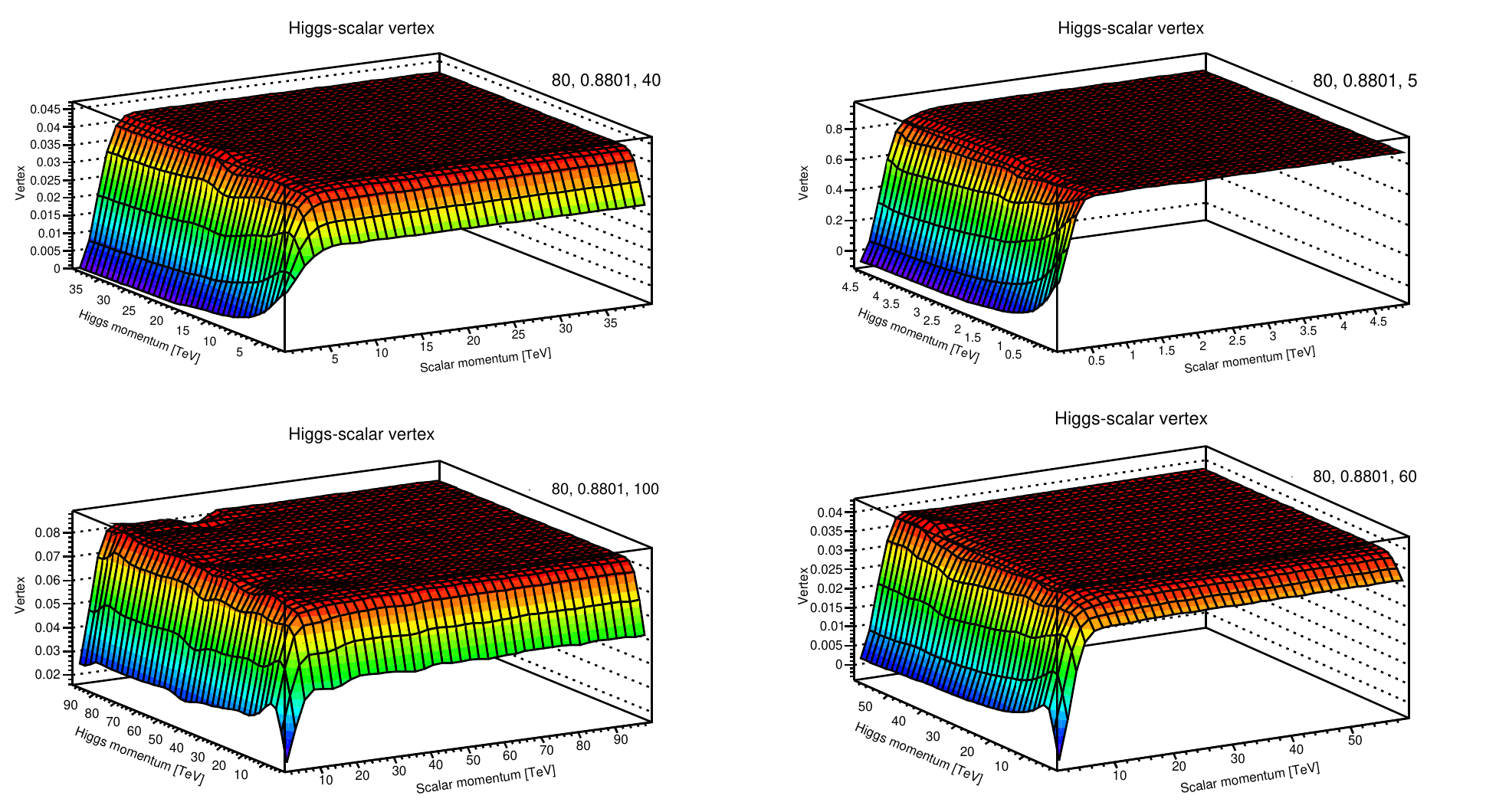}
\caption{\label{fig:vl800p8801} Higgs-scalar vertex for $m_{h}=125.09$ GeV, $m_{s}=80$ GeV, coupling $\lambda=0.8801$, for four cutoff values ($\Lambda$) given in TeV. The figures are labelled as $(m_{s},\lambda,\Lambda)$.}
}
\end{figure}
\begin{figure}
\centering
\parbox{1.0\linewidth}
{
\includegraphics[width=\linewidth]{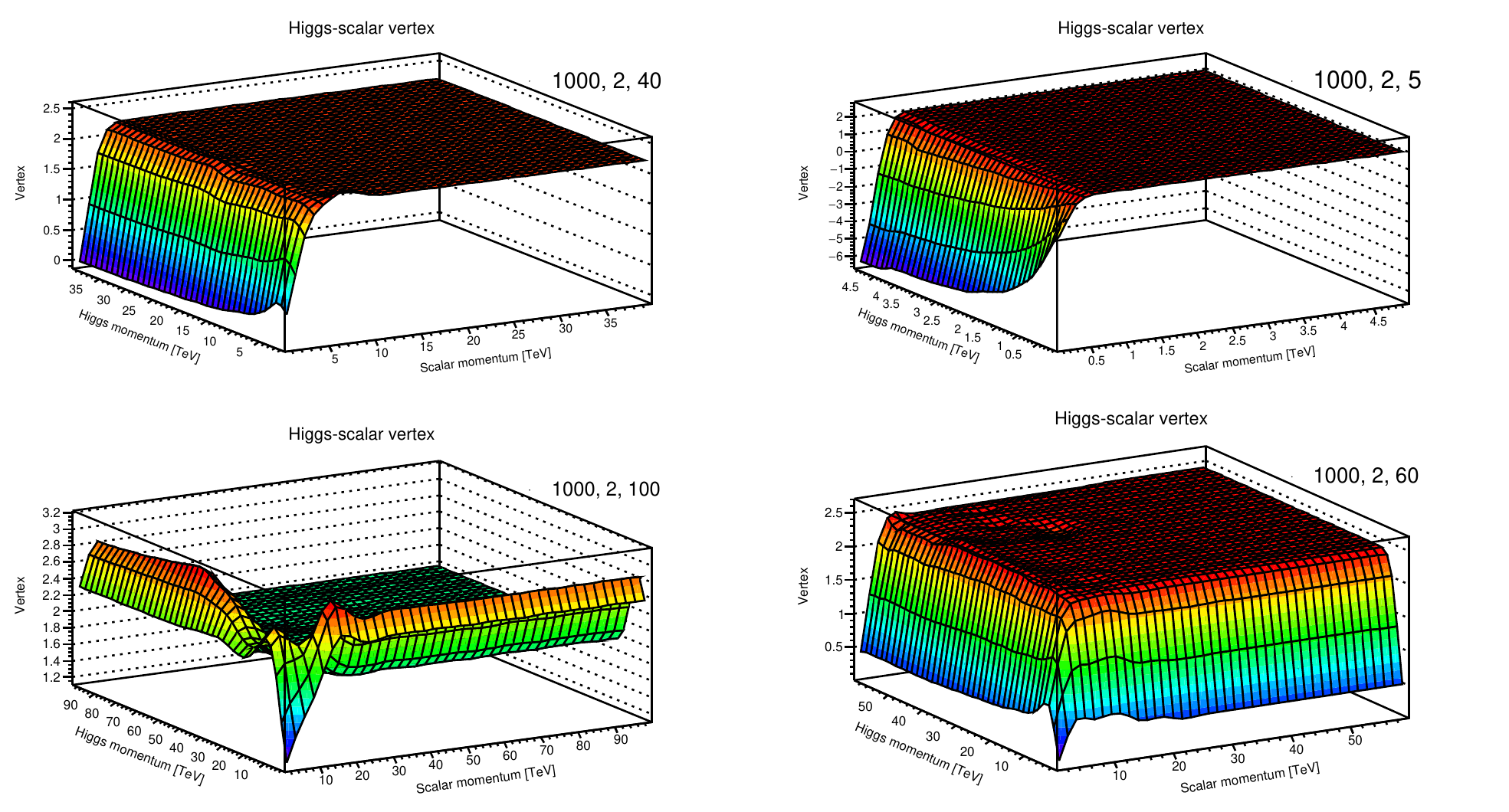}
\caption{\label{fig:vl10002p0} Higgs-scalar vertex for $m_{h}=125.09$ GeV, $m_{s}=1000$ GeV, coupling $\lambda=2.0$, for four cutoff values ($\Lambda$) given in TeV. The figures are labelled as $(m_{s},\lambda,\Lambda)$.}
}
\end{figure}
\begin{figure}
\centering
\parbox{1.0\linewidth}
{
\includegraphics[width=\linewidth]{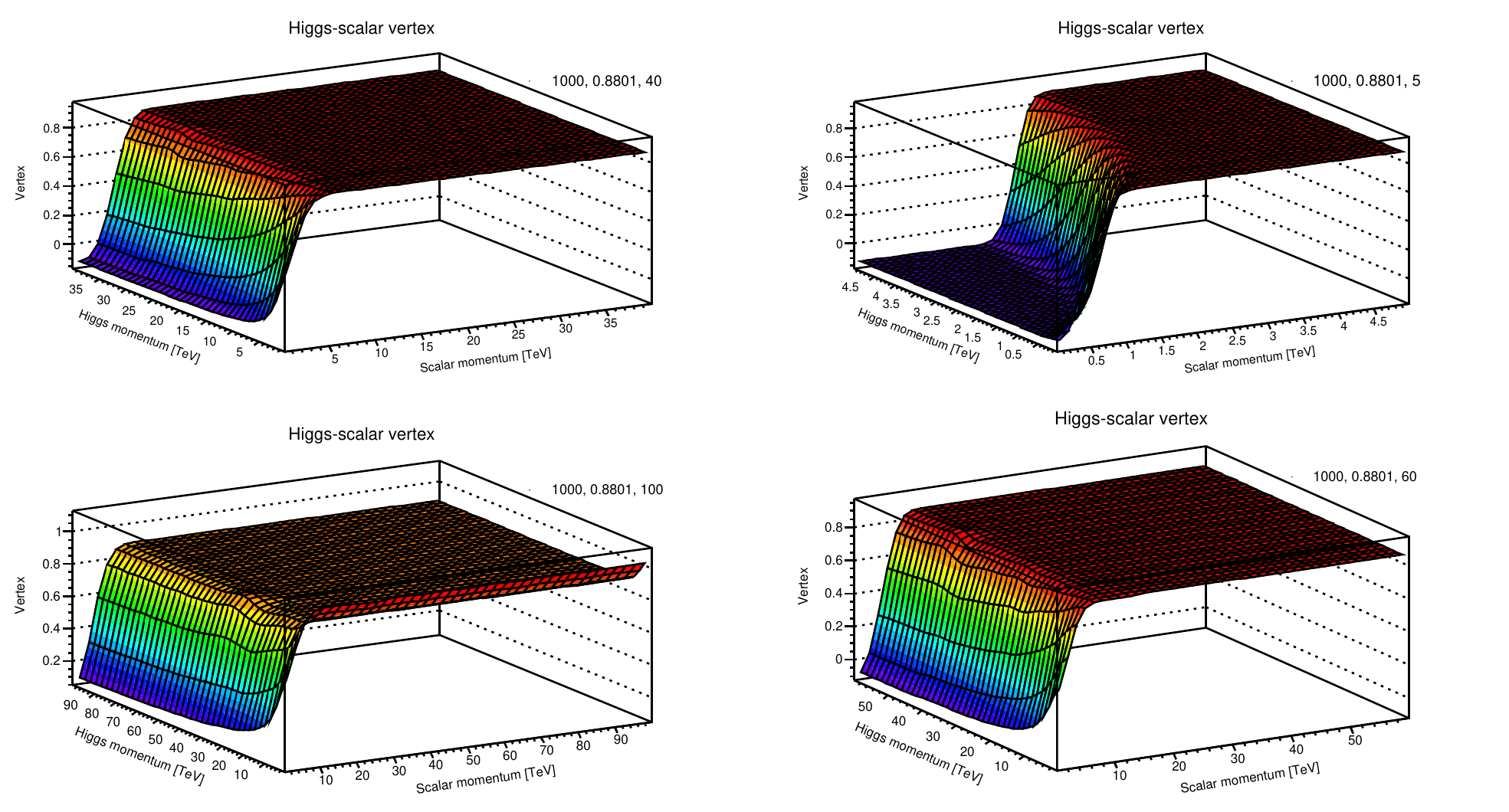}
\caption{\label{fig:vl10000p8801} Higgs-scalar vertex for $m_{h}=125.09$ GeV, $m_{s}=1000$ GeV, coupling $\lambda=0.8801$, for four cutoff values ($\Lambda$) given in TeV. The figures are labelled as $(m_{s},\lambda,\Lambda)$.}
}
\end{figure}
\begin{figure}
\centering
\parbox{1.0\linewidth}
{
\includegraphics[width=\linewidth]{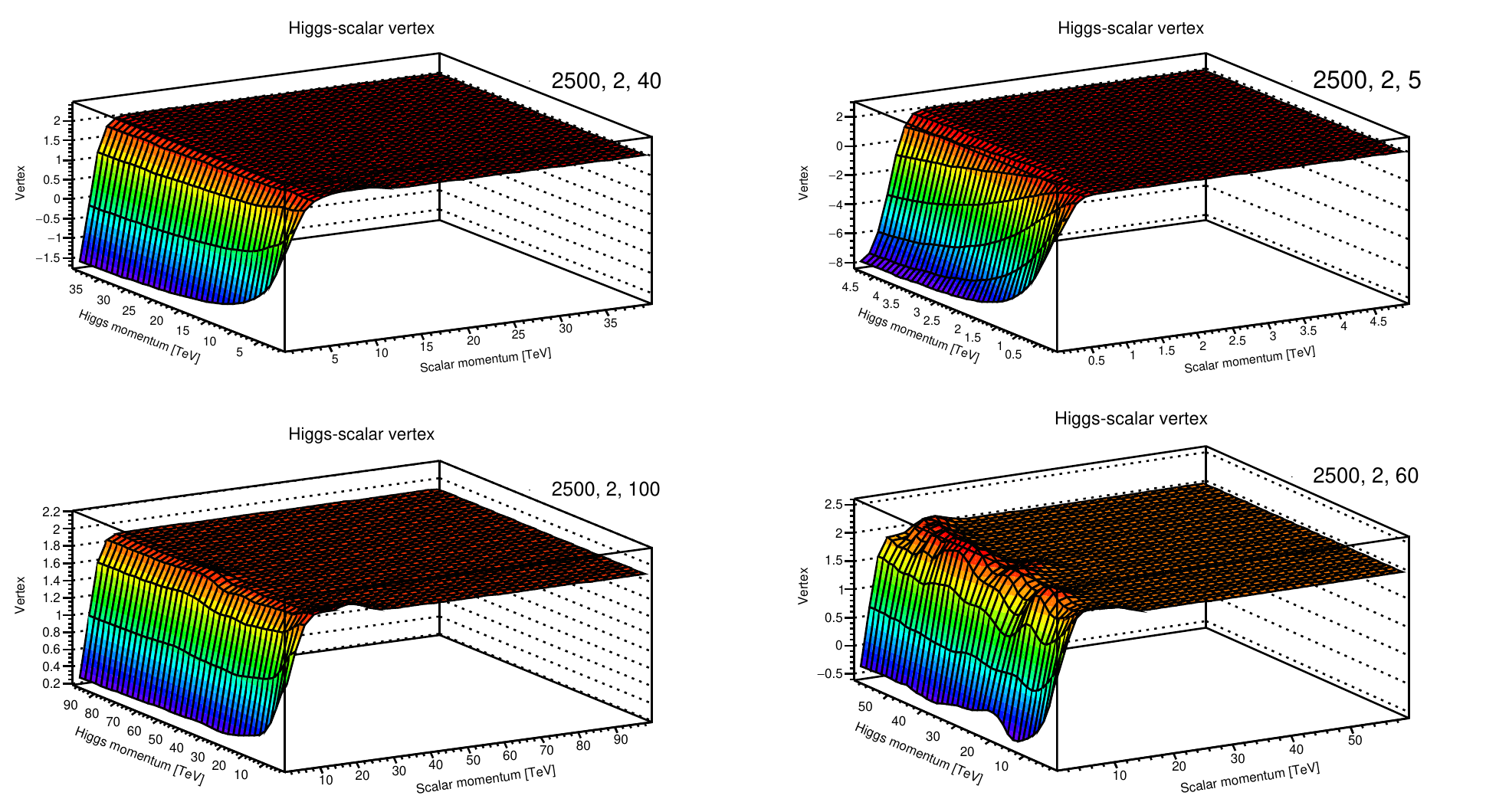}
\caption{\label{fig:vl25002p0} Higgs-scalar vertex for $m_{h}=125.09$ GeV, $m_{s}=2500$ GeV, coupling $\lambda=2.0$, for four cutoff values ($\Lambda$) given in TeV. The figures are labelled as $(m_{s},\lambda,\Lambda)$.}
}
\end{figure}
\begin{figure}
\centering
\parbox{1.0\linewidth}
{
\includegraphics[width=\linewidth]{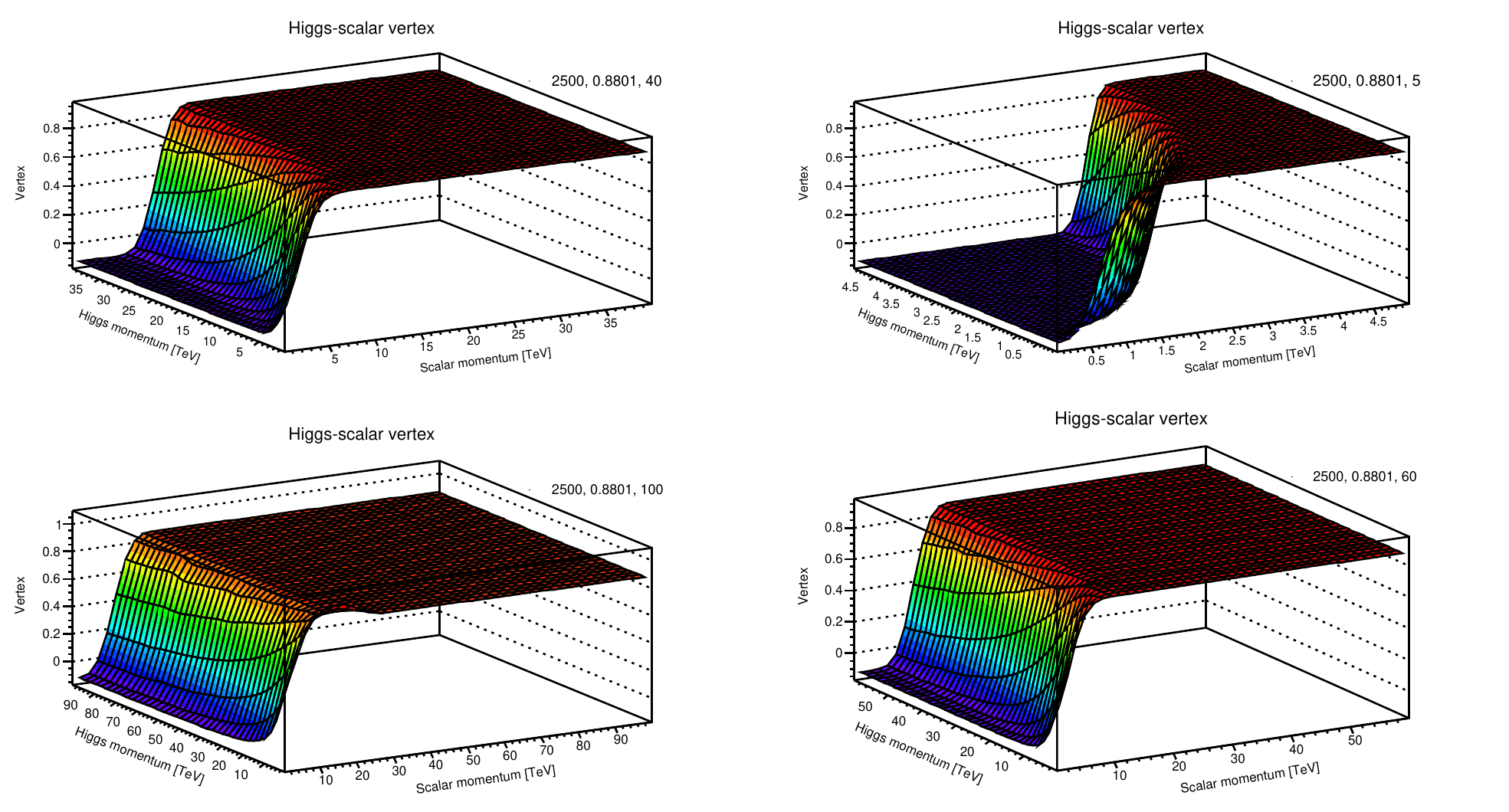}
\caption{\label{fig:vl25000p8801} Higgs-scalar vertex for $m_{h}=125.09$ GeV, $m_{s}=2500$ GeV, coupling $\lambda=0.8801$, for four cutoff values ($\Lambda$) given in TeV. The figures are labelled as $(m_{s},\lambda,\Lambda)$.}
}
\end{figure}
\begin{figure}
\centering
\parbox{1.0\linewidth}
{
\includegraphics[width=\linewidth]{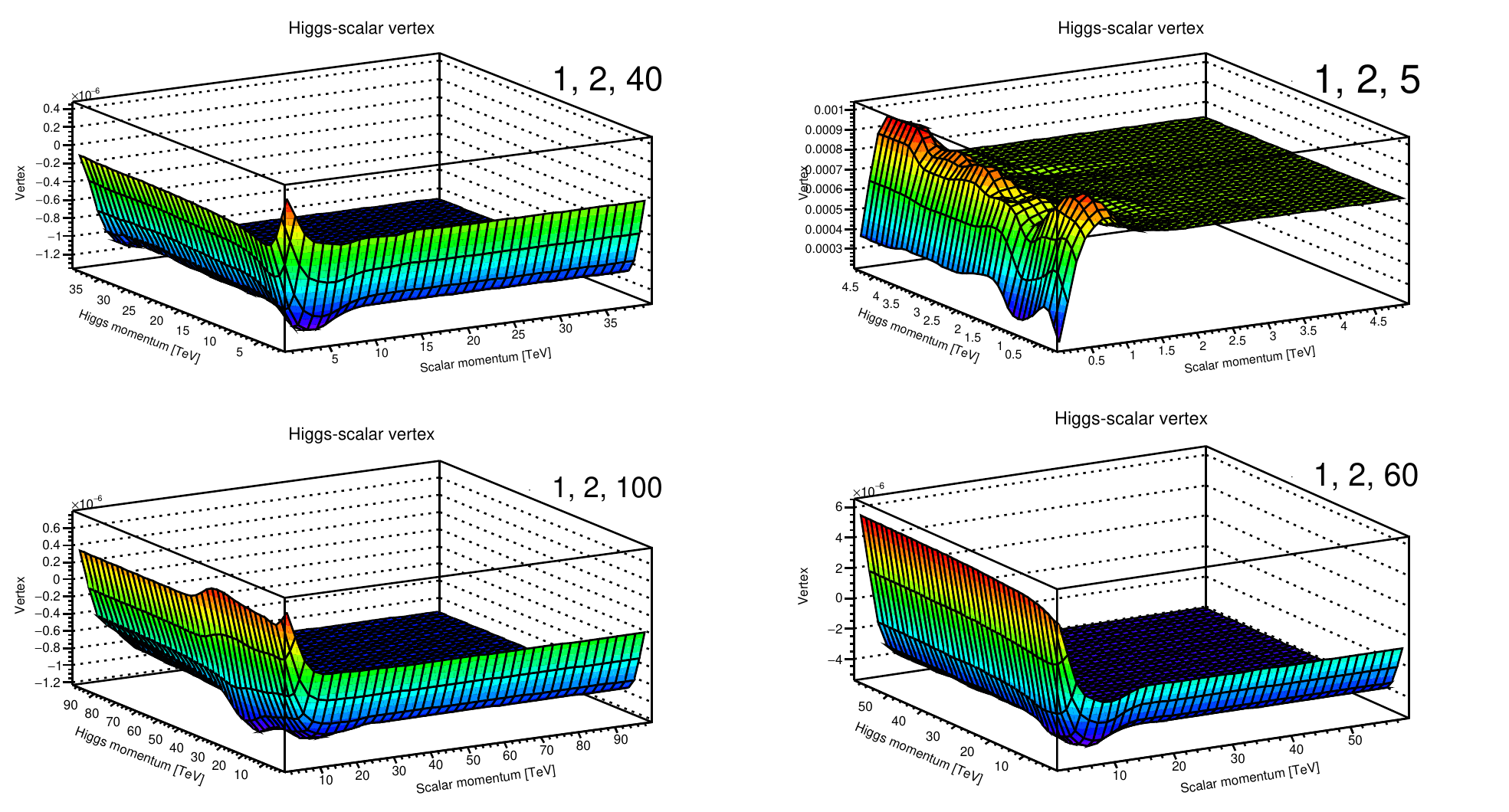}
\caption{\label{fig:vh12p0} Higgs-scalar vertex for $m_{h}=160.0$ GeV, $m_{s}=1$ GeV, coupling $\lambda=2.0$, for four cutoff values ($\Lambda$) given in TeV. The figures are labelled as $(m_{s},\lambda,\Lambda)$.}
}
\end{figure}
\begin{figure}
\centering
\parbox{1.0\linewidth}
{
\includegraphics[width=\linewidth]{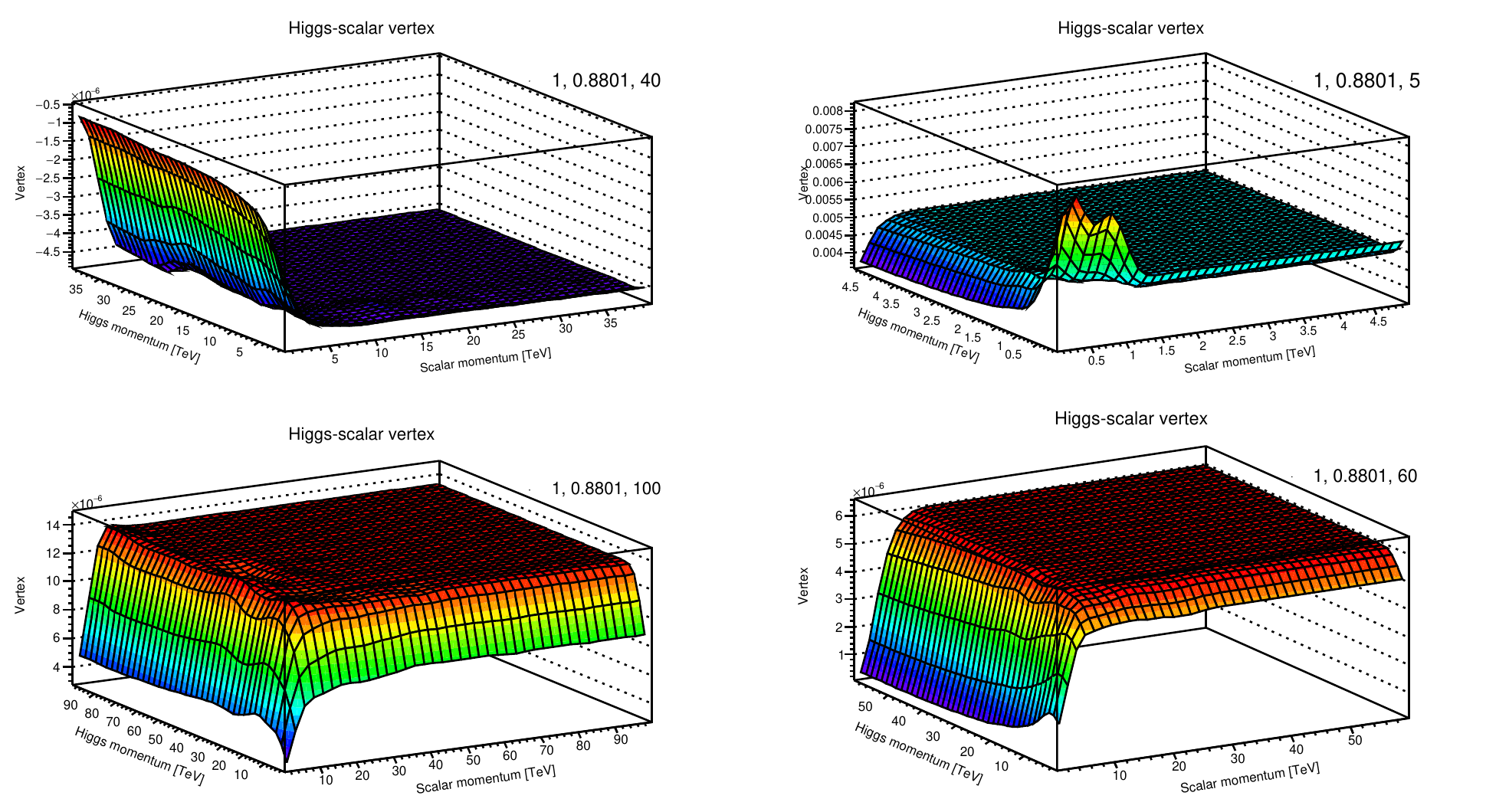}
\caption{\label{fig:vh10p8801} Higgs-scalar vertex for $m_{h}=160.0$ GeV, $m_{s}=1$ GeV, coupling $\lambda=0.8801$, for four cutoff values ($\Lambda$) given in TeV. The figures are labelled as $(m_{s},\lambda,\Lambda)$.}
}
\end{figure}
\begin{figure}
\centering
\parbox{1.0\linewidth}
{
\includegraphics[width=\linewidth]{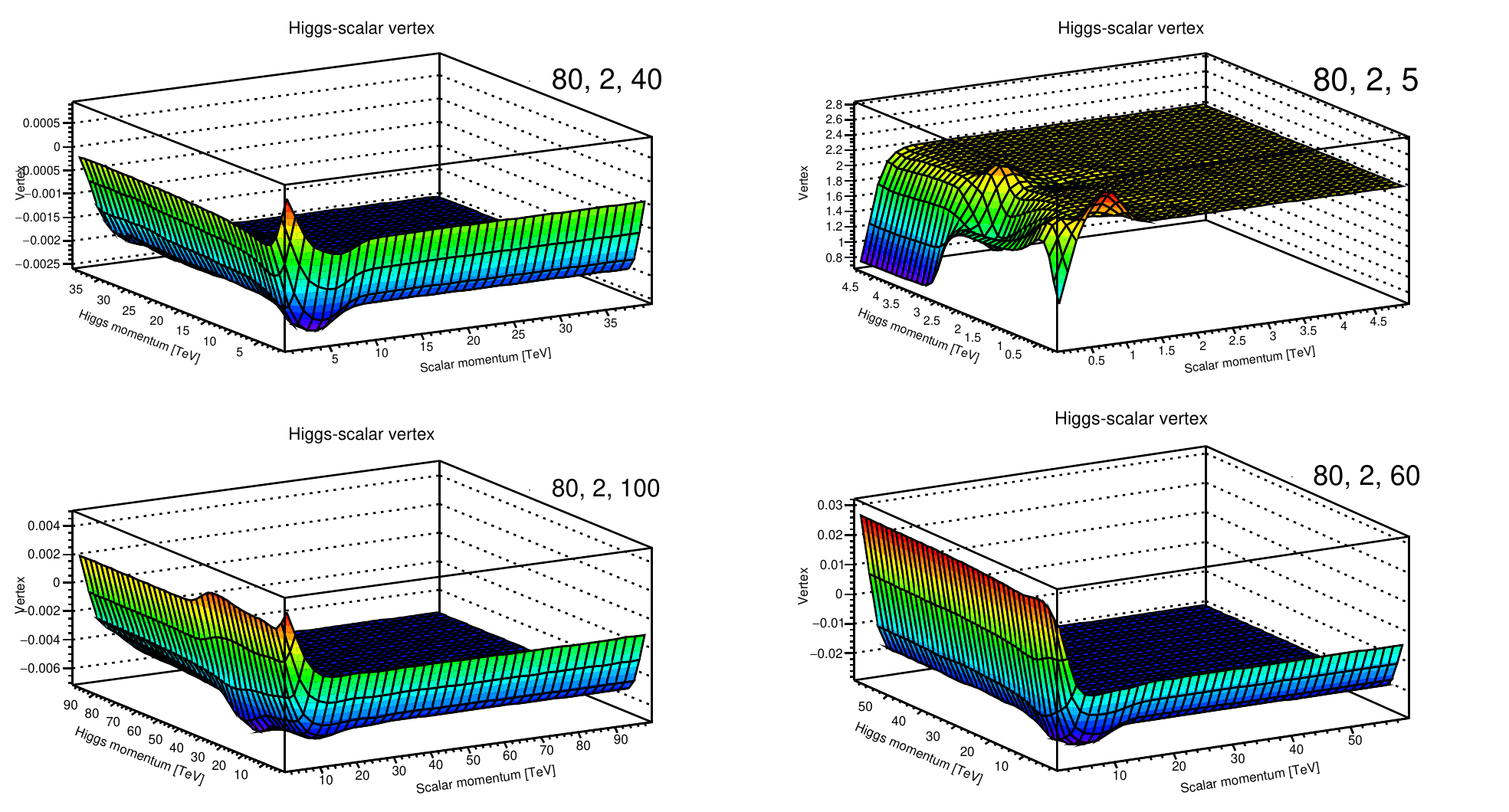}
\caption{\label{fig:vh802p0} Higgs-scalar vertex for $m_{h}=160.0$ GeV, $m_{s}=80$ GeV, coupling $\lambda=2.0$, for four cutoff values ($\Lambda$) given in TeV. The figures are labelled as $(m_{s},\lambda,\Lambda)$.}
}
\end{figure}
\begin{figure}
\centering
\parbox{1.0\linewidth}
{
\includegraphics[width=\linewidth]{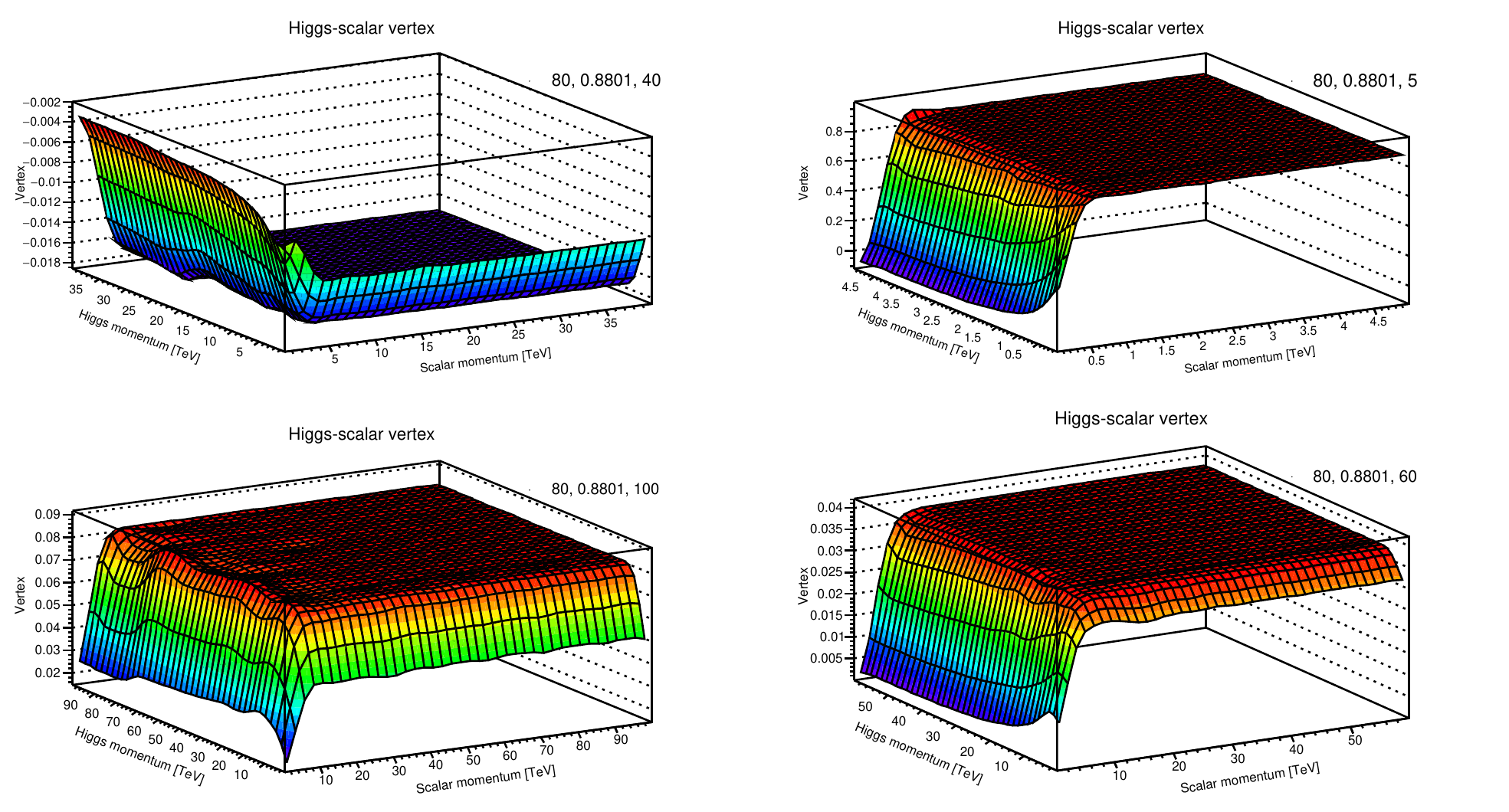}
\caption{\label{fig:vh800p8801} Higgs-scalar vertex for $m_{h}=160.0$ GeV, $m_{s}=80$ GeV, coupling $\lambda=0.8801$, for four cutoff values ($\Lambda$) given in TeV. The figures are labelled as $(m_{s},\lambda,\Lambda)$.}
}
\end{figure}
\begin{figure}
\centering
\parbox{1.0\linewidth}
{
\includegraphics[width=\linewidth]{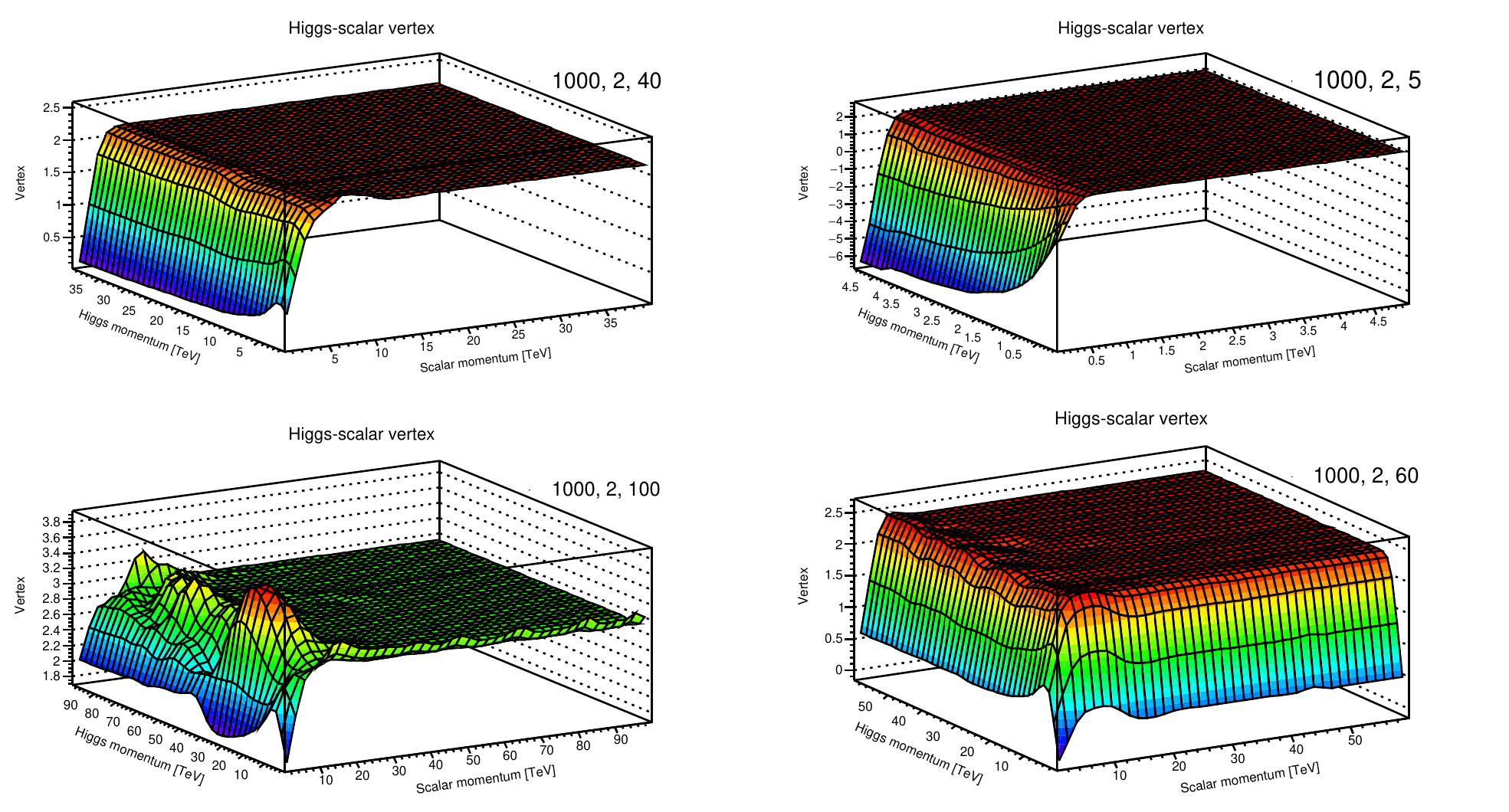}
\caption{\label{fig:vh10002p0} Higgs-scalar vertex for $m_{h}=160.0$ GeV, $m_{s}=1000$ GeV, coupling $\lambda=2.0$, for four cutoff values ($\Lambda$) given in TeV. The figures are labelled as $(m_{s},\lambda,\Lambda)$.}
}
\end{figure}
\begin{figure}
\centering
\parbox{1.0\linewidth}
{
\includegraphics[width=\linewidth]{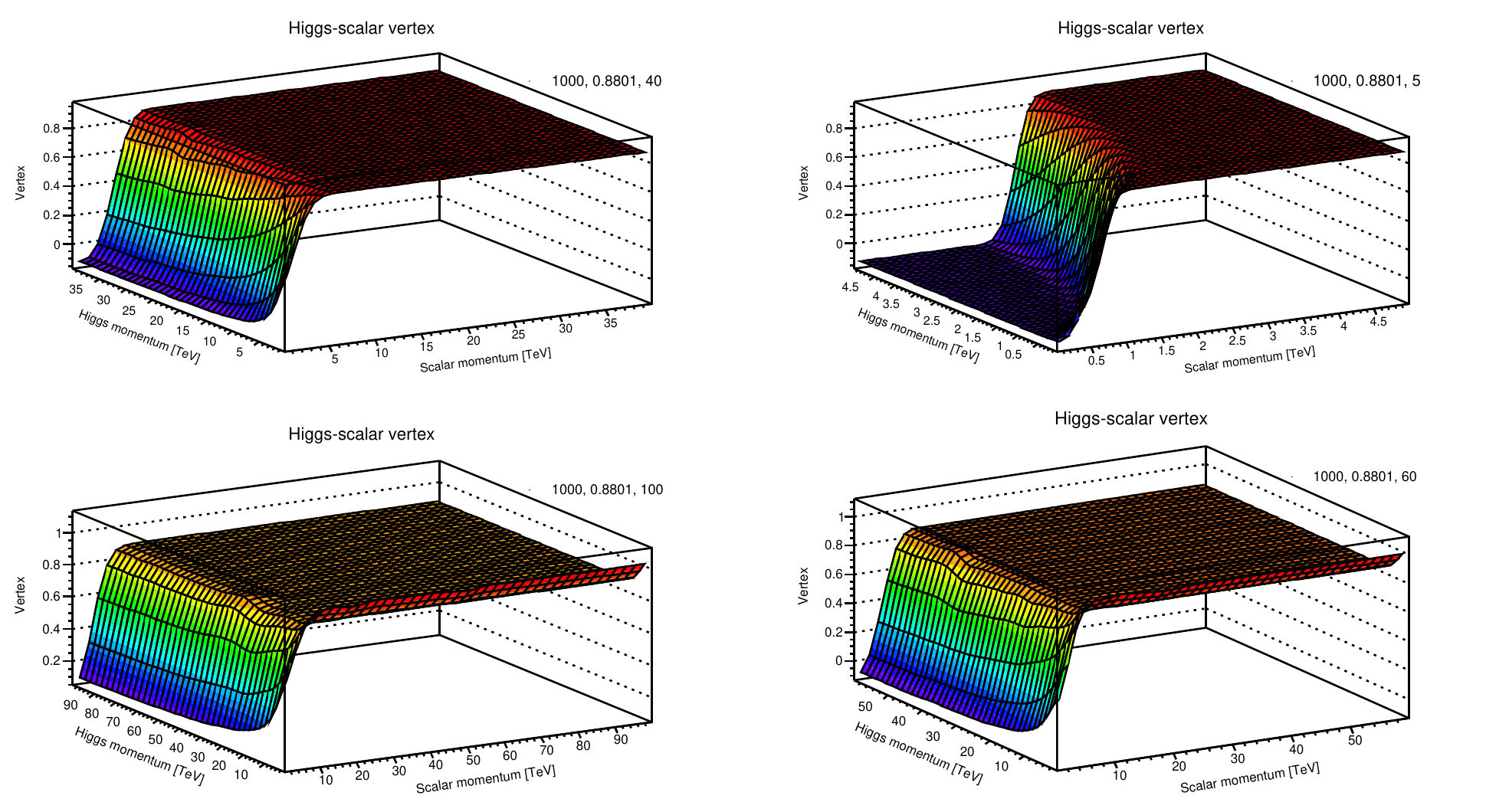}
\caption{\label{fig:vh10000p8801} Higgs-scalar vertex for $m_{h}=160.0$ GeV, $m_{s}=1000$ GeV, coupling $\lambda=0.8801$, for four cutoff values ($\Lambda$) given in TeV. The figures are labelled as $(m_{s},\lambda,\Lambda)$.}
}
\end{figure}
\begin{figure}
\centering
\parbox{1.0\linewidth}
{
\includegraphics[width=\linewidth]{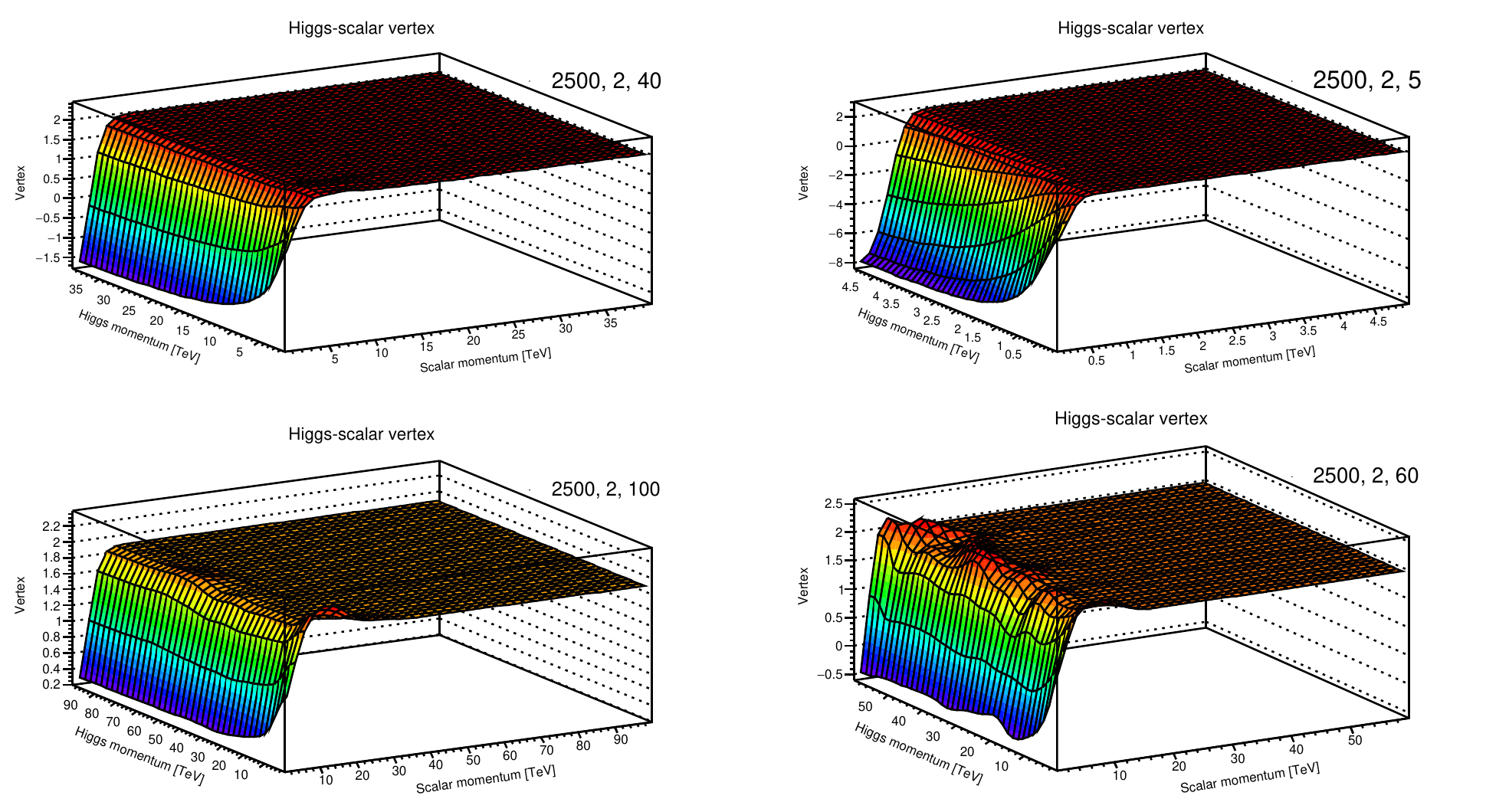}
\caption{\label{fig:vh25002p0} Higgs-scalar vertex for $m_{h}=160.0$ GeV, $m_{s}=2500$ GeV, coupling $\lambda=2.0$, for four cutoff values ($\Lambda$) given in TeV. The figures are labelled as $(m_{s},\lambda,\Lambda)$.}
}
\end{figure}
\begin{figure}
\centering
\parbox{1.0\linewidth}
{
\includegraphics[width=\linewidth]{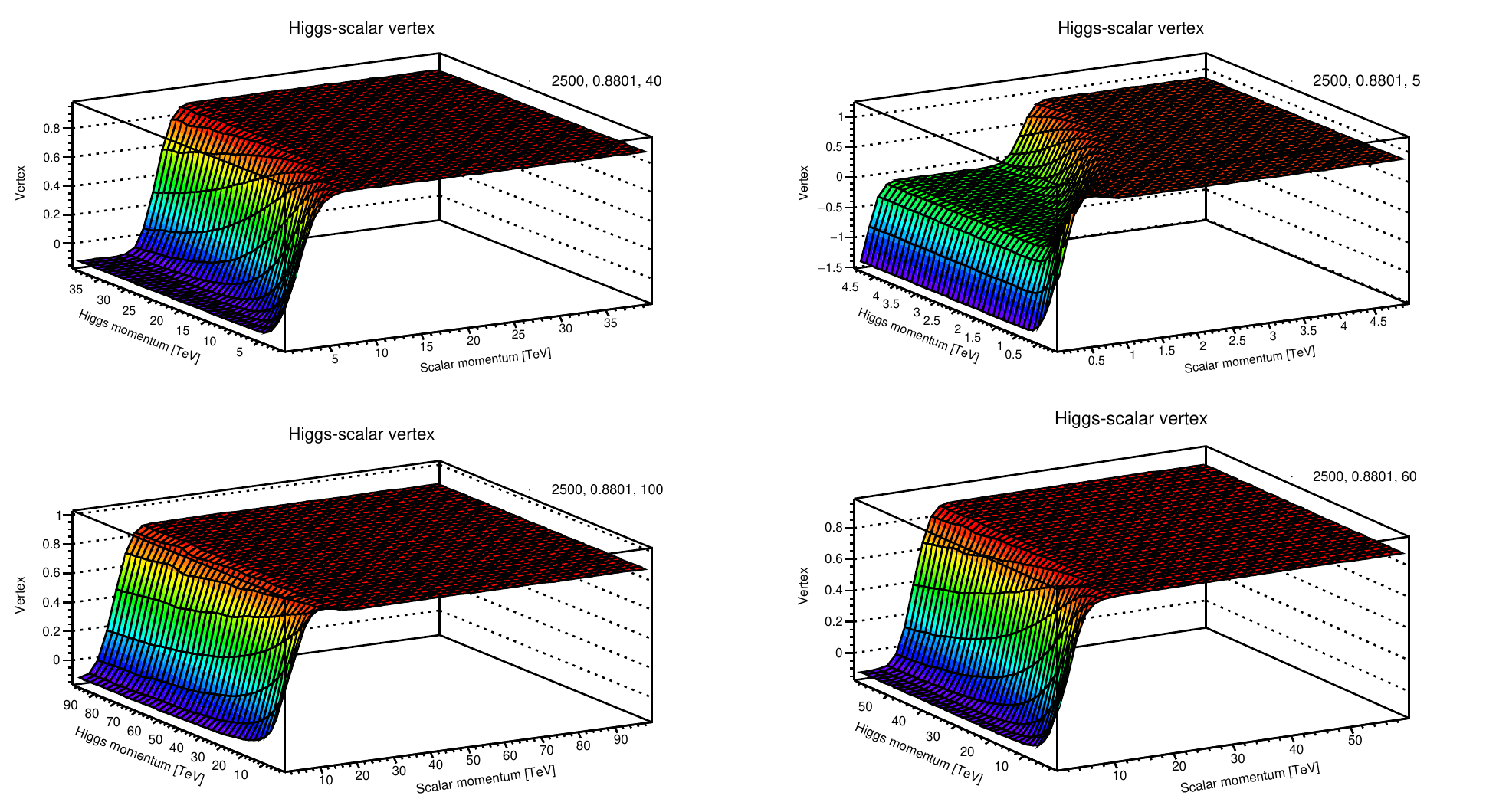}
\caption{\label{fig:vh25000p8801} Higgs-scalar vertex for $m_{h}=160.0$ GeV, $m_{s}=2500$ GeV, coupling $\lambda=0.8801$, for four cutoff values ($\Lambda$) given in TeV. The figures are labelled as $(m_{s},\lambda,\Lambda)$.}
}
\end{figure}
\begin{figure}
\centering
\parbox{1.0\linewidth}
{
\includegraphics[width=\linewidth]{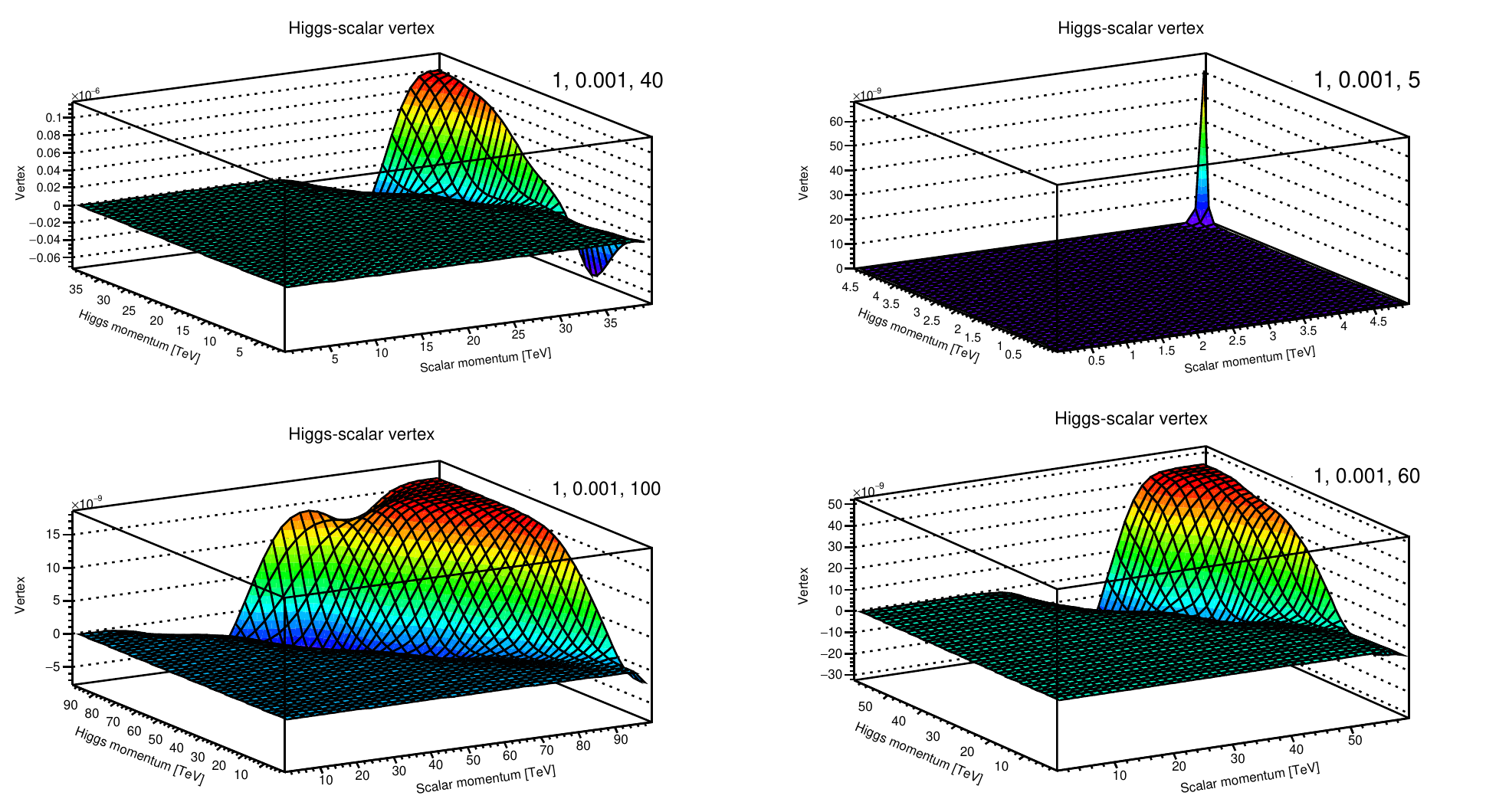}
\caption{\label{fig:vl10p001} Higgs-scalar vertex for $m_{h}=125.09$ GeV, $m_{s}=1$ GeV, coupling $\lambda=0.001$, for four cutoff values ($\Lambda$) given in TeV. The figures are labelled as $(m_{s},\lambda,\Lambda)$.}
}
\end{figure}
\begin{figure}
\centering
\parbox{1.0\linewidth}
{
\includegraphics[width=\linewidth]{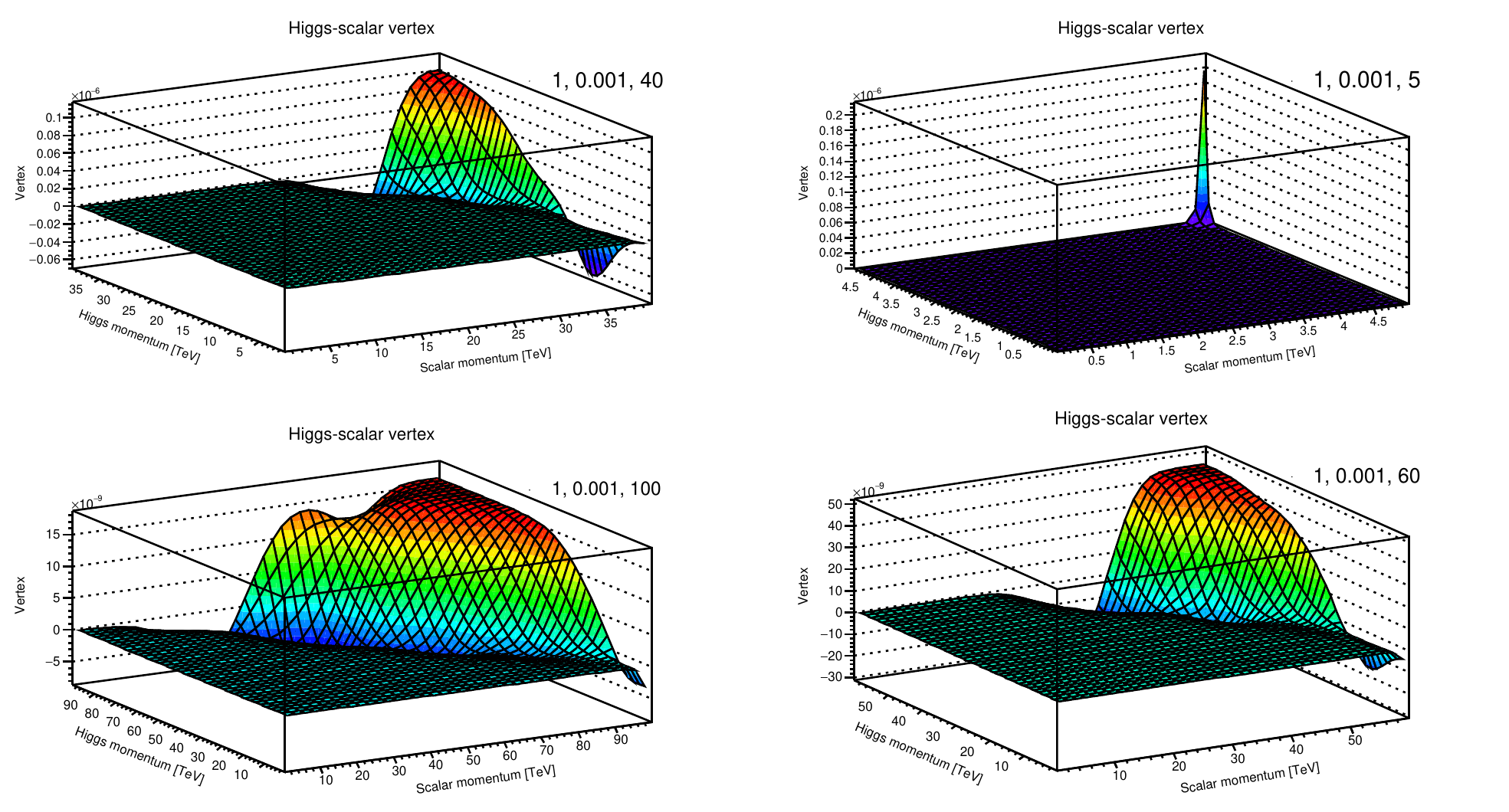}
\caption{\label{fig:vh10p001} Higgs-scalar vertex for $m_{h}=160.0$ GeV, $m_{s}=1$ GeV, coupling $\lambda=0.001$, for four cutoff values ($\Lambda$) given in TeV. The figures are labelled as $(m_{s},\lambda,\Lambda)$.}
}
\end{figure}
\begin{figure}
\centering
\parbox{1.0\linewidth}
{
\includegraphics[width=\linewidth]{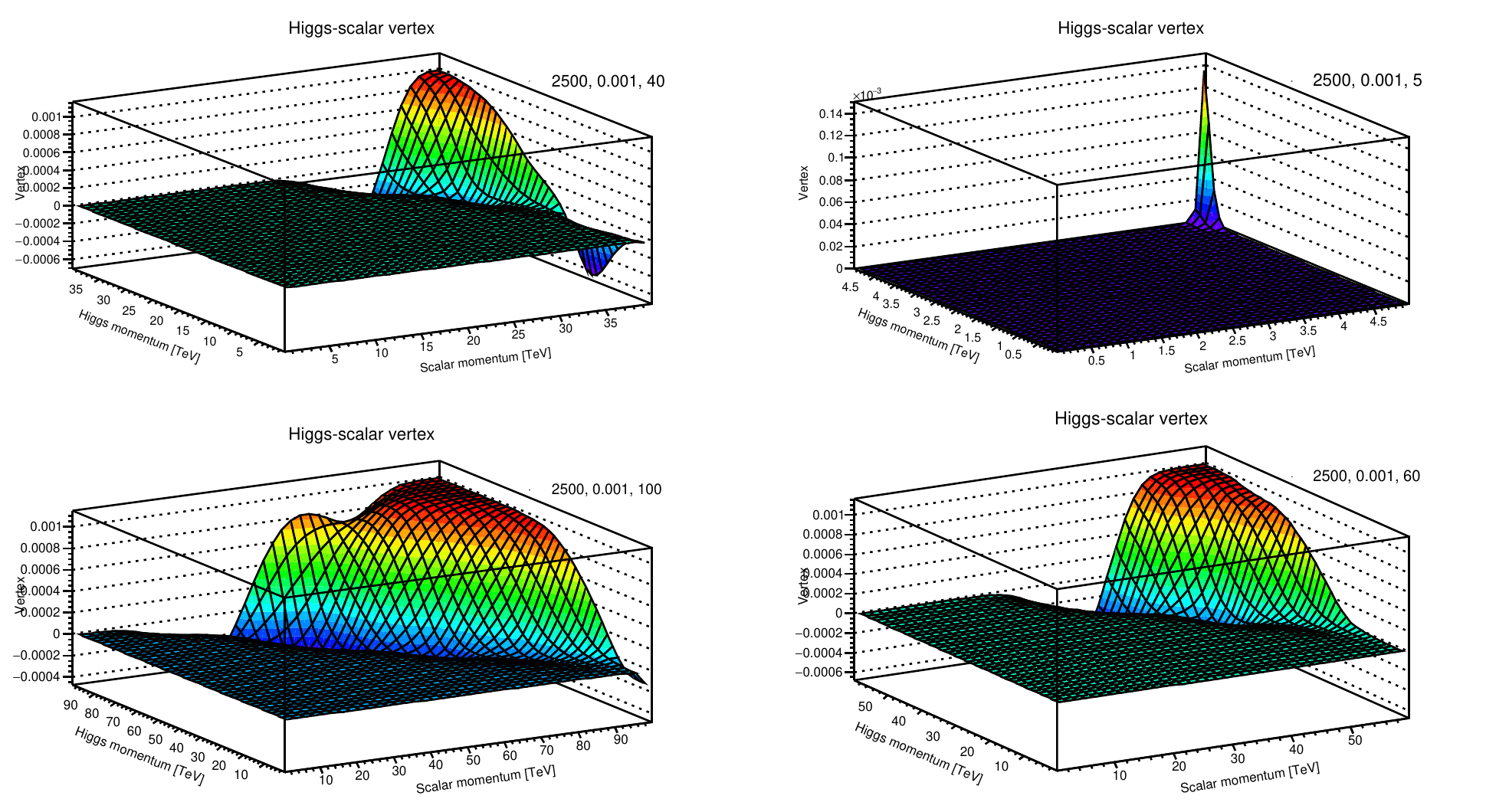}
\caption{\label{fig:vl25000p001} Higgs-scalar vertex for $m_{h}=125.09$ GeV, $m_{s}=2500$ GeV, coupling $\lambda=0.001$, for four cutoff values ($\Lambda$) given in TeV. The figures are labelled as $(m_{s},\lambda,\Lambda)$.}
}
\end{figure}
\begin{figure}
\centering
\parbox{1.0\linewidth}
{
\includegraphics[width=\linewidth]{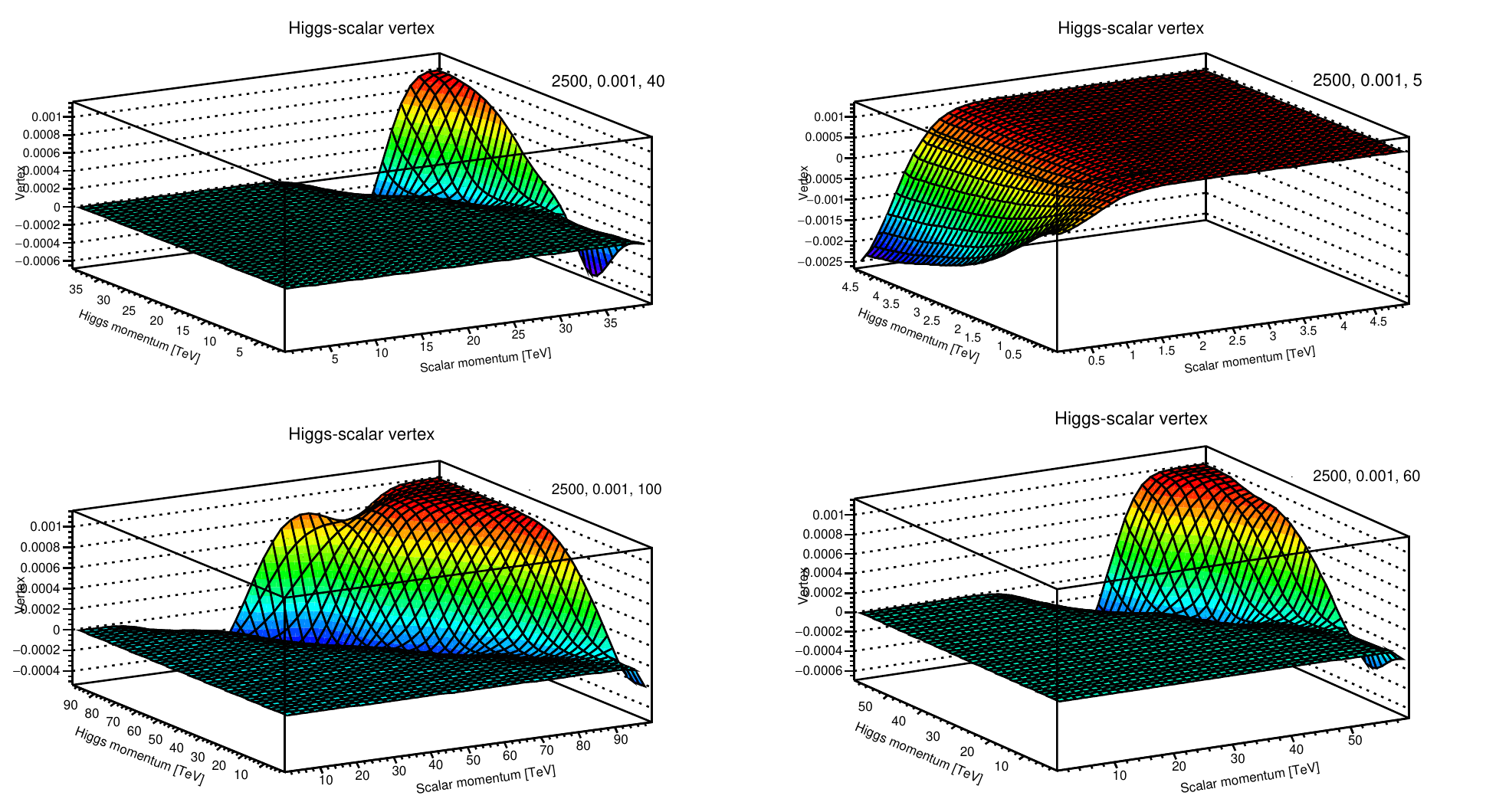}
\caption{\label{fig:vh25000p001} Higgs-scalar vertex for $m_{h}=160.0$ GeV, $m_{s}=2500$ GeV, coupling $\lambda=0.001$, for four cutoff values ($\Lambda$) given in TeV. The figures are labelled as $(m_{s},\lambda,\Lambda)$.}
}
\end{figure}
As Yukawa interaction vertex, defined in equation \ref{vtx:def}, is the only means for Higgs and scalar fields to interact in the Lagrangian, it contains all the non-trivial details regarding interactions in the model considered here. Despite the simplicity of the model, the vertices are found to contain highly non-trivial features and are shown in figures \ref{fig:vl12p0} - \ref{fig:vh25000p001} \footnote{The mild fluctuations in the region of low field momenta are found to have dependence on the density of the momentum points, which is lower than in higher momentum region due to the distribution endowed by Gaussian quadrature algorithm, in the region. Hence, these fluctuations are not to be related to the precision with which the correlation functions are calculated here.}.
\par
The first observation is presence of a major classification in terms of qualitative behavior of the vertices, i.e. either the vertices behave as a combination of two plateaus connected smoothly or a flat function with no significant dependence on field momenta while for very low momentum it exhibits strong dependence. Furthermore, for the later type of the vertices the steep behavior can be descending or ascending as the field momentum is increased which renders the vertices a subclassification.
\par
For the case of mostly flat vertices, the insensitive momentum region is found to have both positive and negative values of the vertices. At this point, it is inconclusive to take such infrared effects as a sign of a certain structure in the parameter space of the model. It is expected that studies of higher vertices may provide further details regarding such an structure, as reported in \cite{Maas:2014pba} for a different model.
\par
An interesting feature is a remarkable suppression of vertices for very low coupling values ($\lambda = 0.001$ in our case) \footnote{For such a low coupling value, the vertices require additional algorithm for suppressing local fluctuations, such as in \cite{Gattringer:2010zz}. Hence, on each momentum point not belonging to any extreme momentum on the boundary surface, the vertex $V(P^{1}_{t},P^{1}_{x})$ is improved to $V(P^{2}_{t},P^{2}_{x})$ twice, for each update during a computation, using the formula given as: $V(P^{2}_{t},P^{2}_{x}) = \frac{1}{5} (V(P^{1}_{t},P^{1}_{x})+V(P^{1}_{t}+\delta P^{1}_{t},P^{1}_{x})+V(P^{1}_{t}-\delta P^{1}_{t},P^{1}_{x})+V(P^{1}_{t},P^{1}_{x}+\delta P^{1}_{x})+V(P^{1}_{t},P^{1}_{x}-\delta P^{1}_{x}))$.}. For these vertices, the higher momentum region is considerably larger in magnitude compared to the suppressed region for lower momentum values, see figures \ref{fig:vl10p001} - \ref{fig:vh25000p001}. However, at this point these vertices can not be taken as a strong sign of trivial vertices as the vertices are many orders higher in magnitude than the preselected precision, and the suppression does not occur over the entire range of momentum values. Qualitatively, these vertices behave similarly, except for the case of very low cutoff, which is an indication that large cutoff effects in the model are prominent at cutoff of the order of few TeVs for small couplings as shown in the figures.
\par
The cutoff effects mentioned above remain considerable for higher couplings as well, particularly for the case of low scalar bare masses, see \ref{fig:vl12p0}, \ref{fig:vl802p0}, \ref{fig:vl10002p0}, \ref{fig:vh12p0}, \ref{fig:vh802p0}. The two plateau vertices are relatively more abundant for the case of low cutoff, which further suggest presence of cutoff effects in the model. It is found that vertices with infrared suppression is found more abundant for the case of higher scalar bare masses, see figures \ref{fig:vl25002p0}, \ref{fig:vl25000p8801}, \ref{fig:vh25002p0}, \ref{fig:vh25000p8801}, while for lower scalar bare mass vertices with such a behavior is not as abundant.
\par
Furthermore, the vertices are not found to be sensitive to Higgs mass which indicates that, should there be a phase structure in the model, it should have comparably more pronounced dependence on the scalar bare masses and the coupling values than the Higgs bare mass, see figures \ref{fig:vh12p0} - \ref{fig:vh25000p8801}.
\par
Lastly, as several vertices are also observed which do not completely belong of one of the two classifications of the vertices, it suggests a possibility of a second order phase transition in the phase space, if there is indeed a phase structure in the theory. It is not a surprise in physics involving Higgs interactions as such a behavior has already been found in other studies involving Higgs interactions, see \cite{Maas:2014pba} for instance.
\par
Most of the vertices in the theory are found to have least dependence in the ultraviolet end where both Higgs and the scalar propagators are qualitatively similar to their tree level structures, which can also be noticed from momentum dependence of their respective self energy terms. Hence, a picture qualitatively similar to what perturbative approach depicts emerges. However, many of the vertices have infrared region where the vertices seem to smoothly depart from their tree level structure. Overall, such a persistent behavior is taken as a sign of stability of the vertices extracted using the numerical approach explained above.
% *********************************************************************
% *********************************************************************
% *********************************************************************
% ************************ Renormalized masses ************************
% *********************************************************************
% *********************************************************************
% *********************************************************************
\subsection{Renormalized Masses} \label{sec:renmasses}
\begin{figure}
\centering
\parbox{1.0\linewidth}
{
\includegraphics[width=\linewidth]{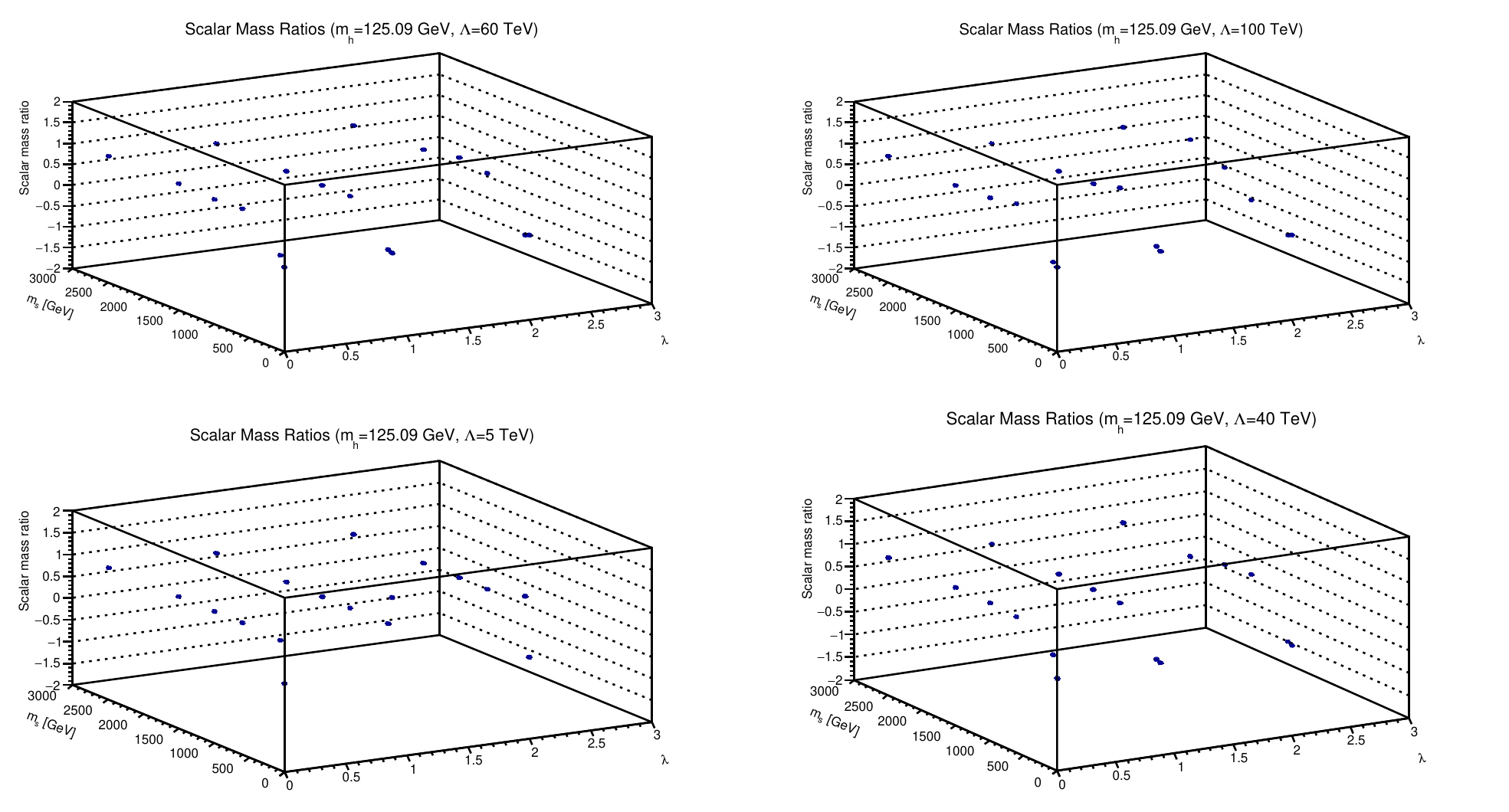}
\caption{\label{fig:slms1} Scalar mass squared ratios ($\frac{m^{2}_{s,r}}{m^{2}_{s}}$) for various couplings and scalar (bare) masses, for different cutoff values, are plotted for $m_{h}=125.09$ GeV.}
}
\centering
\parbox{1.0\linewidth}
{
\includegraphics[width=\linewidth]{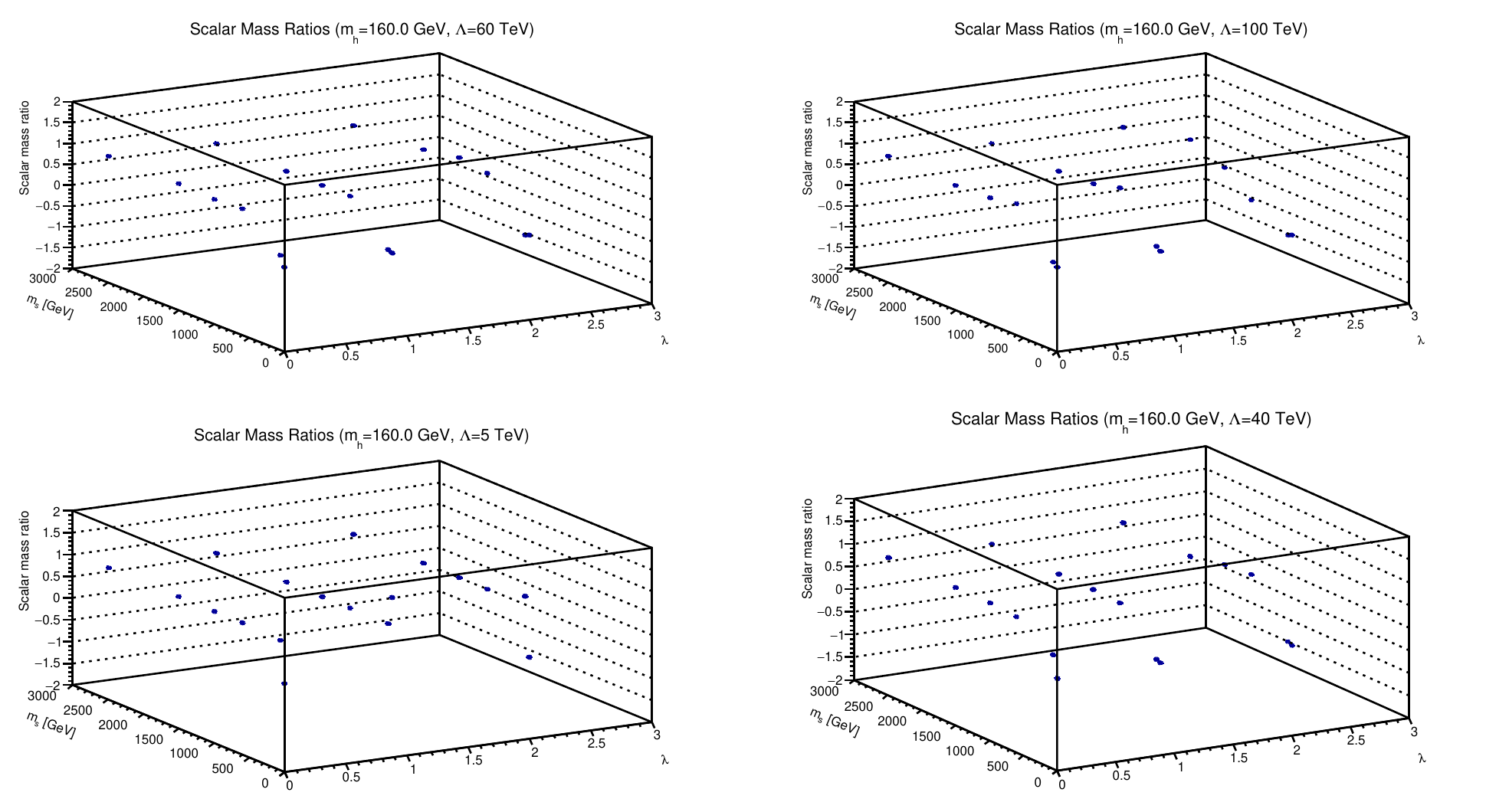}
\caption{\label{fig:shms1} Scalar mass squared ratios ($\frac{m^{2}_{s,r}}{m^{2}_{s}}$) for various couplings and scalar (bare) masses, for different cutoff values, are plotted for $m_{h}=160.0$ GeV.}
}
\end{figure}
From the perspective of phenomenology and measurements, renormalized masses are among the most important quantities in a model. The (squared) renormalized mass for scalar field, using equations \ref{Bpara:dse} for B parameter, is given by
\begin{equation} \label{def:srnmass}
 m^{2}_{s,r}= (1+A)(m^{2}_{s}+2 \lambda \sigma_{s} (1+\alpha))
\end{equation}
while Higgs' mass is kept fixed, as mentioned before.
\par
Results for scalar masses are shown in figures \ref{fig:slms1} and \ref{fig:shms1} for $m_{h}=125.09$ GeV and $m_{h}=160.0$ GeV, respectively, at various cutoff values, against several scalar bare masses and couplings. Ratios between squared renormalized scalar mass and the corresponding bare squared scalar mass are plotted in the figures.
\par
For the case of $m_{h}=125.09$ GeV, a general observation is that for higher scalar masses the ratio is close to 1, indicating absence of any significant beyond tree level contributions to physical scalar mass. It is understandable as it is already known that heavier masses usually have smaller contributions beyond the bare mass values in quantum field theories. The ratio is not found to have any significant dependence on couplings in this region.
\par
However, suppression of the ratio occurs for smaller values of $m_{s}$ which is the result of significant negative contributions from Yukawa interaction in the model. It supports the speculation made in section \ref{sec:sprs} regarding negative renormalized squared scalar masses. For higher cutoff values the effect is found to be more pronounced towards small values ($m_{s}<80$ GeV) of scalar bare masses.
\par
For heavier Higgs case, the quantitative behavior of the ratio is similar, see figure \ref{fig:shms1}. It suggests that at least the Higgs mass in the vicinity of electroweak scale may not effect scalar masses, and the parameter space may not be as sensitive to Higgs mass as it is for scalar bare mass.
\section{Conclusion} \label{sec:conc}
In this paper, results from a systematic studies of Wick Cutkosky's model with a three point Yukawa interaction among Higgs, Higgs bar, and scalar singlet fields are reported. The investigation is a part of understanding richer renormalizable quantum field theories with flat background and investigating their phenomenology with minimum truncations and ansatz. In the presence of implemented numerical constraints, the vertex is found to be stable over most of the parameter space. Even in the presence of only one three point interaction vertex, the model considered is found to posses a number of non-trivial features which begs for further in-depth study of the theory.
\par
Higgs propagators are found to have the simplest features. Presence of quantitative, as well as qualitative, similarity over the considered region of the parameter space raises an speculation that even if the Higgs mass was not fixed to its experimentally known value, it would not change the renormalized mass drastically in this model. It finds support from the case of the SM and several studies involving Higgs interactions. Hence, the model itself is found to be in harmony to current understanding about Higgs interactions, which also indicates robustness of the rather unconventional numerical approach taken for the study.
\par
However, scalar field is found to have a far more active role than the Higgs. The scalar propagators are found to belong to either of the two regimes in terms of their qualitative behavior. All the terms due to implementation of renormalization are found to have certain dependence in the parameter space. The interesting feature of negative squared scalar masses is a result of sensitivity of the model to the involved parameters. As, it is taken as a sign of symmetry breaking, it makes the model highly interesting for further investigation, particularly the region with extremely small scalar bare masses, where such effects are pronounced, and dynamic mass generation in the model.
\par
Presence of suppressed vertices is another interesting feature of the model. However, the vertices remain many orders of magnitude higher than the preselected precision for computations. Hence, interactions can not be concluded as trivial for the model at this point. However, presence of at least two classifications of the vertices in terms of parameter space and the cutoff values begs for further investigation of the theory, as these signs can also be due to a certain structure in the model's parameter space. As there is no sign of abrupt change in the qualitative behavior, a second order phase transition is expected in the model if there is a indeed an structure in the parameter space.
\par
As the vertices, particularly for higher scalar bare masses, show similarity to the tree level structure up to some constant over the region where perturbative picture develops rather clearly in the other correlation functions, and as there is no undesired fluctuations in the vertices, it increases confidence on the numerical approach used to extract the correlation functions from the model, and hence it also warrants further investigation from this perspective.
\section{Acknowledgment} \label{sec:ackn}
I am extremely thankful to Dr. Shabbar Raza for a number of important advices and valuable discussions throughout the endeavor. I would also like to thank Dr. Babar A. Qureshi for discussions in the early phase of this work, Prof. Holger Gies and Dr. M. Sabieh Anwar for valuable suggestions during writing of the manuscript, and my PhD adviser Prof. Axel Maas for introducing DSE approach to me in the first place.
\par
This work was supported by faculty research grant from Habib University Karachi Pakistan, Lahore University of Management Sciences Pakistan, and partially supported by National Center for Physics Quaid-i-Azam University Islamabad Pakistan.
\bibliographystyle{unsrt}
\bibliography{bib}
\end{document}